%% file: main.tex
\pdfoutput=1

\documentclass{article}
\usepackage{natbib} 
\usepackage[utf8]{inputenc}
\usepackage{gensymb}
\usepackage{hyperref}
\usepackage{graphicx} 
\usepackage{subcaption} 
\usepackage{amsmath, amssymb, bm}
\usepackage{units} 
\usepackage[a4paper]{geometry}
\usepackage[labelfont=bf]{caption}
\usepackage{xcolor}
\usepackage{multirow}
\usepackage{bm} 
\usepackage{authblk} 

\usepackage[flushmargin]{footmisc}

\input Macros

\definecolor{turquoise1}{RGB}{115,221,208}
\definecolor{turquoise2}{RGB}{123,251,253}
\definecolor{pink1}{RGB}{182,38,74}
\definecolor{blue1}{RGB}{51,117,199}

\newcommand{\actlegendact}{
\setlength{\fboxsep}{5pt} 
\setlength{\fboxrule}{1pt} 
\begin{picture}(90,20)
\put(5,5){\small TP}
\put(25,10){\fcolorbox{black}{turquoise1}{\null}}
\put(45,5){\small FN}
\put(65,10){\fcolorbox{black}{black}{\null}}
\end{picture}}

\newcommand{\actlegendnot}{
\setlength{\fboxsep}{5pt} 
\setlength{\fboxrule}{1pt} 
\begin{picture}(90,20)
\put(5,5){\small TN}
\put(25,10){\fcolorbox{black}{white}{\null}}
\put(45,5){\small FP}
\put(65,10){\fcolorbox{black}{red}{\null}}
\end{picture}}

\newcommand{\actlegendapplication}{
\setlength{\fboxsep}{5pt} 
\setlength{\fboxrule}{1pt} 
\begin{picture}(70,15)
\put(0,0){$\boldsymbol{+}$}
\put(15,3){\fcolorbox{black}{yellow}{\null}}
\put(35,0){$\boldsymbol{-}$}
\put(50,3){\fcolorbox{black}{turquoise2}{\null}}
\end{picture}}

\newcommand{\Thetahatk}{\hat\Theta^{(k)}}
\newcommand{\half}{\frac{1}{2}}
\newcommand{\diag}{\text{diag}}

\begin{document}

\title{A spatial template independent component analysis model for subject-level brain network estimation and inference}
\author[1]{Amanda F. Mejia\thanks{Corresponding author, afmejia@iu.edu}}
\author[2]{David Bolin}
\author[3]{Yu Ryan Yue}
\author[1]{Jiongran Wang}
\author[4]{Brian S. Caffo}
\author[5,6]{Mary Beth Nebel}

\affil[1]{{\small Department of Statistics, Indiana University, Bloomington, IN, USA}}
\affil[2]{{\small CEMSE Division, King Abdullah University of Science and Technology, Thuwal, Saudi Arabia}}
\affil[3]{{\small Department of Information Systems and Statistics, Baruch College, The City University of New York, New York, NY, USA}}
\affil[4]{{\small Department of Biostatistics, Johns Hopkins University, Baltimore, MD, USA}}
\affil[5]{{\small Center for Neurodevelopmental and Imaging Research, Kennedy Krieger Institute, Baltimore, MD, USA}}
\affil[6]{{\small Department of Neurology, Johns Hopkins University, Baltimore, MD, USA}}
\date{}

\maketitle


\begin{abstract}
Functional brain networks are spatially segregated regions of the brain that tend to exhibit coordinated function.  Independent component analysis (ICA) is commonly applied to functional magnetic resonance imaging (fMRI) data to extract independent components (ICs) representing such networks. While ICA produces highly reliable results at the group level when applied to data from many subjects, single-subject ICA often produces inaccurate results due to high levels of noise.  Previously, we proposed template ICA (tICA), a hierarchical ICA model using empirical population priors derived from large fMRI databases. This approach is fast and results in more reliable subject-level IC estimates than dual regression, a standard alternative. However, this and other hierarchical ICA models assume unrealistically that subject effects, representing deviations from the population mean, are spatially independent. Here, we propose \textit{spatial template ICA} (stICA), which incorporates spatial process priors into the previously proposed tICA framework. This results in greater estimation efficiency of subject ICs and subject effects.  Additionally, the joint posterior distribution can be used to identify areas of significant engagement in the subject ICs or subject effects, using an excursions set approach. By leveraging spatial dependencies and avoiding the need to correct for multiple comparisons, stICA can achieve more power to detect true effects.  Yet stICA introduces computational challenges, as the likelihood does not factorize over locations.  We derive an expectation-maximization (EM) algorithm employing numerical optimization and sparse matrix techniques to obtain maximum likelihood estimates of the model parameters and posterior moments of the ICs and subject effects.  The stICA framework is applied to simulated data using a template consisting of 3 ICs, and to fMRI data from the Human Connectome Project using a template consisting of 16 ICs.  We compare the performance of stICA with two benchmark methods, tICA and dual regression.  To assess the impact of sample size in the fMRI study, scan duration is varied from 400 time points (5 minutes) to 1200 time points (15 minutes).  We find that stICA produces ICs and subject effects that are more accurate and reliable than the benchmark methods and identifies larger and more reliable areas of engagement.  While stICA is more computationally demanding than tICA, the algorithm is quite tractable for realistic datasets: on our data, which consists of approximately 6,000 cortical locations per hemisphere measured over 1200 time points, convergence was achieved within 7 hours per hemisphere. 
\end{abstract}

\section{Introduction}

The functional organization of the human brain is a subject of intense ongoing scientific investigation.  In recent decades, the use of functional magnetic resonance imaging (fMRI) data has rapidly advanced this effort, given its non-invasive nature, relatively high spatial and temporal resolution, and rapid technological evolution. Initial fMRI studies nearly exclusively focused on identifying regions of the brain associated with specific tasks or stimuli.  Task fMRI data can be analyzed in a simple regression framework, with the use of a design matrix based on the task paradigm.  However, in the past 15 years there has been a shift towards studying the functional organization of the brain holistically, in the absence of a particular stimulus paradigm. These so-called resting-state fMRI studies are more challenging to analyze due to the absence of any information about the subject's mental state during the scan on which to base a design matrix.  Therefore, the analysis of resting-state fMRI data typically employs blind source separation algorithms. In particular, independent component analysis (ICA) has been established as a standard and reliable method of analysis for resting-state fMRI at the group level \citep{calhoun2001method, beckmann2005investigations, damoiseaux2006consistent}.

ICA aims to separate fMRI data into a set of spatially independent components (ICs), which are images of the brain representing a spatial source of signal in the data, and a set of corresponding time courses that represent the activity of each IC over time.  ICs may represent sources of neural or artifactual origin.  Neural ICs are called functional brain networks, regions of the brain that tend to activate together.  One challenge is that ICA applied to single-subject fMRI data tends to result in highly noisy IC estimates, due to high noise levels and limited scan duration. Furthermore, single-subject ICs are not easy to match across subjects. Therefore, many studies apply ICA to group-level fMRI data combined across many subjects \citep{calhoun2001method}, and single-subject IC estimates are typically based on post-ICA reconstruction procedures \citep{beckmann2009group, erhardt2011comparison}.  These procedures, while computationally convenient, are not model-based and as such tend to have suboptimal estimation efficiency and do not provide an inferential framework. 

In recent years, hierarchical ICA models have been proposed as a way to simultaneously estimate subject- and group-level ICs \citep{guo2013hierarchical, shi2016investigating}.   These are part of a growing body of literature that uses likelihood-based procedures to perform ICA in a model-based framework \citep{chen2006efficient, eloyan2013likelihood, li2016parcellation, risk2019linear}. There have also been efforts to use existing sets of group-level ICs or ``templates'' to estimate single-subject ICs based on either an optimization framework \citep{lin2010semiblind, du2013group} or a statistical one \citep{mejia2019template}.  These hierarchical and template-based ICA models provide a principled approach to single-subject IC estimation, which has been shown to result in more accurate IC estimates. Formal statistical frameworks such as those proposed by \cite{shi2016investigating} and \cite{mejia2019template} also make inference on subject-level ICs possible and allow estimation and inference of ``subject effects'', deviations of individual subjects' brain networks from group average brain networks.  One drawback of these existing models, however, is that they assume independence of the latent spatial effects across brain locations.  While helpful from a computational perspective, this assumption does not hold up in practice, as brain networks and subject effects tend to be spatially smooth.  Ignoring information shared across neighboring brain locations may result in loss of efficiency in estimated brain networks, as well as reduced power to detect engaged regions in ICs or subject effects.

Indeed, spatial dependencies are nearly universally ignored when modeling fMRI data in general.  Historically, due to the high-dimensional nature and spatial structure of fMRI data, spatial models would have been virtually impossible to estimate without overly drastic simplifications and restrictive assumptions.  Yet in recent years, advances in various domains have made spatial modeling of fMRI feasible.  These include increased computing power, advances in Bayesian computation, and advances in spatial statistics. Additionally and perhaps most importantly, ``cortical surface fMRI'' (cs-fMRI) has grown in popularity and availability since the emergence of the Human Connectome Project \citep{glasser2013minimal, van2013wu}. Historically, one of the major challenges to spatial models for fMRI has been the complex spatial dependence structure within the volume of the brain induced by cortical folding and the presence of multiple tissue classes.  In cs-fMRI, on the other hand, the cortical gray matter, which contains much of the signal of interest, is projected to a 2-dimensional surface manifold \citep{fischl1999cortical}. In this format, the spatial dependence structure is greatly simplified and can be reasonably modeled as a decreasing function of the geodesic distance along the surface \citep{fischl1999cortical}. Additionally, the number of cortical locations in each hemisphere is greatly reduced relative to the number of volume elements (voxels) in the brain, and the data can be resampled to a lower resolution without substantial loss of information. This represents a major computational advantage for analysis of cs-fMRI compared with traditional volumetric fMRI.

Due to the high-dimensional nature of fMRI data, a suitable spatial prior must have a sparse precision structure for computational tractability.  Gaussian Markov random field (GMRF) priors are a popular choice due to their convenient parametric form and simple precision structure \citep{rue2005gaussian}. While many GMRF priors are built on a regular lattice and therefore would not be directly applicable to cs-fMRI data, the somewhat recently developed stochatic partial differential equation (SPDE) GMRF priors are built on a triangular mesh, the format of cs-fMRI data \citep{lindgren2011spde, bolin:aoas11}.  These priors have been used to model task fMRI data \citep{siden2017fast, mejia2019bayesian}, but have never been applied in resting state fMRI analysis. 

Finally, a goal in many ICA-based analyses of resting-state fMRI is to obtain thresholded IC maps representing a set of brain networks.  A common approach to thresholding is to retain locations exceeding a certain z-score or surviving a group-level hypothesis test \citep{allen2011baseline}.  While this may be helpful for visualization purposes, it cannot be used to quantify the size and location of each brain network with any statistical certainty.  Using a statistical framework such as the ones proposed by \cite{shi2016investigating} or \cite{mejia2019template}, one can obtain the marginal posterior variance of each IC at each location and perform the equivalent of a hypothesis test at each location for each IC. However, since a test is performed at every location in the brain, multiple comparisons correction should be performed to control the false positive rate at a desired level. Since spatial dependencies are not accounted for in this framework, power to detect areas engaged in each IC can suffer.  In a spatial Bayesian context, the joint posterior distribution is in theory available and could be used to determine areas of engagement while accounting for spatial dependencies.  This poses some computational challenges, so spatial Bayesian models for task fMRI have traditionally resorted to using the marginal posterior distribution at each location in the brain, followed by multiple comparisons correction.  Recently, \cite{bolin2015excursion} proposed an approach to identification of excursion regions based on the joint posterior distribution, which was adopted by \cite{mejia2019bayesian} to identify activations in task fMRI analysis.  By leveraging information shared across neighboring locations, this approach tends to have greater power to detect true effects, while maintaining strict false positive control.

Here, we propose a novel spatial modeling approach for estimating subject-level ICs based on a set of existing templates, which we call spatial template ICA (stICA).  We develop a computational approach based on expectation-maximization (EM) to obtain maximum likelihood estimates (MLEs) of model parameters and posterior mean and variance estimates of the latent fields.  We adopt an excursions set approach to identify areas of engagement in each IC and areas of deviation (from the group average) in each subject effect map. We address several computational challenges arising from the large scale of the problem: a typical number of brain networks of interest may range from 10 to 100, and the number of cortical locations in each hemisphere, based on a reasonable amount of resampling, is 5,000 to 10,000 or more. This results in computations involving matrices ranging from 50,000$^2$ to 1,000,000$^2$ in size.  As such computations are only tractable with sparse matrices, we must take care when performing the necessary steps to obtain the MLEs and posterior quantities of interest. One additional challenge is the need for numerical optimization to obtain the MLE of one of the model parameters, which could be very slow without a careful approach to computation.  We describe our computational approach in detail below, and find through both simulation studies and real fMRI data analysis that time and memory requirements tend to be reasonable.

The remainder of this paper is organized as follows. Section \ref{sec:methods} describes the stICA model, the EM-based estimation procedure, and the excursions set approach used to identify areas of engagement. Section \ref{sec:simulation} presents a simulation study designed to assess the accuracy of the proposed methods and their power to detect engaged regions. We compare the performance of stICA with that of two baseline methods: template ICA with a spatial independence assumption (tICA) and dual regression, a popular but ad-hoc method to obtain subject-level ICs based on a set of existing group ICs.  Section \ref{sec:application} presents a study of resting-state fMRI data from the Human Connectome Project (HCP) \citep{van2013wu}.  We apply the proposed and benchmark methods to two sessions of data from one randomly selected subject and assess the agreement of the estimated subject-level features across sessions using each method.  We also present the results visually, focusing on one IC representing an attention brain network, to provide a qualitative illustration of the benefits of stICA over existing methods. Finally, we conclude with a brief discussion in Section \ref{sec:discussion}.

\section{Methods}
\label{sec:methods}

Consider a set of $L$ functional brain networks of interest.  These are identified by the researcher a-priori by applying ICA with a specified model order $\geq L$ to data from a group of subjects, then identifying the ICs that represent brain networks of interest. This process requires scientific expertise and often involves adjusting the model order to achieve a desired granularity of the brain networks.  Since we are able to estimate these brain networks with a high degree of accuracy using data from a large group of subjects, we assume the population average ICs to be known.  The between-subject variance within each IC can also be estimated accurately using data from a large group of subjects as in \cite{mejia2019template}; we therefore also assume the between-subject variance of each IC at each brain location to be known.  We refer to the set of known population mean and between-subject variance maps as a \textit{template}.  Here, we describe spatial template ICA (stICA), which models subject-specific ICs as random deviations from the known population averages, and assumes spatial priors on these deviations to leverage their inherent smoothness. Below, we present the stICA model, detail the proposed EM-based estimation approach, and finally describe the excursions set approach used to identify regions of ``engagement'' in each IC or ``deviation'' in the corresponding subject effect maps.  Note that subject-level fMRI data may also contain additional spatial sources of neural or artifactual origin, which combine with the template brain networks to form the observed data; we describe how these are accounted for in Section \ref{sec:Preprocessing}. 

\subsection{The Spatial Template ICA Model}
Let $\bfs(v)=[s_{1}(v),\dots,s_L(v)]'$ 
denote the intensity of the subject-level ICs at brain location $v$, $v=1,\dots V$, which we wish to estimate, and let $\bfs_0(v)$ denote the corresponding known population mean intensities. Let $\dot{\bfy}(v)$ $(T\times 1)$ be the observed fMRI timeseries at location $v$, where $T\geq L$ is the number of time points.  Then the first level of the spatial template ICA (stICA) model is given by
\begin{equation}\label{eqn:eqn1}
    \dot{\bfy}(v) = \dot{\bfM} \bfs(v) + \dot{\bfe}(v),\quad \dot{\bfe}(v) \stackrel{iid}{\sim} N(\bfzero, \nu_0^2 \bfI_T),
\end{equation}

where $\dot{\bfM}$ is the $T\times L$ mixing matrix that describes how each spatial source signal activates over time to result in the overall composite neural activity observed during the fMRI scan.  The standard template ICA (tICA) model proposed by \cite{mejia2019template} assumes that each subject-level IC $\ell=1,\dots,L$ is related to the corresponding population ICs as $s_\ell(v) = s_{0\ell} + \delta_{\ell}(v)$, $\delta_{\ell}(v)\sim N\left(0, \sigma_\ell^2(v)\right)$, where $s_{0\ell}$ and $\sigma_\ell^2(v)$ are known via the template.   The stICA model further assumes that each latent field of deviations $\bfdelta_\ell=(\delta_\ell(1),\dots,\delta_\ell(V))$ exhibits spatial dependence. Therefore, letting $\bfs_\ell=(s_\ell(1),\dots,s_\ell(V))$, the second level of the stICA model for IC $\ell=1,\dots,L$ is defined by 
\begin{equation*}
    \bfs_{\ell} = \bfs_{0\ell} + \bfdelta_\ell, \quad 
    \bfdelta_\ell \sim N\left(\bfzero, \bfD_\ell\bfR_\ell\bfD_\ell\right),
\end{equation*} 
where $\bfs_{0\ell}$ and $\bfD_\ell=\diag\{\sigma_\ell(v):v=1,\dots,V\}$ are known through the template, and $\bfR_\ell$ describes the spatial dependence among the elements of the deviations or subject effects $\bfdelta_\ell$.  More details on the spatial dependence structure of $\bfdelta_\ell$ are provided in Section \ref{sec:Spatial_process_priors} below.

\subsubsection{Preprocessing}
\label{sec:Preprocessing}


Several preprocessing steps must be performed prior to estimation of the stICA model: centering and scaling, removal of nuisance ICs, and dimension reduction. We describe each of these in turn.

We first center the fMRI data $\dot{\bfY}=[\dot{\bfy}(1),\dots,\dot{\bfy}(V)]$ ($T\times V$) across both time and space to remove the mean image and global signal.  This is required for dual regression \citep{beckmann2009group}, a popular ad hoc method for subject-level IC estimation, which we use to obtain parameter starting values. We also (optionally) introduce standard units (as fMRI is unit-less) by scaling by the global image standard deviation, defined as the square root of variance across the image, averaged over time.

We next remove nuisance ICs, defined as ICs that are present in the subject-level data but are not found in the template.  We adapt an approach proposed in \cite{mejia2019template}: first, we use dual regression to obtain an initial estimate of the subject-level template ICs and mixing matrix, which are then subtracted from the data.  Second, we estimate the number of remaining or nuisance ICs using the penalized semi-integrated likelihood (PESEL) method proposed by \cite{sobczyk2017bayesian} using the \texttt{pesel} R package \citep{pesel}.  
Third, we estimate the nuisance ICs using the infomax ICA algorithm of \cite{bell1995information} as implemented in the \texttt{ica} R package \citep{ica}. Fourth, we subtract the estimated nuisance ICs and their mixing matrix from the original centered and scaled data.  This process can be performed iteratively, alternating between template IC estimation and nuisance IC estimation, if adequate computation time is available.






The final preprocessing step is dimension reduction based on singular value decomposition (SVD) of $\dot{\bfY}$ as in \cite{beckmann2003general} and \cite{mejia2019template}. This involves pre-multiplying by a $L\times T$ matrix $\bfH=\bfDelta\bfU$, where $\bfDelta = \text{diag}\{(d_\ell^2-\hat\nu_0^2)^{-1/2}\}_{\ell=1,\dots,L}$, $d_\ell^2$ being the $\ell$th singular value of $Cov(\dot{\bfY})$, $\hat\nu_0^2$ being the mean of the remaining $T-L$ singular values, and $\bfU$ containing its first $L$ singular vectors.  Letting ${\bfy}(v)=\bfH\dot{\bfy}(v)$, ${\bfM}=\bfH\dot{\bfM}$, ${\bfe}(v)=\bfH\dot{\bfe}(v)$, and $\bfC=\bfH\bfH'$, we have 
\begin{equation*}
    {\bfy}(v) = {\bfM} \bfs(v) + {\bfe}(v),\quad {\bfe}(v) \stackrel{iid}{\sim} N(\bfzero, \nu_0^2 \bfC).
\end{equation*}

\subsubsection{Spatial process priors}
\label{sec:Spatial_process_priors}

We assume that each $\bfdelta_\ell$, $\ell=1,\dots,L$, is generated from a stochastic partial differential equation (SPDE) spatial process \citep{lindgren2011spde, bolin:aoas11}.  SPDE priors are a class of Gaussian Markov random field priors, which are popular for high-dimensional spatial problems due to their sparse precision structure.  However, SPDE priors offer several additional benefits that make them uniquely well-suited for analysis of fMRI data. They are constructed as discretizations of continuously indexed Mat\'ern Gaussian fields, and as such have parameters with clear interpretations. Specifically, the random field $X(s), s\in \mathcal{D}$ is defined as the solution to
$(\kappa^2 - \Delta)^{\alpha/2} (\tau X) = \mathcal{W}$, where the equation is posed on the spatial domain $\mathcal{D}$, $\mathcal{W}$ is Gaussian white noise, and $\Delta$ is the Laplacian. The positive parameters $\alpha, \kappa,$ and $\tau$ respectively control the smoothness, correlation range, and variance of $X$. Note that we use the term \textit{smoothness} to refer to the degree of spatial dependence as a function of distance, as it is commonly used in neuroimaging applications, rather than to the differentiability of the field. In this work, we fix $\alpha=2$. The variance of $X$ for $\mathcal{D} = \mathbb{R}^d$ is then $\sigma^2=c_1(\kappa^2\tau^2)^{-1}$, where $c_1=1/(4\pi)$.\footnote{The variance of a general SPDE spatial process is $\Gamma(\nu)\left[\Gamma(\alpha)(4\pi)^{d/2}\kappa^{2\nu}\tau^2 \right]^{-1}$, where $\nu=\alpha-d/2$.}

The discrete model is constructed through a finite element discretization of the SPDE using a basis defined through a triangular mesh on $\mathcal{D}$. Since both the model and the discretization method can be formulated on quite general domains $\mathcal{D}$, the priors are applicable to both cortical surface fMRI data, which has a triangular mesh structure in itself, as well as cerebellar and subcortical volumetric regions, where a mesh can be constructed based on the voxels. 
Because of the possibility of extending the SPDE to generate more general models, and because of the flexibility in how the mesh is constructed, the SPDE priors are more flexible than many other classes of Gaussian Markov random fields built on a regular lattice.  SPDE priors have been previously used successfully for modeling spatial dependence of activation fields in task fMRI studies \citep{siden2017fast, mejia2019bayesian, siden2019spatial}. 

The triangular mesh consists of $N$ vertices representing the $V$ original data locations, plus additional locations that may be added to satisfy certain constraints and to control boundary effects.  Let $\bfA$ ($V\times N$) be a matrix projecting the $V$ original data locations to $N\geq V$ mesh locations.  Then we can represent $\bfdelta_{\ell}=\bfD_\ell\bfA\bfx_\ell$, where $\bfx_\ell$ follows a zero-mean, unit-variance SPDE prior. The precision matrix for the SPDE prior is sparse and takes the form $\tau^2(\kappa^4\bfF+2\kappa^2\bfG+\bfG\bfF^{-1}\bfG)$, where $\bfF$ is a known diagonal matrix and $\bfG$ is a known sparse matrix, which contains non-zero locations in cells corresponding to neighboring vertices in the triangular mesh.    Here, since $\bfx_\ell$ is assumed to have unit variance, $\tau^2=c_1/\kappa^2$. Therefore, $\bfx_\ell\sim N(\bfzero, \bfQ_\ell^{-1})$, where $\bfQ_\ell = c_1(\kappa_\ell^2\bfF+2\bfG+\kappa_\ell^{-2}\bfG\bfF^{-1}\bfG)$.

The second level of the stICA model for each IC $\ell=1,\dots,L$ is therefore given by 
\begin{equation*}
    \bfs_{\ell} = \bfs_{0\ell} + \bfdelta_\ell, \quad 
    \bfdelta_\ell \sim N\left(\bfzero, \bfD_\ell\bfR_\ell\bfD_\ell=\bfD_\ell\bfA\bfQ_\ell^{-1}\bfA'\bfD_\ell\right),
\end{equation*} 
where $\bfs_{0\ell}$ and $\bfD_\ell$ are known through the template, and $\bfA$ is known.  The spatial precision matrix of $\bfdelta_\ell$ is therefore given by $\bfSigma_\ell^{-1}=\bfD_\ell^{-1}\bfR_\ell^{-1}\bfD_\ell^{-1}$, where $\bfR_\ell = \bfA\bfQ_\ell^{-1}\bfA'$.  Note that $\bfR_\ell^{-1}$ is sparse and easily computable (see Appendix \ref{app:inverse}). 

\subsubsection{Full vector form of model}

Given that the model factorizes over locations $v$ at the first level and over ICs $\ell$ at the second level, we rewrite the model in terms of $\bfs=\left[\bfs_{1}',\dots,\bfs_L'\right]'$ ($VL\times 1$).  Let $\bfy=\left[\bfy(1)',\dots,\bfy(V)'\right]'$ and ${\bfe}=\left[{\bfe}(1)',\dots,{\bfe}(V)')\right]'$. Note that $\bfs$ is grouped by ICs, but for convenience in the first level of the model, we can regroup the elements by location using a permutation matrix $\bfP$ so that $\bfP\bfs=\left[\bfs(1)',\dots,\bfs(V)'\right]'$, where $\bfs(v)=\left[s_1(v),\dots,s_L(v)\right]'$.

Letting $\bfM_{\otimes}:=(\bfI_V \otimes {\bfM})$ and $\bfC_{\otimes}:=\bfI_V \otimes \bfC$, where $\otimes$ denotes the Kronecker product, the first level of the stICA model is given by

\begin{equation}\label{eqn:mod_full_1}
    {\bfy} = \bfM_{\otimes} \bfP \bfs + {\bfe} = (\bfI_V \otimes {\bfM}) \bfP \bfs + {\bfe},\quad {\bfe} \stackrel{iid}{\sim} N(\bfzero, \nu_0^2 \bfC_{\otimes}),
\end{equation}

and the second level is given by
\begin{equation}\label{eqn:mod_full_2}
\bfs = \bfs_0 + \bfdelta,\quad \bfdelta\sim N\left(\bfzero,\bfD\bfR\bfD\right),
\end{equation}

where $\bfs_{0}=\left[\bfs_{01}',\dots,\bfs_{0Q}'\right]'$, $\bfD=\diag\{\bfD_1,\cdots,\bfD_L\}$, and  $\bfR=\diag\{\bfR_1,\cdots,\bfR_L\}$.

\subsection{EM Algorithm}
\label{sec:EM}

In this section, we derive an expectation maximization (EM) algorithm to iteratively obtain maximum likelihood estimates (MLE) of the model parameters and maximum a posteriori (MAP) estimates of the latent variables. Considering the subject source signals as the latent variables of interest, the unknown parameters in the model are $\Theta=\left\{\bfM, \{\kappa_\ell:\ell=1,\dots,L\}\right\}$.  We will use the estimate of $\nu_0^2$ obtained through dimension reduction as described in Section \ref{sec:Preprocessing}, and hence treat it as known.

Below, we derive the expected log likelihood and the posterior moments of the latent variables (E-step), which has an explicit closed form.  We then derive the maximum likelihood estimator for $\bfM$ and present a numerical optimization approach for maximum likelihood estimation of the $\kappa_\ell$ (M-step). Although many of the matrices involved are very large ($QV\times QV$, where $Q$ is on the order of $10$ to $100$ and $V$ is on the order of 5,000 to 50,000), all necessary quantities are exactly computable using efficient linear algebra operations and taking advantage of sparsity. Therefore, the only approximation error is introduced via the numerical optimization of the $\kappa_\ell$, which is likely minimal.

\subsubsection{Expected log likelihood (E-Step)}

The likelihood can be written
\begin{equation}
    P(\bfy|\Theta) = P(\bfy|\bfs,\Theta)P(\bfs|\Theta) 
    = \left[g(\bfy:\bfM_{\otimes} \bfP \bfs, \nu_0^2 \bfC_{\otimes} )\right] 
      \left[g(\bfs_{}:\bfs_0, \bfD\bfR\bfD)\right],
\end{equation}
where $g(\cdot:\bfmu,\bfSigma)$ is the multivariate Gaussian density with mean vector $\bfmu$ and covariance $\bfSigma$.  Given the parameter estimates at iteration $k$ of the EM algorithm, $\Thetahatk$, the expected log likelihood is
\begin{equation}\label{eqn:Q}
    Q(\Theta|\Thetahatk)=E\left[\log P(\bfy|\Theta)|\bfy,\Thetahatk\right]=
    Q_1(\Theta|\Thetahatk) + Q_2(\Theta|\Thetahatk),
\end{equation}
where
\begin{equation}\label{eqn:Q1}
\begin{split}
    Q_1(\Theta|\Thetahatk)&= E\left[\log g(\bfy :\bfM_{\otimes} \bfP \bfs, \nu_0^2 \bfC_{\otimes} )|\bfy,\Thetahatk \right] \\
    &\propto - E\left[(\bfy-\bfM_{\otimes}\bfP\bfs)'(\nu_0^2 \bfC_{\otimes})^{-1}(\bfy-\bfM_{\otimes}\bfP\bfs)|\bfy,\Thetahatk\right] \\
    &\propto - E\left[ 
            -2 \bfy'\bfC_{\otimes}^{-1}\bfM_{\otimes}\bfP\bfs 
   +\text{Tr}\left(\bfP'\bfM_{\otimes}'\bfC_{\otimes}^{-1}\bfM_{\otimes}\bfP\bfs\bfs'\right)|\bfy,\Thetahatk\right] \\
    &=  2 \bfy'\bfC_{\otimes}^{-1}\bfM_{\otimes}\bfP E\left[\bfs|\bfy,\Thetahatk\right] 
   -\text{Tr}\left(\bfP'\bfM_{\otimes}'\bfC_{\otimes}^{-1}\bfM_{\otimes}\bfP E\left[\bfs\bfs'|\bfy,\Thetahatk\right]\right), \\
\end{split}
\end{equation}
and

\small 
\begin{equation}\label{eqn:Q2}
\begin{split}
    Q_2(\Theta|\Thetahatk)&=\sum_{\ell=1}^L E\left[\log g(\bfs_{\ell}:\bfs_{0\ell}, \bfD_\ell\bfR_\ell\bfD_\ell )|\bfy,\Thetahatk\right] \\
    &\propto \sum_{\ell=1}^L E\left[\log|\bfD_\ell\bfR_\ell\bfD_\ell|^{-\half}-\half(\bfs_{\ell} - \bfs_{0\ell})'(\bfD_\ell\bfR_\ell\bfD_\ell)^{-1}(\bfs_{\ell} - \bfs_{0\ell})|\bfy,\Thetahatk\right]\\
    &\propto\sum_{\ell=1}^L \log|\bfR_\ell^{-1}| - E\left[(\bfs_{\ell} - \bfs_{0\ell})'(\bfD_\ell\bfR_\ell\bfD_\ell)^{-1}(\bfs_{\ell} - \bfs_{0\ell})|\bfy,\Thetahatk\right]\\
    &= \sum_{\ell=1}^L\log|\bfR_\ell^{-1}| - E\left[\text{Tr}\left\{(\bfD_\ell\bfR_\ell\bfD_\ell)^{-1}\bfs_{\ell}\bfs_{\ell}'\right\} - \bfs_{0\ell}'(\bfD_\ell\bfR_\ell\bfD_\ell)^{-1}\left(2\bfs_{\ell} -\bfs_{0\ell}\right)|\bfy,\Thetahatk\right]\\
    &= \sum_{\ell=1}^L \log|\bfR_\ell^{-1}| - \text{Tr}\left\{\bfD_\ell^{-1}\bfR_\ell^{-1}\bfD_\ell^{-1} E\left[\bfs_{\ell}\bfs_{\ell}'|\bfy,\Thetahatk\right]\right\} + \bfs_{0\ell}'\bfD_\ell^{-1}\bfR_\ell^{-1}\bfD_\ell^{-1}\left(2E\left[\bfs_{\ell}|\bfy,\Thetahatk\right] - \bfs_{0\ell}\right), 
\end{split}
\end{equation}
\normalsize
where $\text{Tr}(\cdot)$ denotes the trace operator, and the last lines in (\ref{eqn:Q1}) and (\ref{eqn:Q2}) follow from the invariance of the trace to cyclic permutations and the linearity of the trace operator.

\subsubsection{Posterior moments of latent variables (E-Step)}
\label{sec:E_step}

To determine the MLEs of the unknown parameters, the expected log likelihood is to be maximized. For this, we must determine the form of the posterior moments appearing in $Q_1(\Theta|\Thetahatk)$ and $Q_2(\Theta|\Thetahatk)$, namely $E[\bfs|\bfy,\Thetahatk]$, $E[\bfs\bfs'|\bfy,\Thetahatk]$, $E[\bfs_{\ell}|\bfy,\Thetahatk]$ and $E[\bfs_{\ell}\bfs_{\ell}'|\bfy,\Thetahatk]$.  We therefore derive the posterior distribution of $\bfs$, conditional on the current estimates of the parameters, $\Thetahatk$, which is Gaussian.  

\small
\begin{equation}\label{eqn:post_s}
\begin{split}
    p\left(\bfs \middle| \bfy,\Thetahatk\right) 
    &\propto p\left(\bfy \middle| \bfs,\Thetahatk\right)
             p\left(\bfs \middle| \Thetahatk\right) \\
    &=g\left(\bfy:\bfM_{\otimes}\bfP\bfs,\nu_0^2\bfC_{\otimes}\right)
      g\left(\bfs:\bfs_0,\bfD\bfR\bfD\right) \\
    &\propto \exp\left\{-\half\Big((\bfy-\bfM_{\otimes}\bfP\bfs)'(\nu_0^2\bfC_{\otimes})^{-1}(\bfy-\bfM_{\otimes}\bfP\bfs)
    +(\bfs-\bfs_0)'(\bfD\bfR\bfD)^{-1}(\bfs-\bfs_0)\Big)\right\}\\
    &\propto \exp\left\{-\half\Big(-2\bfs'\bfP'\bfM_{\otimes}'(\nu_0^2\bfC_{\otimes})^{-1}\bfy - 2\bfs'(\bfD\bfR\bfD)^{-1}\bfs_0 +
    \bfs'(\bfD\bfR\bfD)^{-1}\bfs + \bfs'\bfP'\bfM_{\otimes}'(\nu_0^2\bfC_{\otimes})^{-1}\bfM_{\otimes}\bfP\bfs\Big)\right\}\\
    &=\exp\left\{-\half\Big(-2\bfs'\big[\bfP'\bfM_{\otimes}'(\nu_0^2\bfC_{\otimes})^{-1}\bfy + (\bfD\bfR\bfD)^{-1}\bfs_0\big] +
    \bfs'\big[(\bfD\bfR\bfD)^{-1}+\bfP'\bfM_{\otimes}'(\nu_0^2\bfC_{\otimes})^{-1}\bfM_{\otimes}\bfP\big]\bfs\Big)\right\} \\
    &\propto g\big(\bfs:\bfmu_{s|y},\bfSigma_{s|y}\big),
\end{split}
\end{equation}

\normalsize
where, defining $\bfOmega:=\bfR^{-1}+\bfD\bfP'\bfM_{\otimes}'(\nu_0^2\bfC_{\otimes})^{-1}\bfM_{\otimes}\bfP\bfD$ and $\bfm:=\bfD\bfP'\bfM_{\otimes}'(\nu_0^2\bfC_{\otimes})^{-1}\bfy +
\bfR^{-1}\bfD^{-1}\bfs_0$, 
\begin{equation}
\begin{split}
\bfSigma_{s|y} &= \big(\bfD^{-1}\bfR^{-1}\bfD^{-1}+\bfP'\bfM_{\otimes}'(\nu_0^2\bfC_{\otimes})^{-1}\bfM_{\otimes}\bfP\big)^{-1} \\
&=\bfD\big(\bfR^{-1}+\bfD\bfP'\bfM_{\otimes}'(\nu_0^2\bfC_{\otimes})^{-1}\bfM_{\otimes}\bfP\bfD\big)^{-1}\bfD \\
&=\bfD\bfOmega^{-1}\bfD \\
\bfmu_{s|y} &= \bfSigma_{s|y}\big[\bfP'\bfM_{\otimes}'(\nu_0^2\bfC_{\otimes})^{-1}\bfy + \bfD^{-1}\bfR^{-1}\bfD^{-1}\bfs_0\big] \\
&= \bfD\bfOmega^{-1}\big[\bfD\bfP'\bfM_{\otimes}'(\nu_0^2\bfC_{\otimes})^{-1}\bfy +
\bfR^{-1}\bfD^{-1}\bfs_0\big] \\
&= \bfD\bfOmega^{-1}\bfm
\end{split}
\end{equation}

We can compute $\bfm$, since $\bfD$, $\bfP$ and $\bfC_{\otimes}$ are known sparse matrices, and $\bfR^{-1}$ and $\bfM_{\otimes}$ are fixed sparse matrices given fixed values of the parameters.  We can then compute $\bfOmega^{-1}\bfm$ by solving for $\bfx$ in the system of linear equations $\bfOmega\bfx = \bfm$, and premultiply by $\bfD$ to obtain $\bfmu_{s|y}$.

The first and second posterior moments of $\bfs$ appearing in $L(\Theta|\Thetahatk)$ are given by
\begin{equation}\label{eqn:post_sv}
    \begin{split}
        E[\bfs|\bfy,\Thetahatk] &= \bfmu_{s|y} \\
        E[\bfs\bfs'|\bfy,\Thetahatk] &= \bfmu_{s|y}\bfmu_{s|y}' + \bfSigma_{s|y}.
    \end{split}
\end{equation}

The posterior moments of $\bfs_\ell$ can then be obtained by defining $\bfE_\ell$ as a $V\times VL$ matrix that equals zero everywhere except the $\ell$th column block of size $V$, which equals $\bfI_V$, so that $\bfs_\ell = \bfE_\ell\bfs$ and
\begin{equation}\label{eqn:post_svq}
    \begin{split}
        E[\bfs_\ell|\bfy,\Thetahatk] &= \bfE_\ell\bfmu_{s|y} \\
        E[\bfs_\ell\bfs_\ell'|\bfy,\Thetahatk] &= \bfE_\ell\left(\bfmu_{s|y}\bfmu_{s|y}' + \bfSigma_{s|y}\right)\bfE_\ell'.
    \end{split}
\end{equation}
                              
Computation of $\bfSigma_{s|y}$ would involve inverting a $VL\times VL$ matrix, which is not computationally feasible. However, computation of $\bfSigma_{s|y}$ is not required to obtain the posterior means and parameter estimates. In the next section, we derive MLEs of each of the parameters and describe the computational strategy for computing terms involving large matrix inverses.

\subsubsection{MLEs of parameters (M-Step)}

We now derive the maximum likelihood estimators (MLEs) of the model parameters $\Theta=\left\{\bfM, \{\kappa_\ell\}_{\ell=1,\dots,L}\right\}$.  Note that from equations (\ref{eqn:Q1}) and (\ref{eqn:Q2}), $\bfM$ only appears in $Q_1(\Theta|\Thetahatk)$, while the $\kappa_\ell$ only appear in $Q_2(\Theta|\Thetahatk)$ (via $\bfR_\ell$).  

\textit{Mixing matrix.} First, note that $Q_1(\Theta|\Thetahatk)$ factorizes over locations $v=1,\dots,V$, since $\bfM_{\otimes}=\bfI_V\otimes\bfM$ and $\bfC_{\otimes}=\bfI_V\otimes\bfC$ are block diagonal matrices with $V$ blocks.  Letting $\bft=\bfP E\left[\bfs|\bfy,\Thetahatk\right]$ and $\bfT=\bfP E\left[\bfs\bfs'|\bfy,\Thetahatk\right]\bfP'$, we can rewrite $Q_1(\Theta|\Thetahatk)$ as:

\begin{equation}
\begin{split}
    Q_1(\Theta|\Thetahatk)
    &\propto  2 \bfy'\bfC_{\otimes}^{-1}\bfM_{\otimes}\bfP E\left(\bfs|\bfy,\Thetahatk\right)
   -\text{Tr}\left\{\bfM_{\otimes}'\bfC_{\otimes}^{-1}\bfM_{\otimes}\bfP E\left[\bfs\bfs'|\bfy,\Thetahatk\right]\right\} \\
   &= \sum_{v=1}^V \Big[ 2 \bfy(v)'\bfC^{-1}\bfM\bft(v) 
   -\text{Tr}\left\{\bfM'\bfC^{-1}\bfM\bfT(v,v)\right\}
   \Big],
\end{split}
\end{equation}

where $\bft(v)$ denotes the $v$th set of $L$ entries of $\bft$ (those corresponding to location $v$), and $\bfT(v,v)$ denotes the $v$th diagonal block of $\bfT$ of size $L\times L$.

Then, setting 
\begin{equation*}
\begin{split}
    \frac{\partial Q_1}{\partial\bfM} &=
    \sum_{v=1}^V \Big[
    2 \bfC^{-1} \bfy(v)\bft(v)' 
    - 2 \bfC^{-1} \bfM\bfT(v,v)
    \Big]\\
\end{split}    
\end{equation*}
to zero, we have
\begin{equation}\label{eqn:Mhat}
    {\hat\bfM} = \Big(\sum_{v=1}^V\bfy(v)\bft(v)'\Big)
    \Big(\sum_{v=1}^V \bfT(v,v) \Big)^{-1}.
\end{equation}

See Appendix \ref{app:Mhat} for details on the computation of $\hat\bfM$.

\textit{SPDE Parameters.} Turning our attention to $\kappa_\ell$, let 
$$
Q_2(\Theta|\Thetahatk) = Q_2^{(1)}(\Theta|\Thetahatk) - Q_2^{(2)}(\Theta|\Thetahatk) + Q_2^{(3)}(\Theta|\Thetahatk),
$$
where
\begin{equation}
    \begin{split}
Q_2^{(1)}(\Theta|\Thetahatk) &= \sum_{\ell=1}^L \log|\bfR_\ell^{-1}| \\
Q_2^{(2)}(\Theta|\Thetahatk) &= \sum_{\ell=1}^L
\text{Tr}\left\{\bfD_\ell^{-1}\bfR_\ell^{-1}\bfD_\ell^{-1} E\left[\bfs_{\ell}\bfs_{\ell}'|\bfy,\Thetahatk\right]\right\} \\
Q_2^{(3)}(\Theta|\Thetahatk) &= \sum_{\ell=1}^L \bfs_{0\ell}'\bfD_\ell^{-1}\bfR_\ell^{-1}\bfD_\ell^{-1}\left(2E\left[\bfs_{\ell}|\bfy,\Thetahatk\right] - \bfs_{0\ell}\right).
    \end{split}
\end{equation}

For ease of notation, let $\bfb_\ell$ represent the $\ell$th set of $V$ elements of a $VL$-vector $\bfb$, and $\bfB_{\ell,\ell}$ represent the $\ell$th $V\times V$ diagonal block of a $VL\times VL$ matrix $\bfB$.  Let $\hat\bfSigma_{s|y}$, $\hat\bfmu_{s|y}$, $\hat\bfOmega$ and $\hat\bfm$ represent the values of each term as defined in Section \ref{sec:E_step} plugging in the current parameter estimates $\Thetahatk$.  Define $\hat\bfW = \hat\bfOmega^{-1}\hat\bfm\hat\bfm'\hat\bfOmega^{-1}$, $\bfu=\bfD^{-1}\bfs_{0}$ and $\hat\bfv=2\hat\bfOmega^{-1}\hat\bfm - \bfD^{-1}\bfs_{0}$.  Note that $\hat\bfOmega^{-1}\hat\bfm$ and $\bfD^{-1}\bfs_0$ are computed in the E-step, so $\hat\bfW$,  $\bfu$ and $\bfv$ are all available.  Then we can write 
\begin{equation}
    \begin{split}
Q_2^{(2)}(\Theta|\Thetahatk) 
&= \text{Tr}\left\{\bfD^{-1}\bfR^{-1}\bfD^{-1} E\left[\bfs\bfs'|\bfy,\Thetahatk\right]\right\} \\
&= \text{Tr}\left\{\bfD^{-1}\bfR^{-1}\bfD^{-1} \left[\hat\bfSigma_{s|y}+\hat\bfmu_{s|y}\hat\bfmu_{s|y}'\right]\right\} \\
&= \text{Tr}\left\{\bfD^{-1}\bfR^{-1}\bfD^{-1} \left[\bfD\hat{\bfOmega}^{-1}\bfD+\bfD\hat\bfOmega^{-1}\hat\bfm\hat\bfm'\hat\bfOmega^{-1}\bfD\right]\right\} \\
&= \text{Tr}\left\{\bfR^{-1} \left[\hat{\bfOmega}^{-1}+\hat\bfW\right]\right\} \\
&= \sum_{\ell=1}^L
\text{Tr}\left\{\bfR_\ell^{-1} (\hat\bfOmega^{-1})_{\ell,\ell}\right\} + \text{Tr}\left\{\bfR_\ell^{-1} \hat\bfW_{\ell,\ell}\right\},
    \end{split}
\end{equation}
and
\begin{equation}
    \begin{split}
Q_2^{(3)}(\Theta|\Thetahatk) 
&= \bfs_{0}'\bfD^{-1}\bfR^{-1}\bfD^{-1}\left(2E\left[\bfs|\bfy,\Thetahatk\right] - \bfs_{0}\right) \\
&= \bfs_{0}'\bfD^{-1}\bfR^{-1}\left(2\hat\bfOmega^{-1}\hat\bfm - \bfD^{-1}\bfs_{0}\right) \\
&= \bfu'\bfR^{-1}\hat\bfv \\
&= \sum_{\ell=1}^L \bfu'_\ell\bfR_\ell^{-1}\hat\bfv_\ell.
    \end{split}
\end{equation}
 
Recall that $\bfR_\ell^{-1}$ is sparse and easily computable, as described in Appendix \ref{app:inverse}.
For each $\kappa_\ell$, $\ell=1,\dots,L$, the function to be maximized is

\begin{equation}\label{eqn:opt_kappa_l}
f_\ell(\kappa_\ell|\Thetahatk):=
\log|\bfR_\ell^{-1}| -
\text{Tr}\left\{\bfR_\ell^{-1} (\hat\bfOmega^{-1})_{\ell,\ell}\right\} - \text{Tr}\left\{\bfR_\ell^{-1} \hat\bfW_{\ell,\ell}\right\}
+ \bfu'_\ell\bfR_\ell^{-1}\hat\bfv_\ell.
\end{equation}

To optimize $f_\ell(\kappa_\ell|\Thetahatk)$ numerically, the only difficulty lies in computing the trace terms involving $(\hat\bfOmega^{-1})_{\ell,\ell}$ and $\hat\bfW_{\ell,\ell}$, both of which are dense $V\times V$ matrices.  However, we do not need to compute these matrices, but rather the trace of their product with $\bfR_\ell^{-1}$, a sparse matrix.  Note that the sparsity pattern of $\bfR_\ell^{-1}$ is common across $\ell$ and does not depend on the value of $\kappa_\ell>0$.  The computational strategy for these terms is detailed in Appendix \ref{app:kappa-hat}.

To accelerate the convergence of the EM algorithm, we adopt the squared iterative method (SQUAREM) proposed by \citep{varadhan2008simple}, which is implemented in the R package SQUAREM \cite{SQUAREM}; we use version 2020.2 with the default control parameters.  The algorithm is repeated until the Euclidean norm of the change in all model parameters falls below a given tolerance.

\subsubsection{The Common Smoothness Model and EM Algorithm}
\label{sec:common_smoothness}
It may be reasonable to assume that the $L$ spatial source signals have similar degrees of smoothness, which is controlled by the $\kappa$ parameter in the SPDE precision matrix.  Each source signal represents a functional brain area, which tend to be similar in their spatial distribution on the cortical surface at a given dimensionality $L$.  In a standard ICA setting, this may not be a reasonable assumption due to the presence of noise-related source signals capturing variability in the data related to subject head motion, respiration and other types of artifacts. The spatial properties of such components is quite distinct from those of functional brain areas. However, here such nuisance signals are not included in the model but are instead estimated and removed from the data a-priori, as described in Section \ref{sec:Preprocessing}.  Furthermore, we observe empirically that the MLEs for $\kappa_\ell$ obtained through the EM algorithm described above tend to be similar across ICs. Therefore, we may consider a simplified version of the model assuming a common $\kappa$ parameter across the $L$ ICs, which can be estimated more efficiently using the EM algorithm.

The spatial ICA model with common smoothness is given by equations (\ref{eqn:mod_full_1}-\ref{eqn:mod_full_2}), except that now $\bfR_\ell\equiv\bfR_*$.  The latent variables and unknown model parameters in this model, now just $\Theta=\{\bfM,\kappa\}$, can again be estimated through EM. The expected log likelihood given in equations (\ref{eqn:Q}-\ref{eqn:Q2}) and the posterior moments of the latent variables given in equation (\ref{eqn:post_sv}) are unchanged, except that each instance of $\bfR_\ell$ is replaced by $\bfR_*$. The MLEs of $\bfM$ is unchanged, and the MLE of $\kappa$ is now the value that maximizes 
\begin{equation}\label{eqn:opt_kappa}
    \sum_{\ell=1}^L f_\ell(\kappa_\ell|\Thetahatk)=
    Q\log|\bfR_*^{-1}| 
     - \sum_{\ell=1}^L
    \left[ \text{Tr}\left\{\bfR_*^{-1} (\hat\bfOmega^{-1})_{\ell,\ell}\right\} + \text{Tr}\left\{\bfR_*^{-1} \hat\bfW_{\ell,\ell}\right\}\right] 
     + \sum_{\ell=1}^L \bfu'_\ell\bfR_*^{-1}\hat\bfv_\ell.
\end{equation}

This can again be done through numerical optimization using the techniques described in Appendix \ref{app:kappa-hat}.

\subsection{Identifying Regions of Engagement and Deviation}
\label{sec:excursions}
After the model parameters have been estimated, the next step is to identify brain locations that are engaged in each brain network represented by an IC. Similarly, we may wish to identify brain locations in a subject IC that deviate from the population mean. The simplest way of testing whether a given location is engaged would be to perform a standard hypothesis test for the value of each field at that location. The disadvantage with this is that it would lead to a massive multiple comparisons problem. Fortunately, an advantage of the spatial Bayesian framework is that the posterior distribution of $\bfs$ (or equivalently, $\bfdelta$) accounts for spatial dependence and can be used to identify engaged brain locations in a way that avoids massive multiple comparisons. 

We adopt the excursions set method proposed by \cite{bolin2015excursion}, which efficiently estimates the set of locations exceeding a threshold $\gamma$ with joint posterior probability at least $1-\alpha$. An excursion set of a function $f(v)$ is a set of locations where $f(v)>\gamma$ for some given threshold $\gamma$, i.e., a set $A_\gamma^+(\bm{f}) = \{ w\in\Omega; f(v)> \gamma \}$. In the case when $\bm{f}$ is an indirectly observed random field, such as $\bfs$, we define the excursion set as a region where it with some (high) probability exceeds the threshold. Specifically, the excursion set with probability $\alpha$, $E_{\gamma,\alpha}^{+}(\bfs)$, is defined as 
\begin{equation}\label{def:E}
E_{\gamma,\alpha}^+(\bfs) = \arg\max_{D}\left\{|D|:P\left(D\subset A_\gamma^+(\bfs)\right)\geq 1 - \alpha \right\}.
\end{equation}
That is, it is the largest set for which the level $\gamma$ is exceeded at all locations in the set with probability $1-\alpha$ . \cite{bolin2015excursion} proposed a computationally efficient approach for finding these sets, which is implemented in the \texttt{excursions} R package \citep{bolin2016calculating}. 

In our context, we have multiple random fields $\bfs_\ell$, and can thus compute an excursion set for each of these fields in one of two ways. One approach is to compute a joint excursion set for all fields, so that the joint probability over all fields is maintained at $1-\alpha$. This set can be computed based on the joint posterior distribution for the $\bfs_\ell$ fields. An alternative approach is to compute the excursion set for each $\bfs_\ell$ separately, based on the corresponding marginal posterior. We adopt the latter approach, which means that for each latent field, the probability that not all identified locations are engaged is controlled at $\alpha$. These computations can be performed efficiently in \texttt{excursions} without first marginalizing out each $\bfs_\ell$, which would be computationally expensive.

\section{Simulation Study}
\label{sec:simulation}

We perform a simulation study to quantify the accuracy of the proposed spatial template ICA (stICA) approach and to compare its performance with two alternatives: standard template ICA (tICA) without spatial priors \citep{mejia2019template} and dual regression, a fast but ad-hoc method that is commonly used in practice \citep{beckmann2009group}. 

First, we give a brief overview of the simulation design before going into detail below. We consider $L=3$ two-dimensional brain network maps.  For each brain network, we use a generating mean and variance map to generate random draws.  Since those draws are independent across space, the random subject effects (difference from population mean) are then spatially smoothed to produce the $L=3$ subject-level ICs for each subject.  Based on the ICs from a set of training subjects, we produce templates using Monte Carlo simulation.  We then simulate fMRI timeseries for 50 test subjects and apply each method (i.e., dual regression, tICA, and stICA) to obtain estimates of the subject-level ICs and the mixing matrix representing the temporal activation profile of each IC.  For stICA and tICA, we also identify areas of engagement.  Finally, as the ``functional connectivity'' between the spatial source signals is also of interest, we compute the correlation of the temporal activation profiles of each pair of ICs to produce an $L\times L$ functional connectivity matrix.  We evaluate the performance of stICA relative to tICA and dual regression based on the accuracy of the estimated ICs, of the functional connectivity matrices, and of the areas of engagement.  

\begin{figure}
\centering
\begin{tabular}{cccc}
& IC 1 & IC 2 & IC 3 \\[4pt]
\begin{picture}(10,90)\put(0,48){\rotatebox[origin=c]{90}{Generating Mean}}\end{picture} & 
\includegraphics[height=1.6in, page=1, trim=7mm 2mm 25mm 20mm, clip]{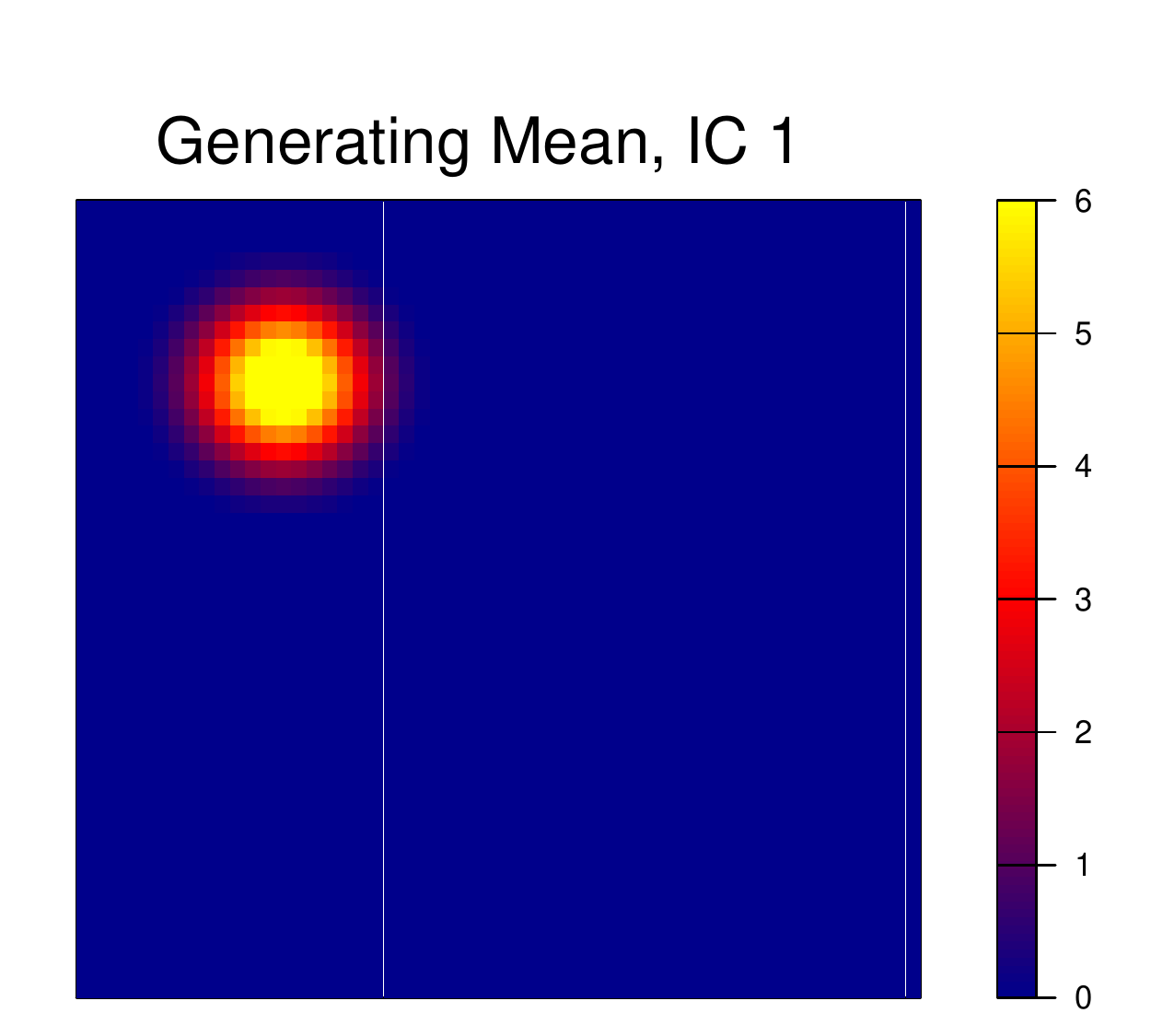} &
\includegraphics[height=1.6in, page=2, trim=7mm 2mm 25mm 20mm, clip]{simulation/plots/generating_means.pdf} &
\includegraphics[height=1.6in, page=3, trim=7mm 2mm 25mm 20mm, clip]{simulation/plots/generating_means.pdf} \\
 & \multicolumn{3}{c}{\includegraphics[width=3in, trim=2mm 5mm 17mm 5mm, clip]{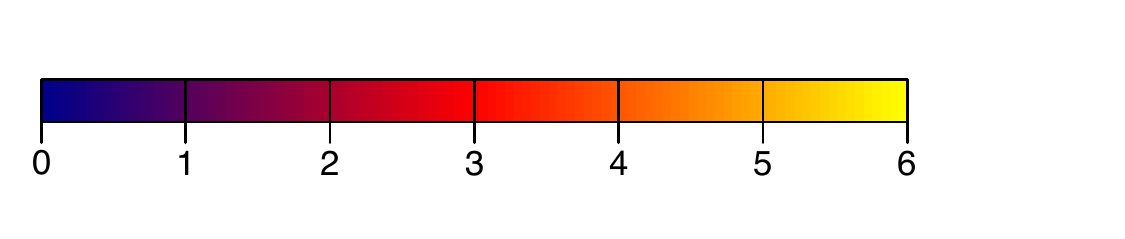}} \\[8pt]
\begin{picture}(10,90)\put(0,48){\rotatebox[origin=c]{90}{Generating Variance}}\end{picture} & 
\includegraphics[height=1.6in, page=1, trim=7mm 2mm 25mm 20mm, clip]{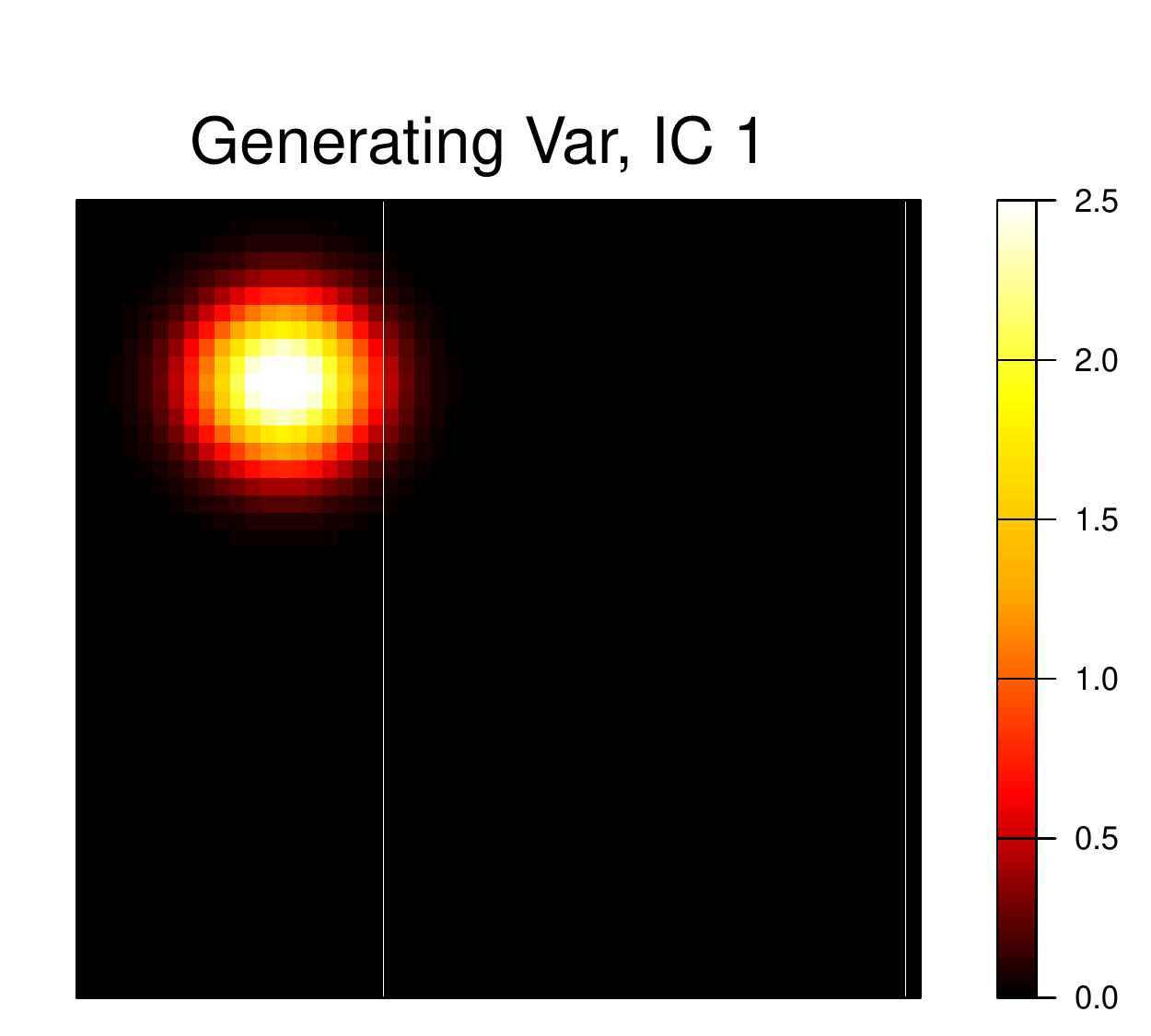} &
\includegraphics[height=1.6in, page=2, trim=7mm 2mm 25mm 20mm, clip]{simulation/plots/generating_vars.pdf} &
\includegraphics[height=1.6in, page=3, trim=7mm 2mm 25mm 20mm, clip]{simulation/plots/generating_vars.pdf} \\
 & \multicolumn{3}{c}{\includegraphics[width=3in, trim=2mm 5mm 17mm 5mm, clip]{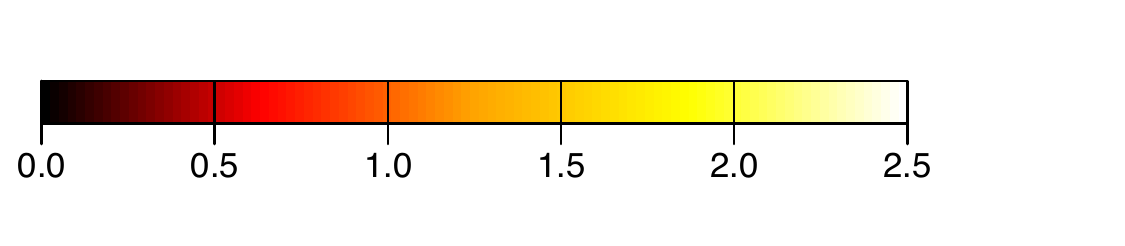}} \\[4pt]
\end{tabular}
\caption{\small Generating mean and variance maps for three ICs in simulation study.\\[40pt] }
\label{fig:sim_generating}
\end{figure}

\begin{figure}
\centering
\begin{tabular}{cccc}
& IC 1 & IC 2  & IC 3  \\[4pt]
\begin{picture}(10,90)\put(0,48){\rotatebox[origin=c]{90}{Subject Effects}}\end{picture} & 
\includegraphics[height=1.6in, page=5, trim=7mm 2mm 25mm 20mm, clip]{simulation/plots/subj1_IC1.pdf} &
\includegraphics[height=1.6in, page=5, trim=7mm 2mm 25mm 20mm, clip]{simulation/plots/subj1_IC2.pdf} &
\includegraphics[height=1.6in, page=5, trim=7mm 2mm 25mm 20mm, clip]{simulation/plots/subj1_IC3.pdf} \\
 & \multicolumn{3}{c}{\includegraphics[width=3in, trim=2mm 5mm 17mm 5mm, clip]{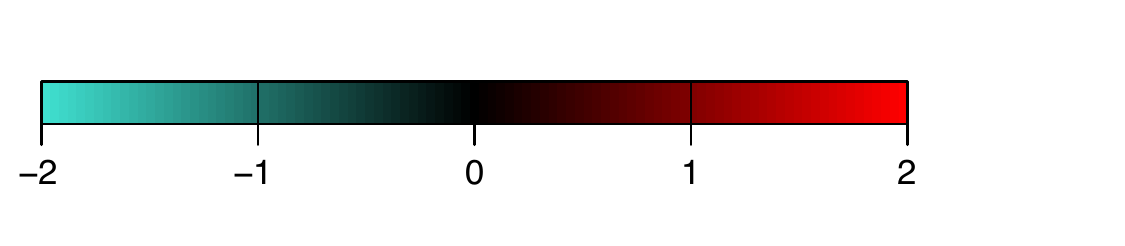}} \\[4pt]
\begin{picture}(10,90)\put(0,48){\rotatebox[origin=c]{90}{Subject ICs}}\end{picture} & 
\includegraphics[height=1.6in, page=1, trim=7mm 2mm 25mm 20mm, clip]{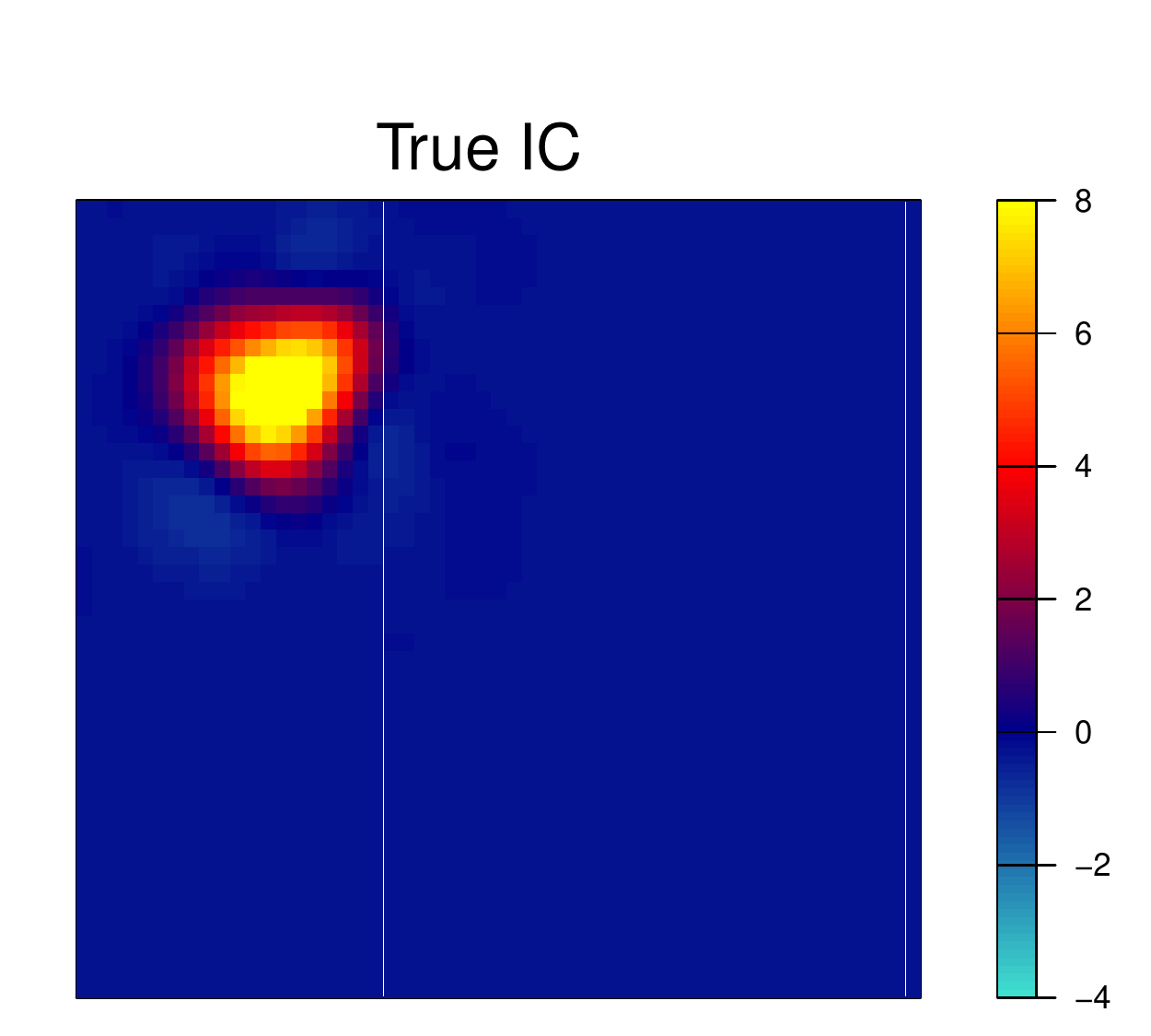} &
\includegraphics[height=1.6in, page=1, trim=7mm 2mm 25mm 20mm, clip]{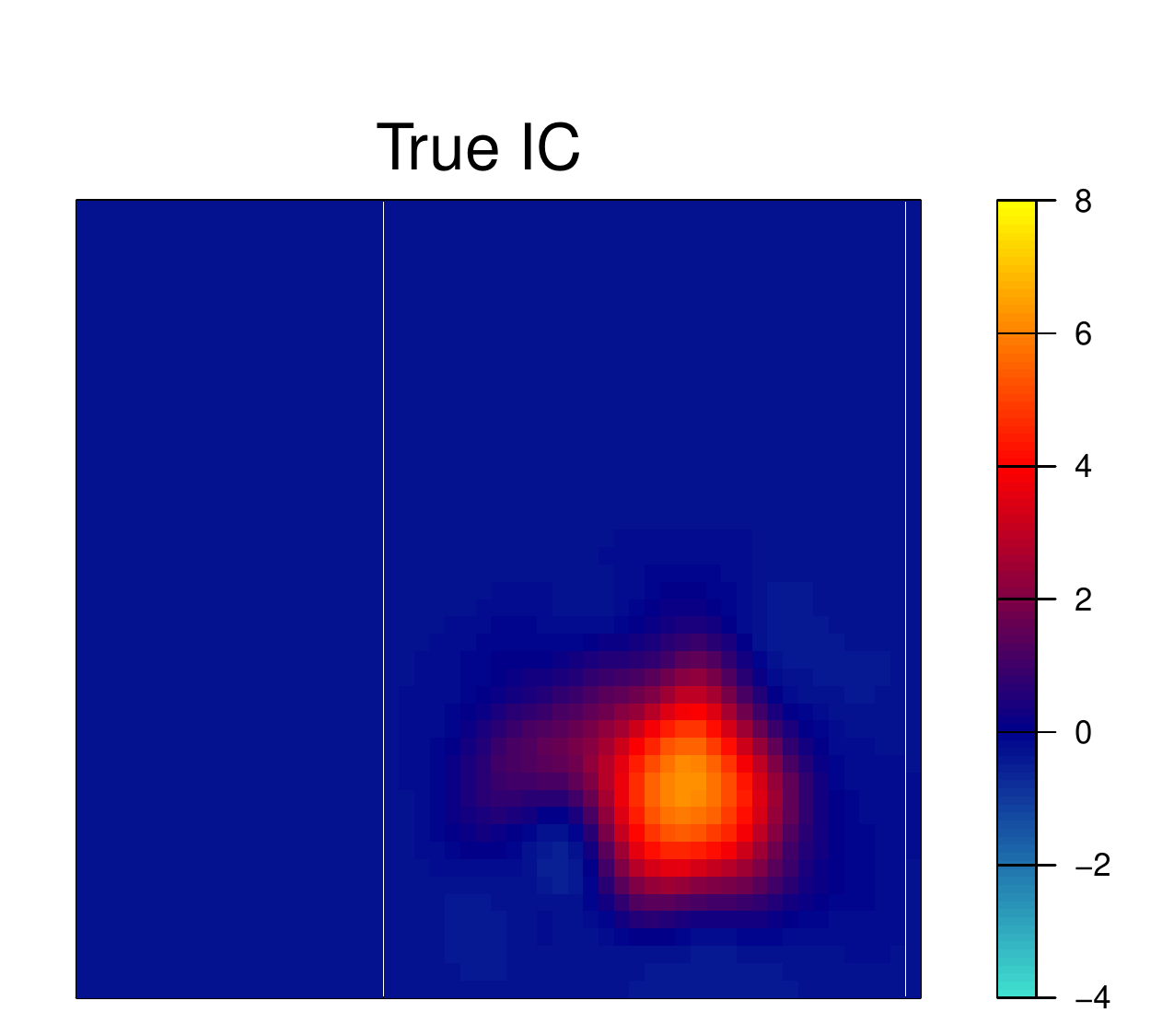} &
\includegraphics[height=1.6in, page=1, trim=7mm 2mm 25mm 20mm, clip]{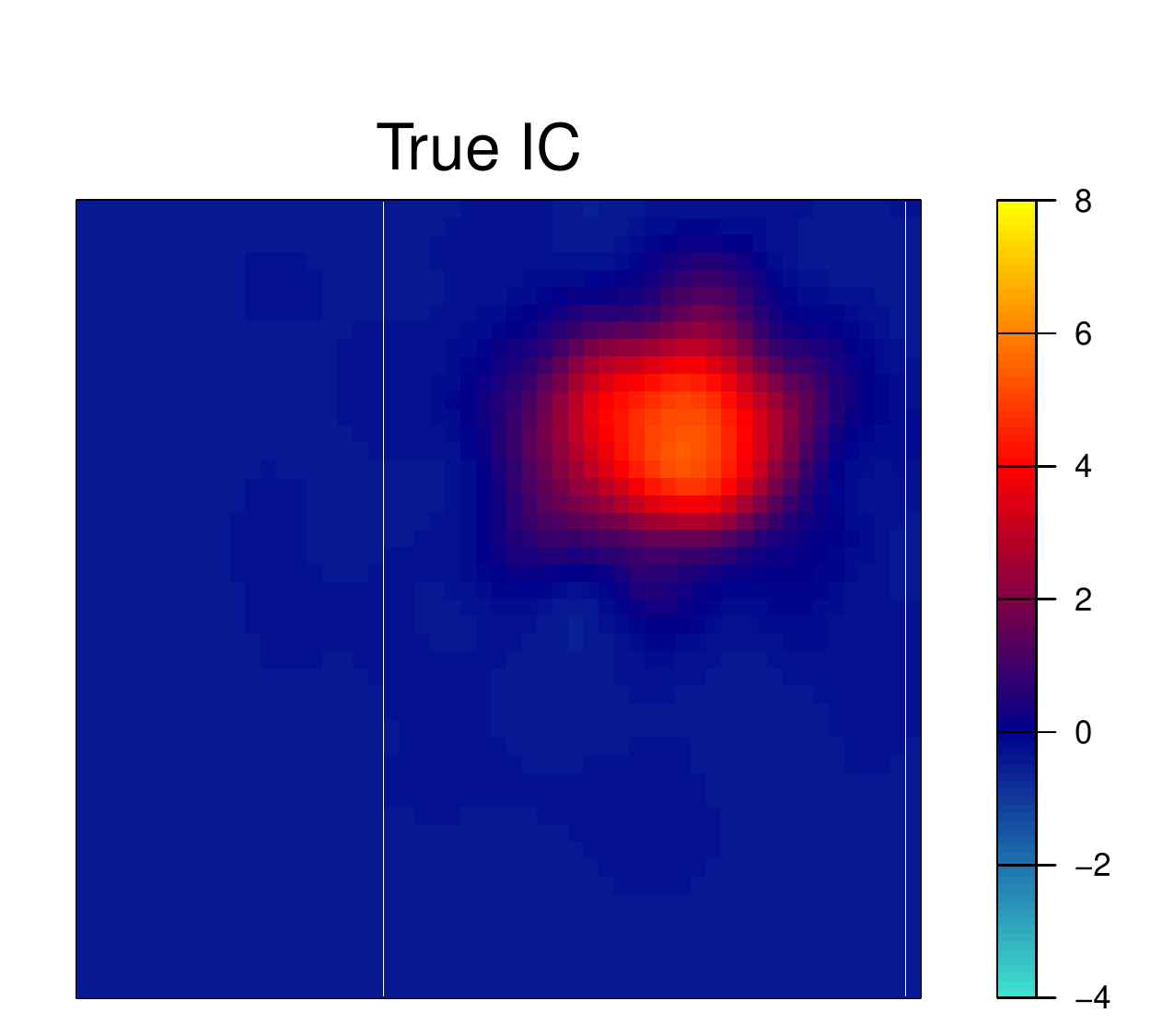} \\
 & \multicolumn{3}{c}{\includegraphics[width=3in, trim=2mm 5mm 17mm 5mm, clip]{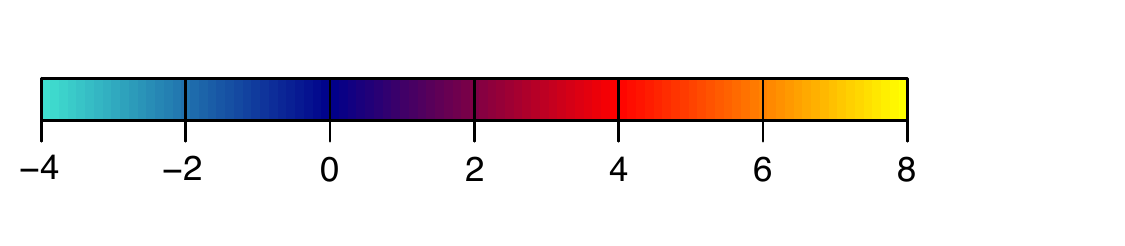}}
\end{tabular}
\caption{\small Subject effects and ICs for one example subject.}
\label{fig:sim_data}
\end{figure}

\textit{Subject-level ICs.} Each IC is a 2-dimensional image consisting of $46\times 55$ isotropic 1mm pixels for a total of $V=2530$ data locations.  The generating mean and variance maps are shown in Figure \ref{fig:sim_generating}. Each mean map is generated by placing a point mass of a given amplitude at a certain location and convolving it with a 2-dimensional Gaussian kernel.  The point mass for each IC is located at $(12,15)$, $(35,40)$ and $(15,40)$, respectively, and the full width at half maximum (FWHM) of the Gaussian kernel is 30, 40 and 45.\footnote{A FWHM of $f$ is equivalent to a standard deviation of $\sigma = f/\sqrt{8\ln(2)}$.} For each IC, the generating variance is proportional to the corresponding mean, reflecting the observation that the maps representing subjects' functional brain networks tend to vary the most in the region where that network is ``engaged'', and tend to vary little over subjects in ``background'' areas. 

\begin{figure}
\centering
\begin{tabular}{cccc}
& IC 1 & IC 2 & IC 3  \\[4pt]
\begin{picture}(10,90)\put(0,48){\rotatebox[origin=c]{90}{Template Mean}}\end{picture} & 
\includegraphics[height=1.6in, page=1, trim=7mm 2mm 25mm 20mm, clip]{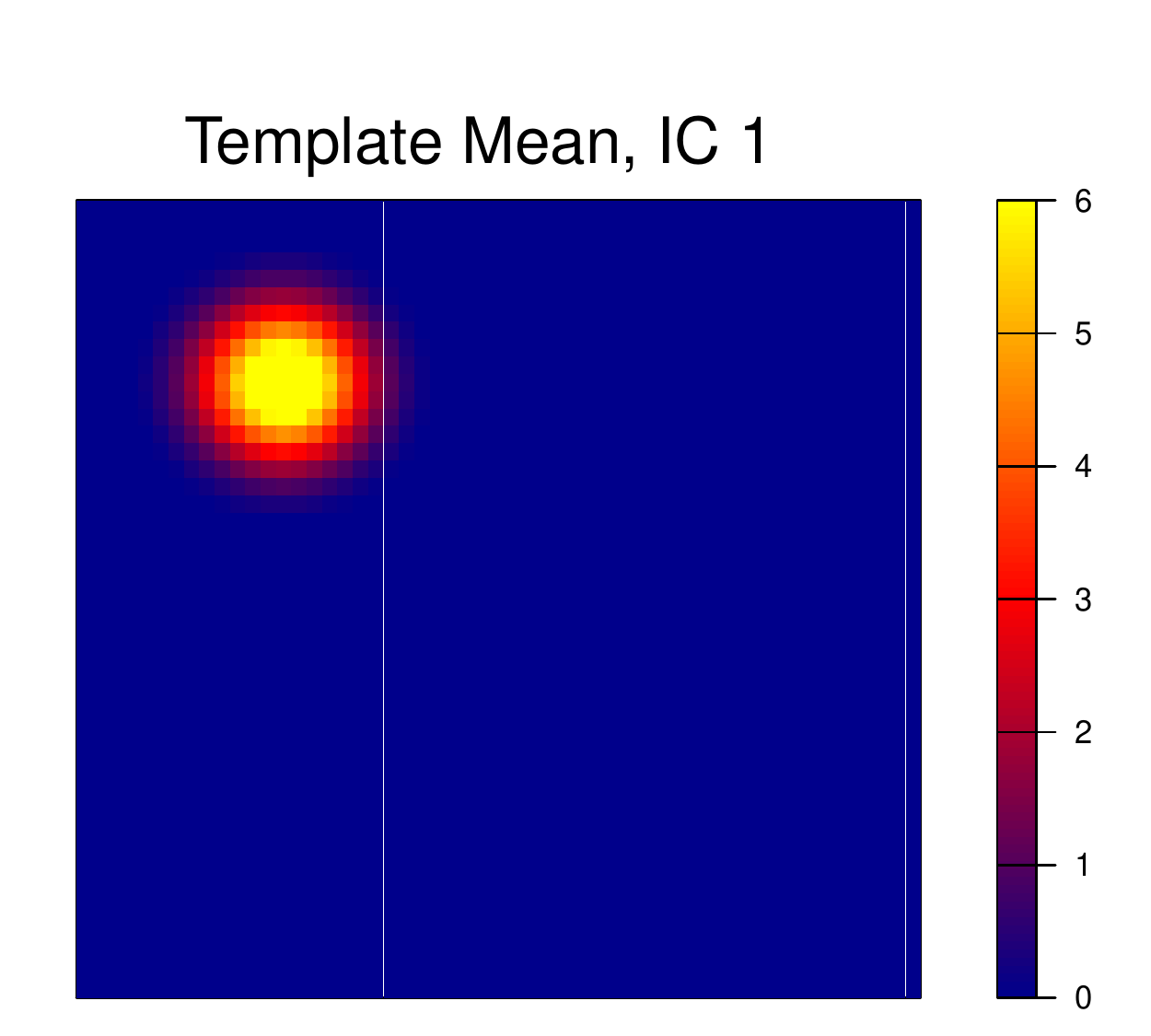} &
\includegraphics[height=1.6in, page=3, trim=7mm 2mm 25mm 20mm, clip]{simulation/plots/templates.pdf} &
\includegraphics[height=1.6in, page=5, trim=7mm 2mm 25mm 20mm, clip]{simulation/plots/templates.pdf} \\
 & \multicolumn{3}{c}{\includegraphics[width=3in, trim=2mm 5mm 17mm 5mm, clip]{simulation/plots/legend_means.pdf}} \\[8pt]
\begin{picture}(10,90)\put(0,48){\rotatebox[origin=c]{90}{Template Variance}}\end{picture} & 
\includegraphics[height=1.6in, page=2, trim=7mm 2mm 25mm 20mm, clip]{simulation/plots/templates.pdf} &
\includegraphics[height=1.6in, page=4, trim=7mm 2mm 25mm 20mm, clip]{simulation/plots/templates.pdf} &
\includegraphics[height=1.6in, page=6, trim=7mm 2mm 25mm 20mm, clip]{simulation/plots/templates.pdf} \\
 & \multicolumn{3}{c}{\includegraphics[width=3in, trim=2mm 5mm 17mm 5mm, clip]{simulation/plots/legend_vars.pdf}} \\[4pt]
\end{tabular}
\caption{\small Template (mean and between-subject variance) maps for each IC, based on $1000$ Monte Carlo samples.}
\label{fig:sim_templates}
\end{figure}

Subject effects represent deviations of the subject ICs from the corresponding population mean ICs. The spatial template ICA model assumes that these deviations are smooth, as they tend to be in practice.  For each subject, we first generate spatially independent (e.g., unsmoothed) subject effects from a mean-zero Normal distribution with variance given by the generating variance at pixel $v$ in IC $\ell$, displayed in the second row of Figure \ref{fig:sim_generating}. We then smooth each subject effect map by convolving it with a Gaussian kernel with FWHM equal to 5mm, equivalent to a standard deviation of 2.12mm.  We apply a global rescaling factor to the smoothed effect maps to match the generating variance at the peak pixel.  Subject ICs are finally obtained by adding the smoothed and rescaled subject effects to the population mean maps (Figure \ref{fig:sim_generating}).  Subject effect maps and ICs for one example subject are shown in Figure \ref{fig:sim_data}.  Based on 1000 Monte Carlo replicates, the templates are defined by the Monte Carlo mean and variance at each location $v$ for each IC $\ell$, denoted $\{s_{0\ell}(v):v=1,\dots,V\}$ and $\{\sigma_{\ell}^2(v):v=1,\dots,V\}$, respectively.  The resulting templates are shown in Figure \ref{fig:sim_templates}.

\begin{figure}
\centering
\includegraphics[width=6in]{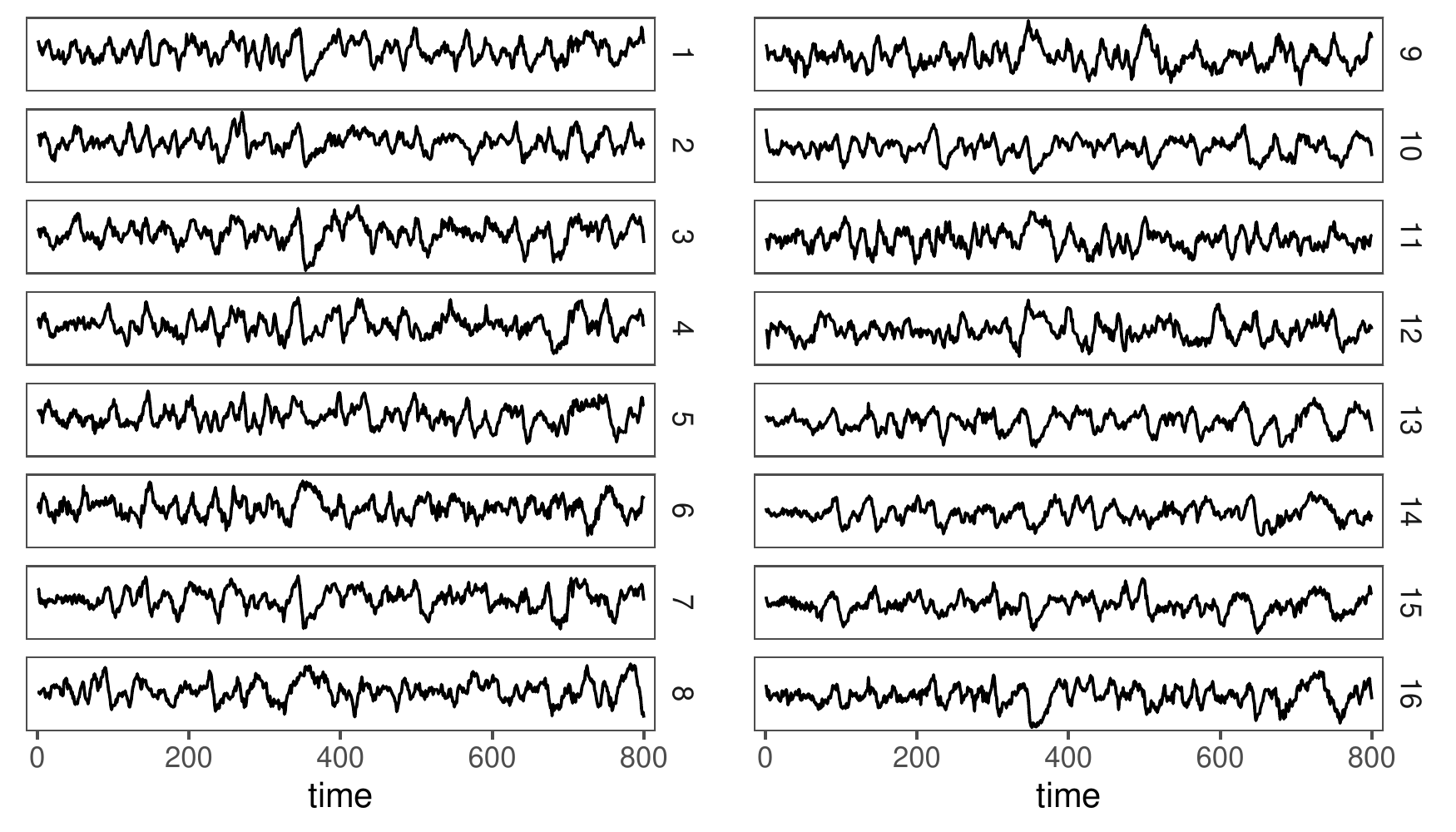}
\caption{\small Sixteen realistic timecourses, from which three were sampled to create the mixing matrix for each subject in the simulation study.  Each timecourse was based on the fMRI signal associated with an ICA-based brain network in one subject from the Human Connectome Project.  As fMRI data is collected in arbitrary units, no y-axis values are displayed.  For mixing matrix construction, each timecourse was centered and scaled to have mean zero and standard deviation one.}
\label{fig:sim_timecourses}
\end{figure}

\textit{fMRI Timeseries.} Using the subject-level ICs resulting from the process described above, we generate fMRI timeseries for $n=50$ test subjects based on the model in equation (\ref{eqn:eqn1}) as follows.  We set $T=800$, representing approximately $10$ minutes assuming a time resolution of $0.72$ seconds, as in the Human Connectome Project \cite{van2013wu}. This duration and temporal resolution are fairly typical of modern resting-state fMRI acquisitions.  We first generate the temporal activation profiles of each IC, which form the $T \times L$ mixing matrix, and a set of white noise residuals generated as independent $N(0,\nu_0^2)$ samples.  We set $\nu_0$ to $11.2$ to yield a signal-to-noise (SNR) of 0.5, similar to real fMRI data.  For the temporal activation profiles, we draw without replacement from a set of $16$ mean-zero timecourses that are based on real fMRI data from the Human Connectome Project (see Figure \ref{fig:sim_timecourses}).  Each timecourse is normalized to have mean zero and standard deviation 1.  

Note that while the mixing matrix itself may not be of scientific interest, what is often of interest is the pairwise similarity or coherence (i.e., Pearson correlation) of the timecourses, known as the functional connectivity (FC) between the brain networks represented by each IC \citep{joel2011relationship}. We therefore assess each method's ability to accurately estimate the FC matrix for each subject, in addition to the IC maps.

\begin{figure}
\centering
\begin{subfigure}[b]{0.48\textwidth}
\includegraphics[height=3in, trim = 2cm 2cm 1cm 2cm, clip]{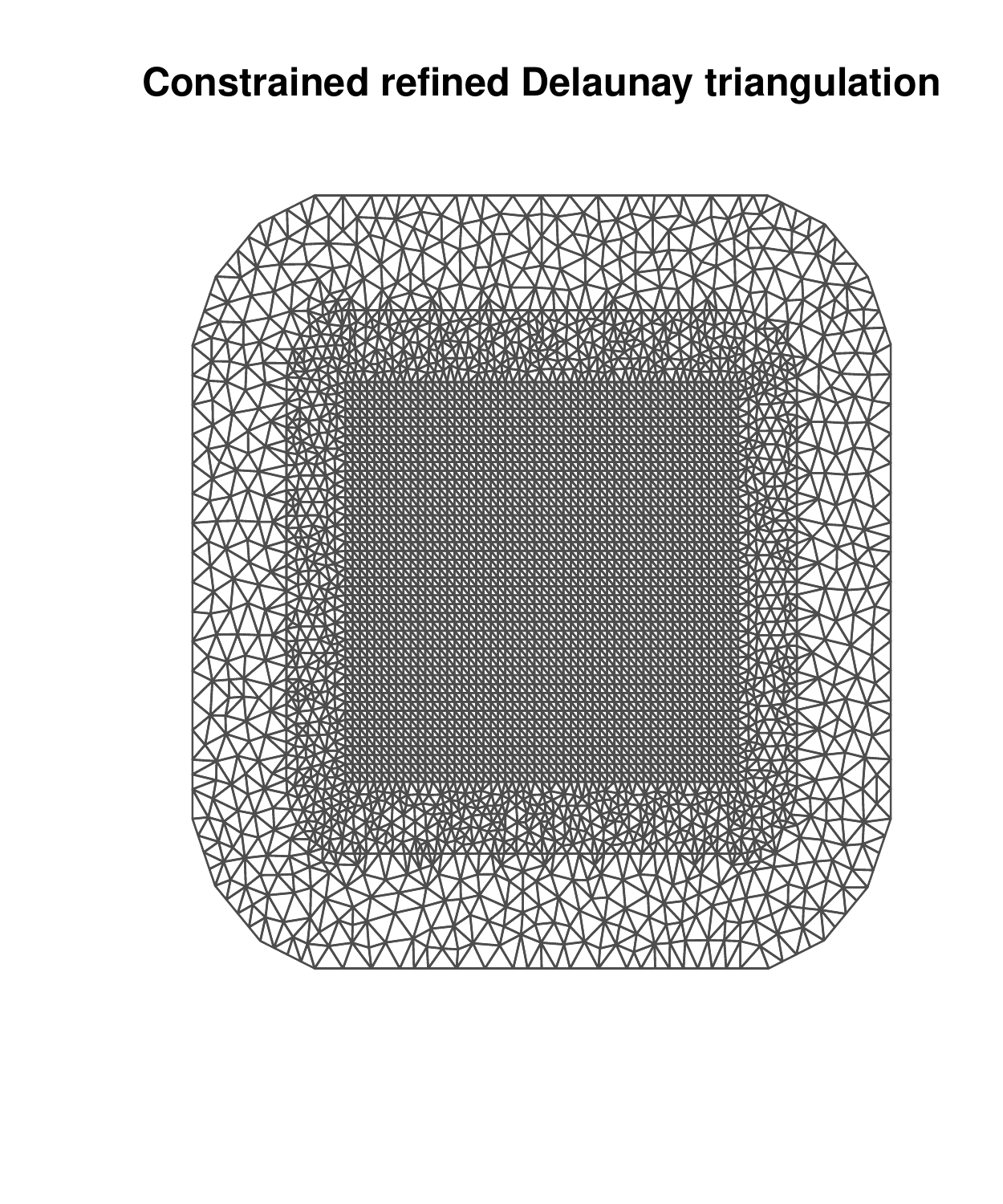}
\caption{\small Triangular mesh used in the simulation study}
\end{subfigure}
\begin{subfigure}[b]{0.48\textwidth}
\includegraphics[height=3in]{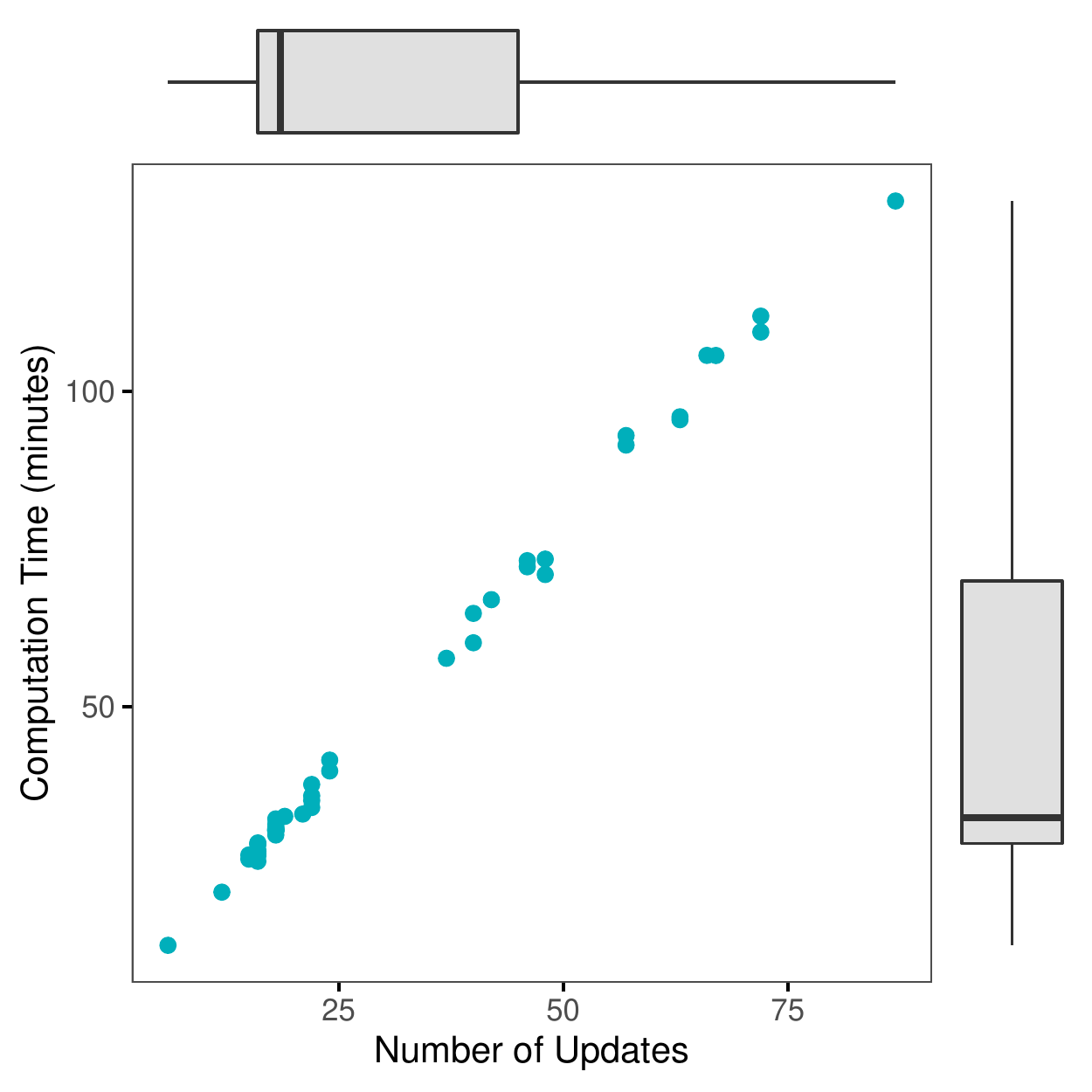}
\caption{Computation time for simulation study}
\end{subfigure}
\caption{For the simulation study, panel (a) shows the triangular mesh used to model the observed data using an SPDE spatial process.  The vertices in the middle square of smaller triangles correspond to the original data locations, while the two boundary layers of larger triangles are added to improve performance along the data boundary.  Panel (b) shows the number of parameter updates and computation time required for each test subject in the simulation study.  Each point represents one subject, and the marginal boxplots show the distribution across subjects.}
\label{fig:sim_mesh_comptime}
\end{figure}

\textit{Model Estimation.}   Based on the original $2530$ data locations, we construct a mesh with the \texttt{inla.mesh.2d} from the INLA R package \citep{lindgren2015bayesian}. The triangular mesh, consisting of 4,214, vertices is shown in Figure \ref{fig:sim_mesh_comptime}(a). It shows a middle square of smaller triangles connecting vertices corresponding to the observed data, surrounded by two boundary layers of larger triangles, which are added to improve performance along the data boundary.  To estimate the ICs and mixing matrix of each simulated test subject using stICA, we apply the EM algorithm with common smoothness assumption, as described in Section \ref{sec:EM}. We run the algorithm until convergence with a tolerance of 0.001, up to 100 parameter updates. The number of updates and computation time for each subject is displayed in Figure \ref{fig:sim_mesh_comptime}(b).  The median computation time for model estimation is 32 minutes over 18 parameter updates.  After model estimation, we identify areas of engagement for each subject and IC via the excursions set approach described in Section \ref{sec:excursions}. We use an engagement threshold of $\gamma=1$ (this avoids very small values induced by the smoothing used to generate subject ICs) and a significance level of $\alpha=0.1$. For comparison purposes, we use the variance estimates produced by tICA to identify areas of engagement by performing a one-sided t-test at each location.  We obtain engagement maps based on Bonferroni correction for familywise error rate (FWER) control at $\alpha=0.1$.  Note that the probability of not committing a familywise error (defined as having at least one false positive pixel) is analogous to the probability that all locations labelled as significant are truly engaged, which is the basis of the excursions set approach used in stICA. The difference is that stICA employs the joint posterior distribution of the latent fields, while tICA employs the marginal posterior distribution at every location. As such, using stICA we would expect to discover larger areas of engagement versus tICA, since the positive dependence between neighboring locations is accounted for.

\textit{Performance Measures.} We compute the following measures to compare the performance of stICA with tICA and dual regression: mean squared error (MSE) maps of the estimated ICs; correlation between the estimated and true ICs for each subject; the accuracy of engagement maps in terms of false positive rate (percentage of truly non-engaged pixels labelled as engaged) and power (percentage of truly engaged pixels labelled as engaged); and the MSE of functional connectivity.

\subsection{Simulation Results}

\begin{figure}
    \centering
    \begin{tabular}{cccc}
    & Subject 1 & Subject 2  & Subject 3  \\[4pt]
    \begin{picture}(10,90)\put(0,48){\rotatebox[origin=c]{90}{True IC}}\end{picture} & 
    \includegraphics[height=1.6in, page=1, trim=7mm 2mm 25mm 20mm, clip]{simulation/plots/subj1_IC1.pdf} &
    \includegraphics[height=1.6in, page=1, trim=7mm 2mm 25mm 20mm, clip]{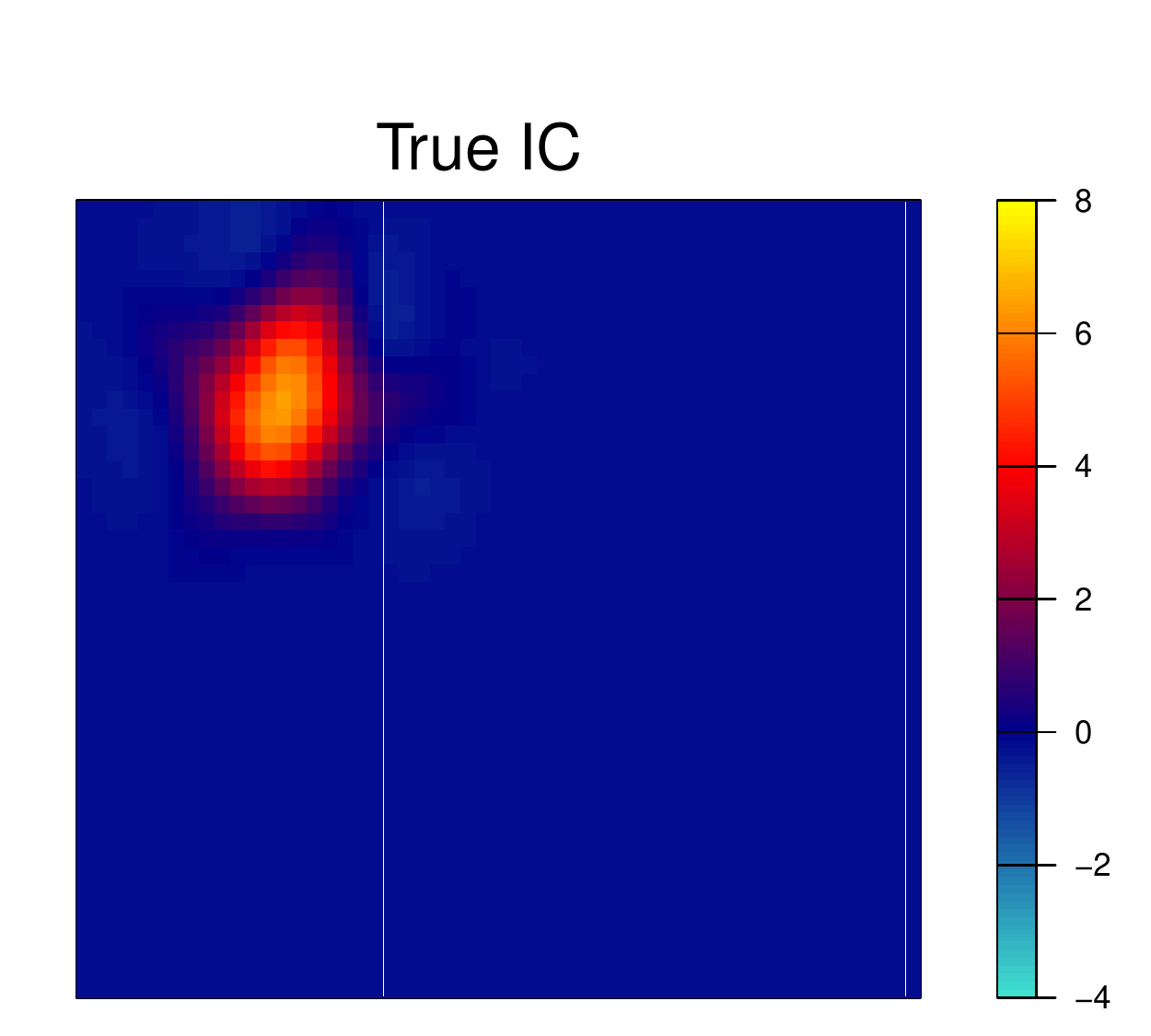} &
    \includegraphics[height=1.6in, page=1, trim=7mm 2mm 25mm 20mm, clip]{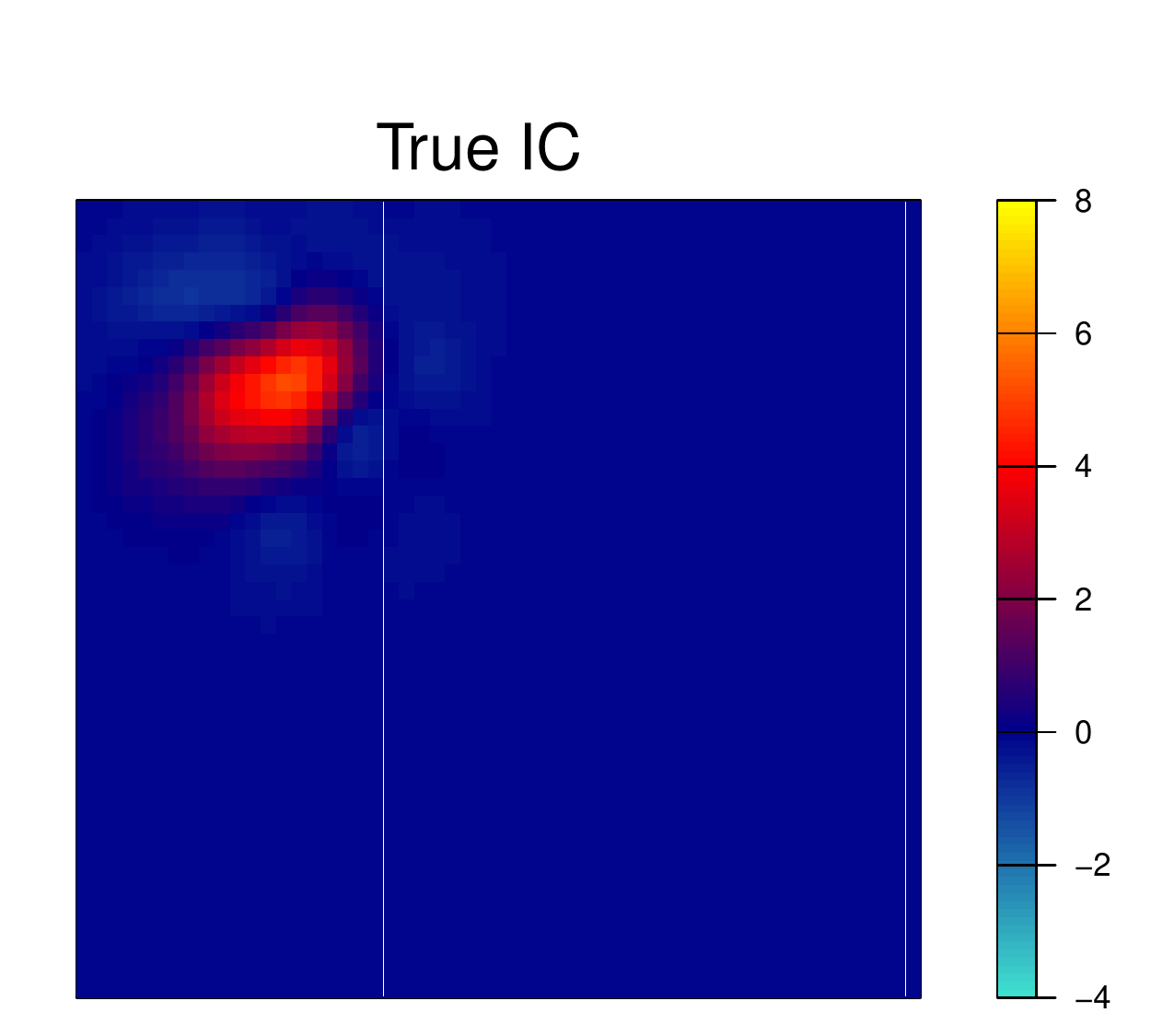} \\[4pt]
    \begin{picture}(10,90)\put(0,48){\rotatebox[origin=c]{90}{stICA}}\end{picture} & 
    \includegraphics[height=1.6in, page=2, trim=7mm 2mm 25mm 20mm, clip]{simulation/plots/subj1_IC1.pdf} &
    \includegraphics[height=1.6in, page=2, trim=7mm 2mm 25mm 20mm, clip]{simulation/plots/subj2_IC1.pdf} &
    \includegraphics[height=1.6in, page=2, trim=7mm 2mm 25mm 20mm, clip]{simulation/plots/subj3_IC1.pdf}  \\[4pt]
    \begin{picture}(10,90)\put(0,48){\rotatebox[origin=c]{90}{tICA}}\end{picture} & 
    \includegraphics[height=1.6in, page=3, trim=7mm 2mm 25mm 20mm, clip]{simulation/plots/subj1_IC1.pdf} &
    \includegraphics[height=1.6in, page=3, trim=7mm 2mm 25mm 20mm, clip]{simulation/plots/subj2_IC1.pdf} &
    \includegraphics[height=1.6in, page=3, trim=7mm 2mm 25mm 20mm, clip]{simulation/plots/subj3_IC1.pdf} \\[4pt]
    \begin{picture}(10,90)\put(0,48){\rotatebox[origin=c]{90}{Dual Regression}}\end{picture} & 
    \includegraphics[height=1.6in, page=4, trim=7mm 2mm 25mm 20mm, clip]{simulation/plots/subj1_IC1.pdf} &
    \includegraphics[height=1.6in, page=4, trim=7mm 2mm 25mm 20mm, clip]{simulation/plots/subj2_IC1.pdf} &
    \includegraphics[height=1.6in, page=4, trim=7mm 2mm 25mm 20mm, clip]{simulation/plots/subj3_IC1.pdf} \\[4pt]
    &   \multicolumn{3}{c}{\includegraphics[width=3in, trim=2mm 5mm 17mm 5mm, clip]{simulation/plots/legend_subjIC.pdf}}
    \end{tabular}
    \caption{True and estimated maps of IC 1 for three randomly selected simulation subjects. The stICA estimates appear very similar to the true ICs. The tICA estimates are similarly accurate in background areas but are slightly more noisy in engaged areas. In contrast to stICA which incorporates spatial priors and empirical population priors, tICA only uses population priors, and those priors tend to have higher variance in engaged areas. Dual regression is more noisy across the image compared with tICA or stICA.}
    \label{fig:sim_estimates}
\end{figure}

\begin{figure}
\centering
\begin{subfigure}[b]{1\textwidth}
\centering
\begin{tabular}{cccc}
& IC 1  & IC 2  & IC 3   \\[4pt]
\begin{picture}(10,90)\put(0,48){\rotatebox[origin=c]{90}{stICA}}\end{picture} & 
\includegraphics[height=1.4in, page=3, trim=7mm 2mm 25mm 20mm, clip]{simulation/plots/MSE_IC1.pdf} &
\includegraphics[height=1.4in, page=3, trim=7mm 2mm 25mm 20mm, clip]{simulation/plots/MSE_IC2.pdf} &
\includegraphics[height=1.4in, page=3, trim=7mm 2mm 25mm 20mm, clip]{simulation/plots/MSE_IC3.pdf} \\[4pt]
\begin{picture}(10,90)\put(0,48){\rotatebox[origin=c]{90}{tICA}}\end{picture} & 
\includegraphics[height=1.4in, page=2, trim=7mm 2mm 25mm 20mm, clip]{simulation/plots/MSE_IC1.pdf} &
\includegraphics[height=1.4in, page=2, trim=7mm 2mm 25mm 20mm, clip]{simulation/plots/MSE_IC2.pdf} &
\includegraphics[height=1.4in, page=2, trim=7mm 2mm 25mm 20mm, clip]{simulation/plots/MSE_IC3.pdf} \\[4pt]
\begin{picture}(10,90)\put(0,48){\rotatebox[origin=c]{90}{Dual Regression}}\end{picture} & 
\includegraphics[height=1.4in, page=1, trim=7mm 2mm 25mm 20mm, clip]{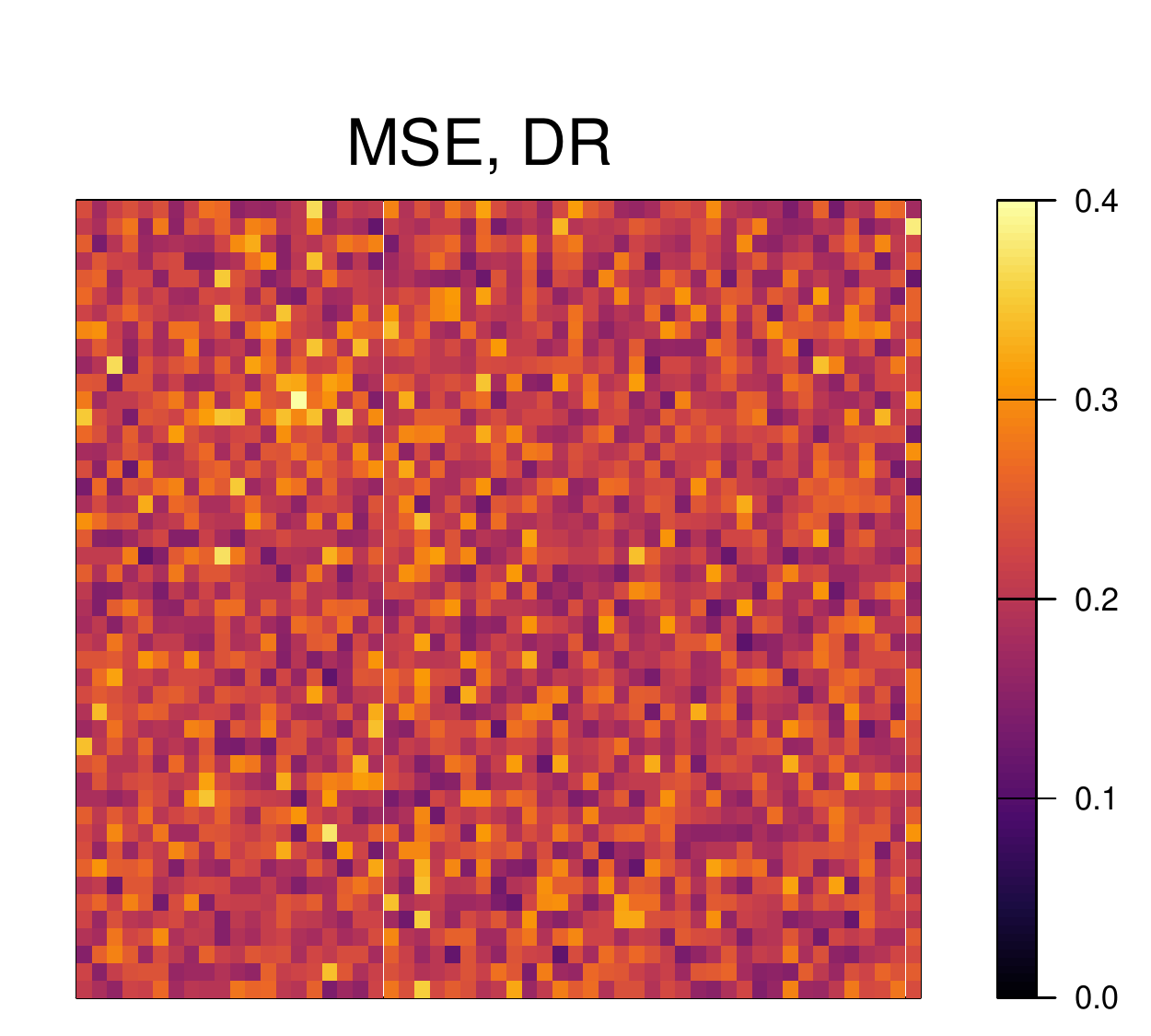} &
\includegraphics[height=1.4in, page=1, trim=7mm 2mm 25mm 20mm, clip]{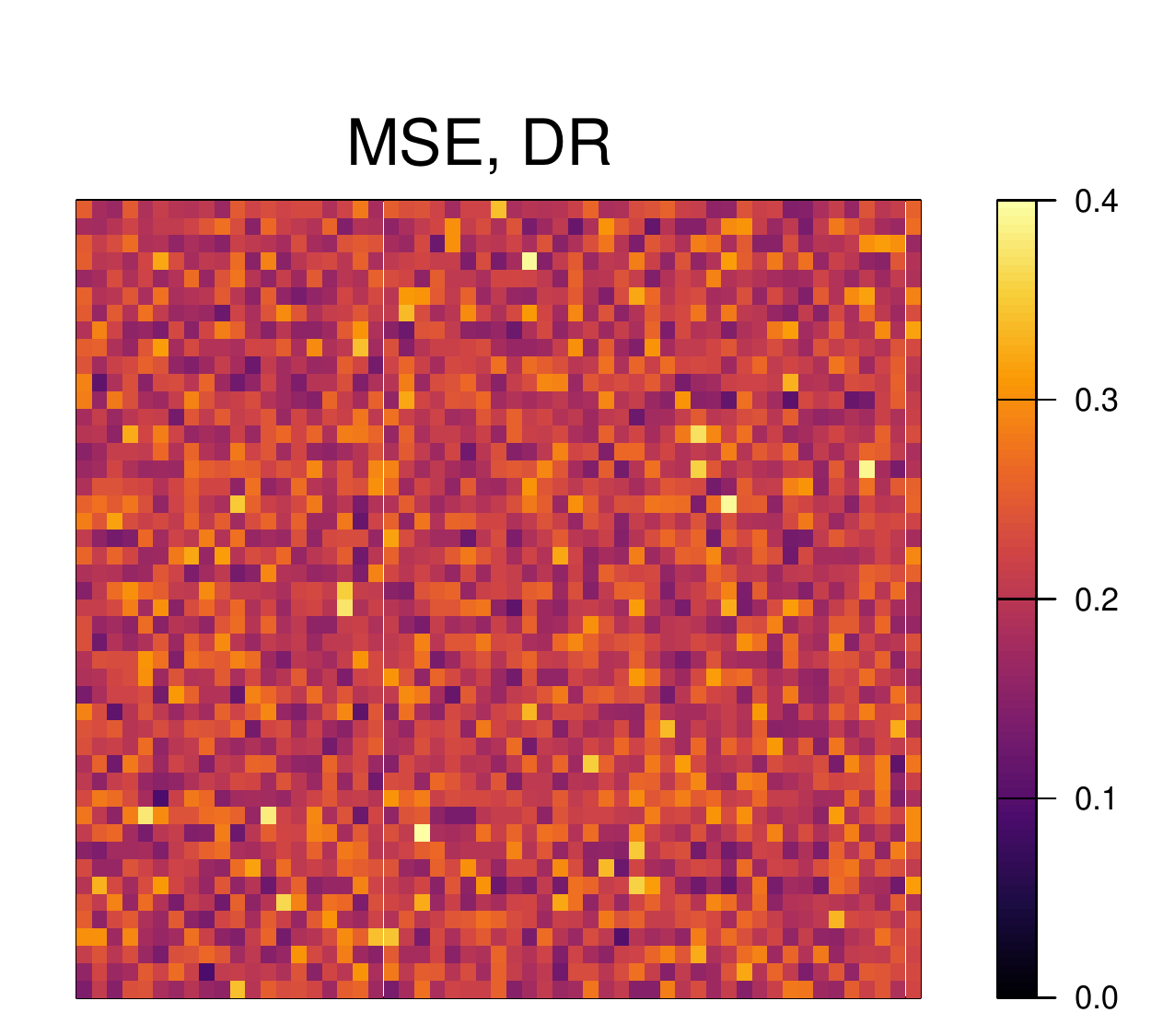} &
\includegraphics[height=1.4in, page=1, trim=7mm 2mm 25mm 20mm, clip]{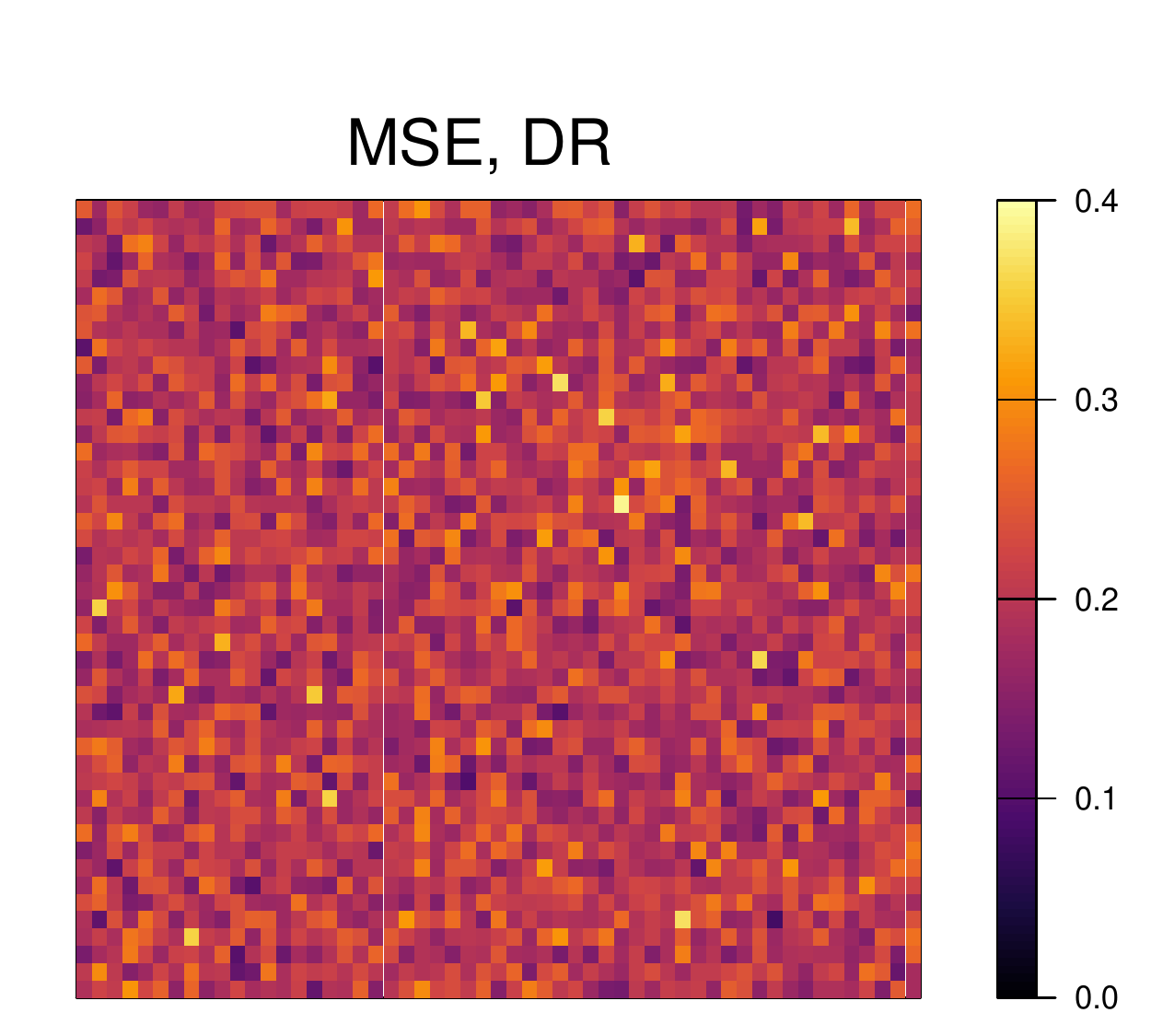} \\[4pt]
& \multicolumn{3}{c}{\includegraphics[width=3in, trim=2mm 5mm 17mm 5mm, clip]{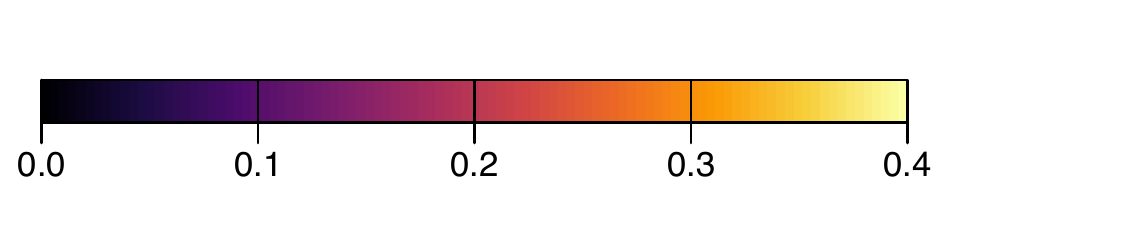}}
\end{tabular}
\caption{\small MSE of the estimated ICs.}
\end{subfigure}
\begin{subfigure}[b]{1\textwidth}
\includegraphics[page=1, height=1.9in, trim=0 7mm 0 0, clip]{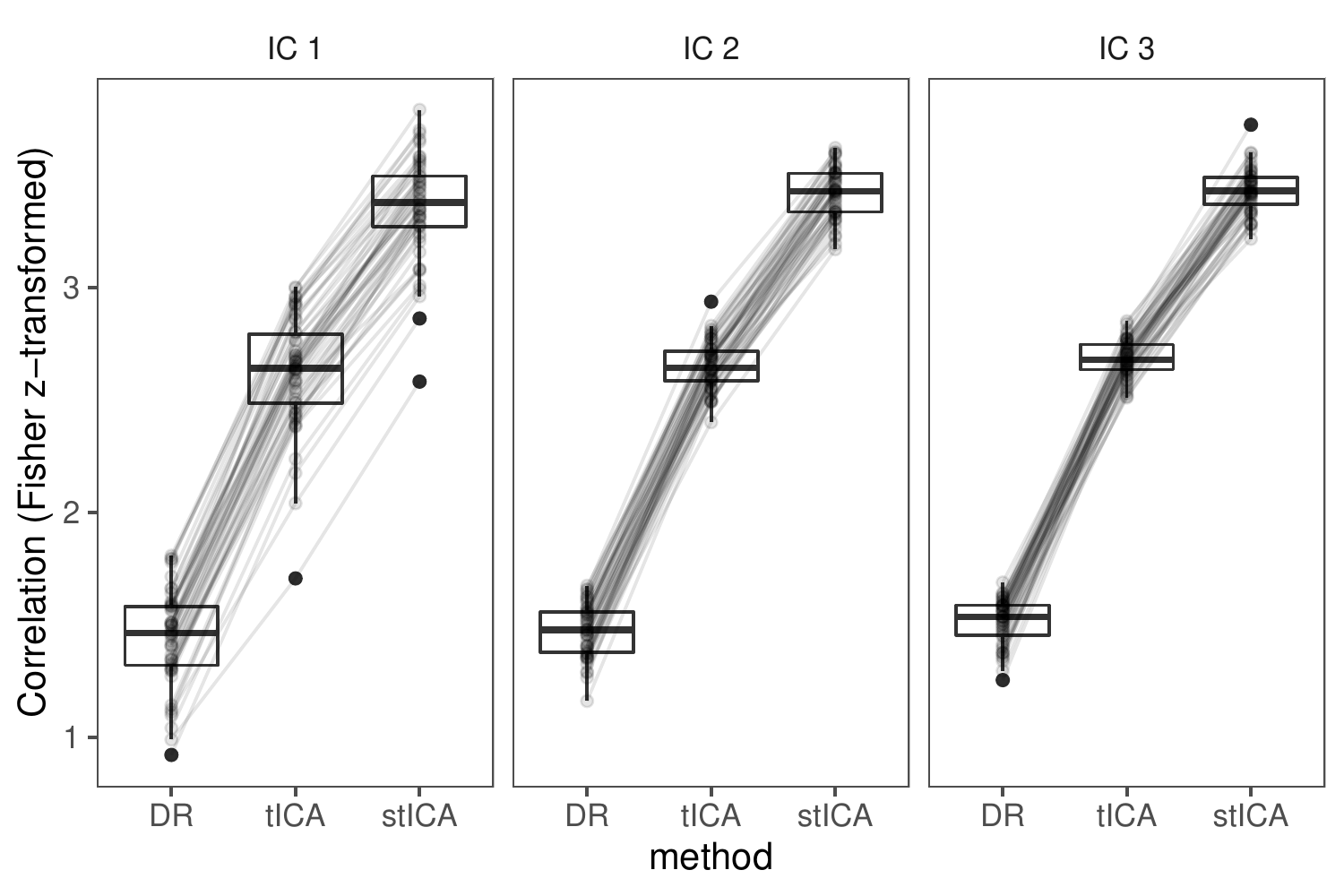} 
\includegraphics[page=2, height=1.9in, trim=0 7mm 0 0, clip]{simulation/plots/cor_cat.pdf}
\caption{\small Correlation between the estimated and true ICs.}
\end{subfigure}
\caption{\small \textbf{Accuracy of estimated ICs in the simulation study.} (a) \textit{MSE of estimated ICs using stICA, tICA and dual regression.}  stICA yields highly accurate IC estimates across the entire image. tICA outperforms dual regression across the image but is less accurate than stICA in the areas of engagement for each IC.  This illustrates the benefit of including both population and spatial priors in the stICA model. Dual regression, which is commonly used in practice, results in highly noisy IC estimates.  (b) \textit{Pearson correlation and correlation-at-the-top (CAT) of the estimated and true ICs.} Each line represents an individual simulation subject, and boxplots are displayed to summarize the distribution across subjects. Correlations are Fisher $z$-transformed (i.e., $z(r)=\tfrac{1}{2}\ln((1+r)/(1-r))$) to enable better comparison of high values. These results show that stICA strongly outperforms tICA, and both stICA and tICA outperform dual regression. This again illustrates the benefit of incorporating both population and spatial priors into the stICA model.}
\label{fig:sim_ICs}
\end{figure}

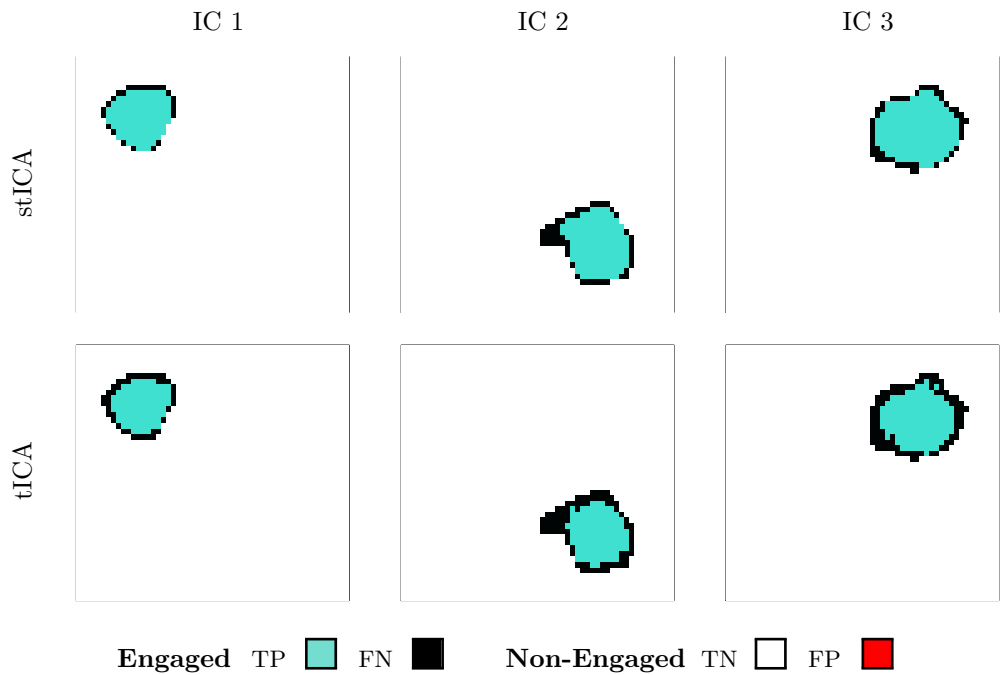
\begin{figure}
    \centering
   \begin{tabular}{cccc}
    & IC 1 & IC 2 & IC 3  \\[4pt]
    \begin{picture}(10,90)\put(0,48){\rotatebox[origin=c]{90}{stICA}}\end{picture} & 
    \includegraphics[height=1.4in, page=2, trim=7mm 2mm 2cm 20mm, clip]{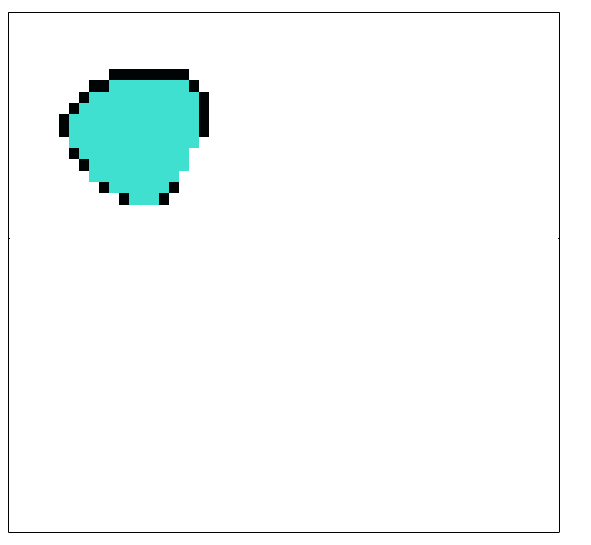} &
    \includegraphics[height=1.4in, page=2, trim=7mm 2mm 2cm 20mm, clip]{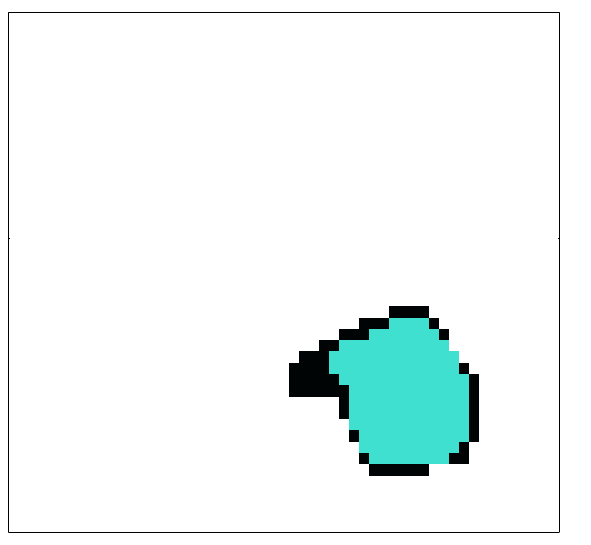} &
    \includegraphics[height=1.4in, page=2, trim=7mm 2mm 2cm 20mm, clip]{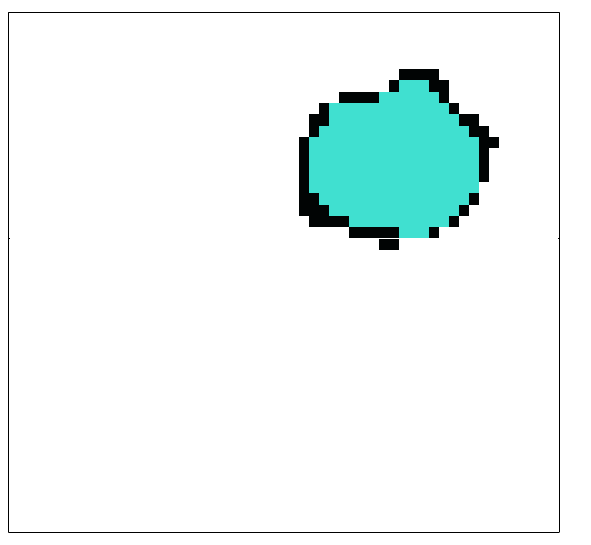} \\[4pt]
    \begin{picture}(10,90)\put(0,48){\rotatebox[origin=c]{90}{tICA}}\end{picture} & 
    \includegraphics[height=1.4in, page=4, trim=7mm 2mm 2cm 20mm, clip]{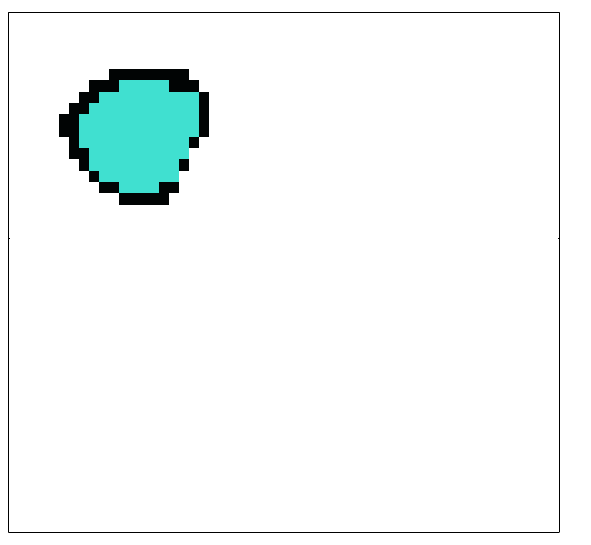} &
    \includegraphics[height=1.4in, page=4, trim=7mm 2mm 2cm 20mm, clip]{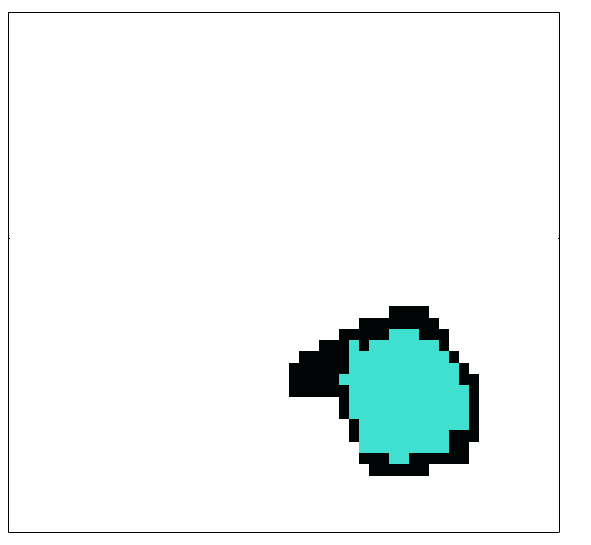} &
    \includegraphics[height=1.4in, page=4, trim=7mm 2mm 2cm 20mm, clip]{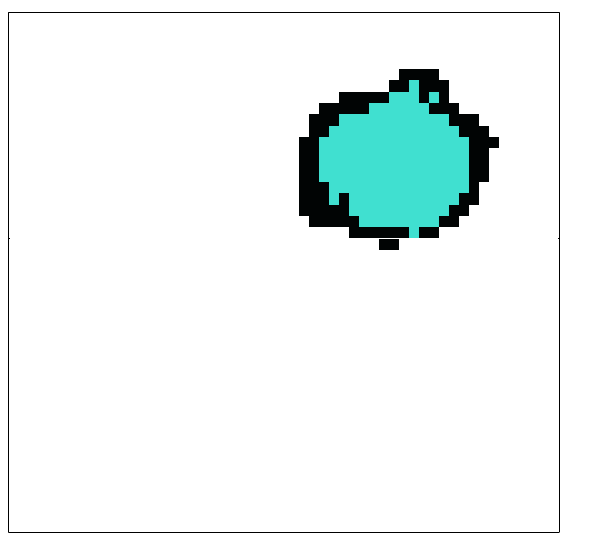} \\[4pt]
    & \multicolumn{3}{c}{
    \begin{picture}(45,20)\put(13,5){\textbf{Engaged}}\end{picture}
    \actlegendact  
    \begin{picture}(65,20)\put(10,5){\textbf{Non-Engaged}}\end{picture}
    \actlegendnot \hspace{1cm} }\\[4pt]
    \end{tabular}
    \caption{\small \textbf{Example areas of engagement in the simulation study.}  For one randomly selected simulation subject, truly engaged pixels are shown in turquoise (true positive, TP) and black (false negative, FN).  Non-engaged pixels are shown in white (true negative, TN) and red (false positive, FP).  In this case, there are no false positive pixels in any of the maps.}
    \label{fig:sim_activations}
\end{figure}

\begin{figure}
    \centering
    \includegraphics[width=4.5in, trim=0 1cm 0 0, clip]{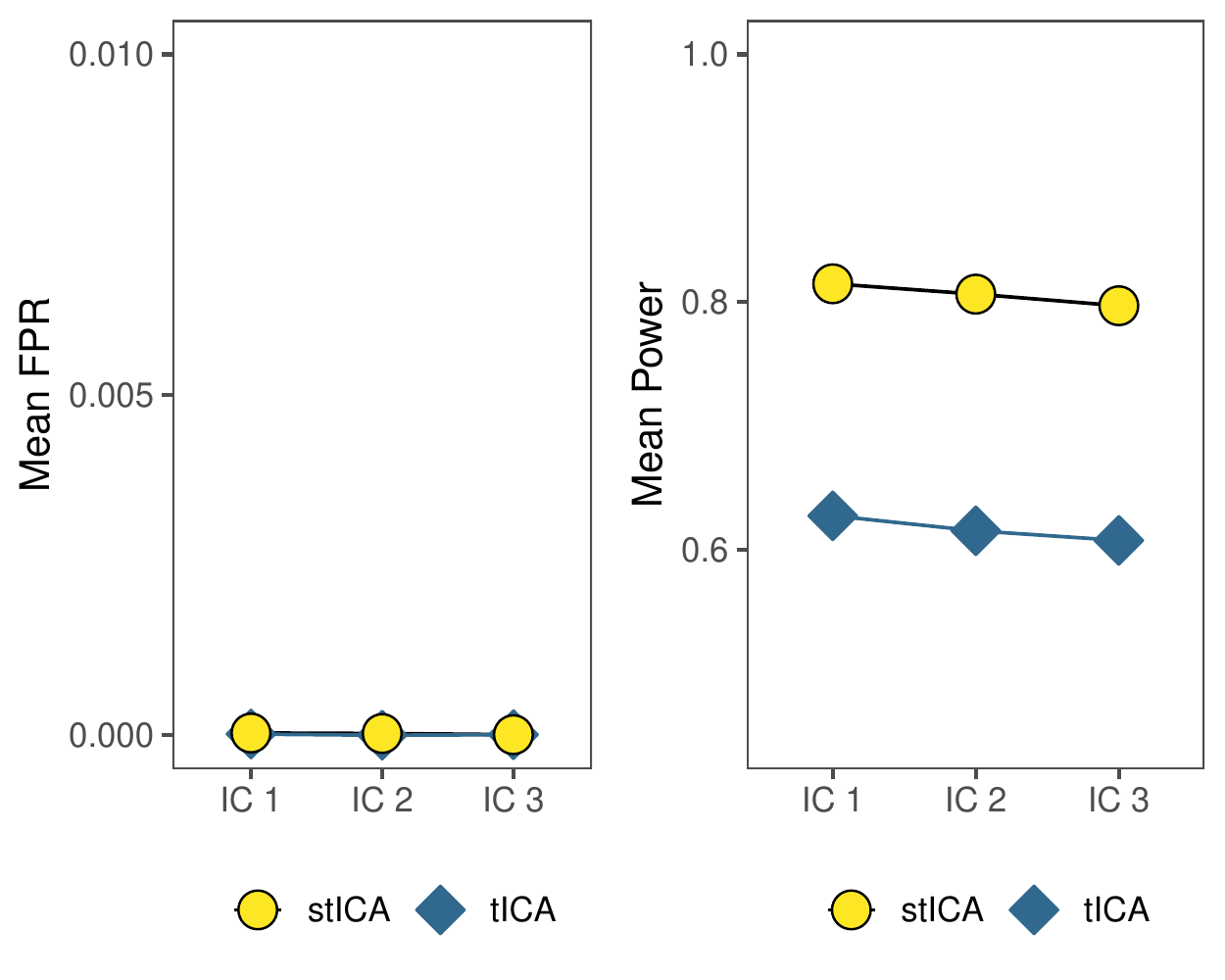}\\
    \caption{\small \textbf{Accuracy of areas of engagement in the simulation study. } stICA achieves higher power to detect true engagements compared with tICA, while maintaining strict false positive control.}
    \label{fig:sim_FPR_FNR}
\end{figure}

\begin{figure}
\centering
\includegraphics[height=3in, trim=0 0 0 7mm, clip]{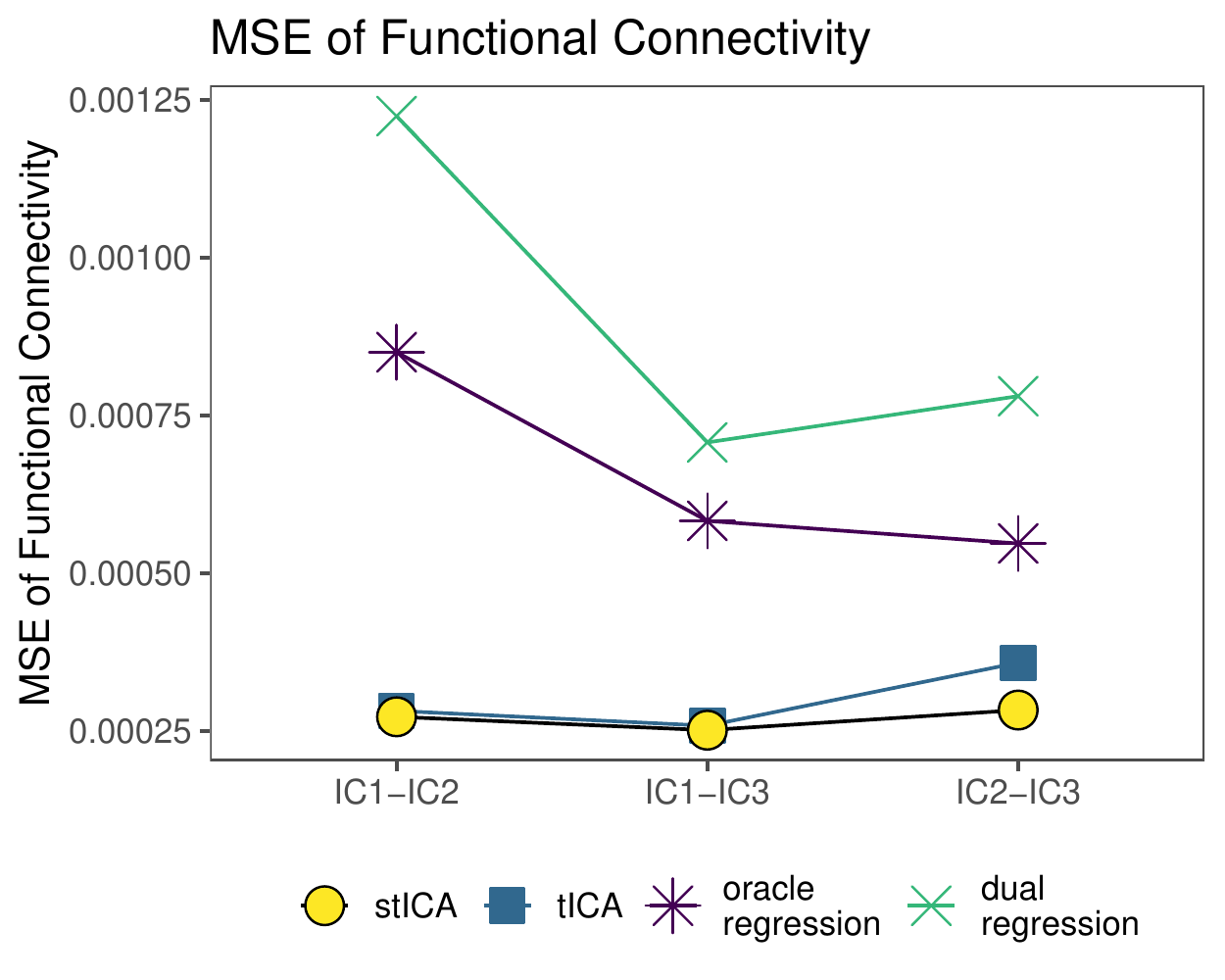} 
\caption{\small\textbf{ MSE of functional connectivity (FC) between each pair of ICs in the simulation study.}  The most accurate FC estimates are achieved with stICA.  Notably, stICA and tICA both strongly outperform oracle regression, which is based on regression of the true ICs against the fMRI timeseries data.  tICA and stICA both result in substantially more accurate estimation of FC compared with dual regression.}
\label{fig:sim_FC}
\end{figure}

In this section, we present the performance of stICA alongside two competing methods, tICA and dual regression, in accurately recovering the true subject-level ICs and FC matrices and identifying areas of engagement in each IC.  For FC matrix estimation, we also include an oracle regression estimator, which is based on regressing the true subject-level IC maps against the fMRI timeseries data. This represents the best possible performance of dual regression for FC matrix estimation, since dual regression is based on such an approach, but without access to the true, unknown ICs.

We first consider the ability of each method to estimate the true subject-level IC maps.  Figure \ref{fig:sim_estimates} displays the true values and estimates of IC 1 for the first three simulation subjects. The stICA estimates appear highly similar to the true ICs. The tICA estimates are similarly accurate in background areas, but are slightly more noisy in areas of engagement. This is because while stICA incorporates spatial priors and empirical population priors, tICA only uses population priors, which tend to have higher variance in areas of engagement. Dual regression is less accurate across the image compared with tICA or stICA.  

Figure \ref{fig:sim_ICs} displays two measures of IC accuracy: mean squared error (MSE) across subjects (a) and Pearson correlation across pixels (b).  To compute MSE, the estimated maps from stICA and tICA were first rescaled to match the scale of the true ICs, since when the ICs and mixing matrix are simultaneously estimated, the ICs are only identifiable up to a scaling factor. Panel (a) shows that stICA yields highly accurate IC estimates across the entire image. tICA outperforms dual regression across the image but is less accurate than stICA in the areas of engagement for each IC.  This illustrates the benefit of including both population and spatial priors in the stICA model. Dual regression, which is commonly used in practice, results in highly noisy IC estimates.  Panel (b) displays two Pearson correlation-based measures for each subject and IC: correlation across all pixels (left) and correlation-at-the-top (CAT) (right), which is computed only based on locations with true IC value over 1, the threshold assumed for identification of areas of engagement. CAT therefore conveys the ability of each method to accurately estimate the areas of most scientific interest.  In terms of both correlation and CAT, stICA strongly outperforms tICA, and both stICA and tICA outperform dual regression. This again illustrates the benefit of incorporating both spatial priors into the stICA model, in addition to the empirical population priors used in tICA.

Figure \ref{fig:sim_activations} displayes true and estimated areas of engagement for IC 1 for the first three simulation subjects.  The stICA areas of engagement are based on the joint posterior distribution of each IC, as described in Section \ref{sec:excursions}.  The tICA areas of engagement are based on performing a $t$-test at every location and correcting for multiple comparisons through Bonferroni correction to control the FWER.  The stICA areas are somewhat larger and more similar to the true areas.  The FWER-corrected tICA areas are more conservative. Figure \ref{fig:sim_FPR_FNR} displays the false positive rate (FPR) of both methods, averaged over subjects, along with the power to detect true engagements.  Both stICA and tICA enforce stringent false positive control, but stICA has substantially higher power to detect true effects, achieving over 80\% mean power for each IC.

Finally, assess the accuracy of stICA, along with competing methods, for estimating functional connectivity (FC).  Figure \ref{fig:sim_FC} displays the MSE of FC between each pair of ICs. The most accurate FC estimates are achieved with stICA.  Notably, stICA and tICA both strongly outperform oracle regression, which is based on regression of the true ICs against the fMRI timeseries data.  tICA and stICA both result in substantially more accurate estimation of FC compared with dual regression.

\section{Data Analysis}
\label{sec:application}

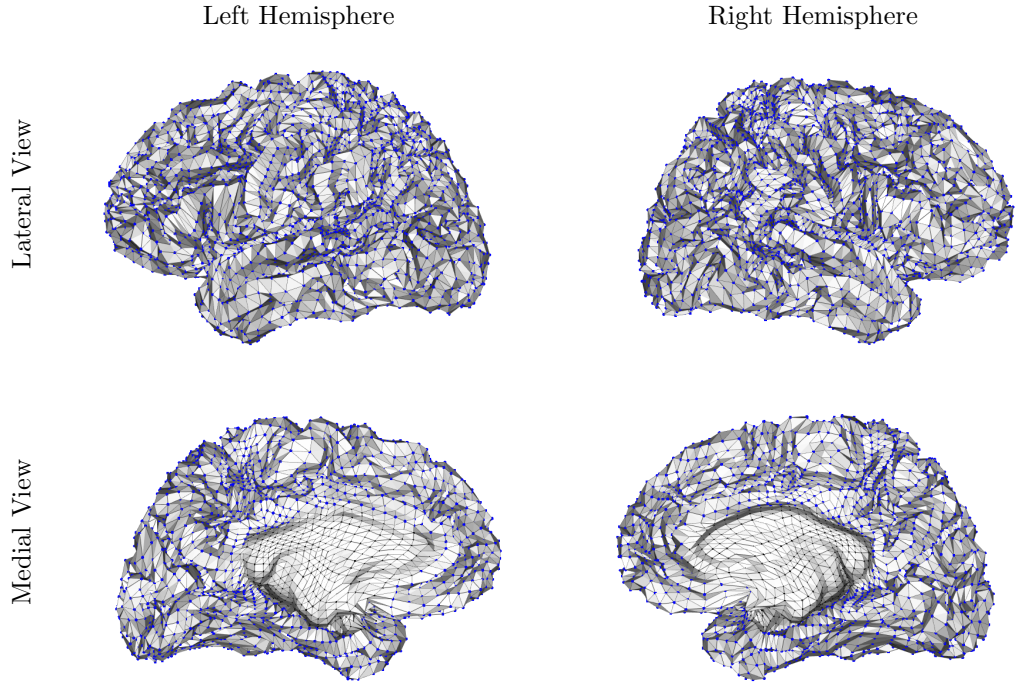
\begin{figure}
    \centering
    \begin{tabular}{ccc}
    & Left Hemisphere  & Right Hemisphere  \\[4pt]
    \begin{picture}(0,120)\put(-5,60){\rotatebox[origin=c]{90}{Lateral View}}\end{picture} & 
    \includegraphics[width=2.5in, trim=10cm 6cm 10cm 6cm, clip]{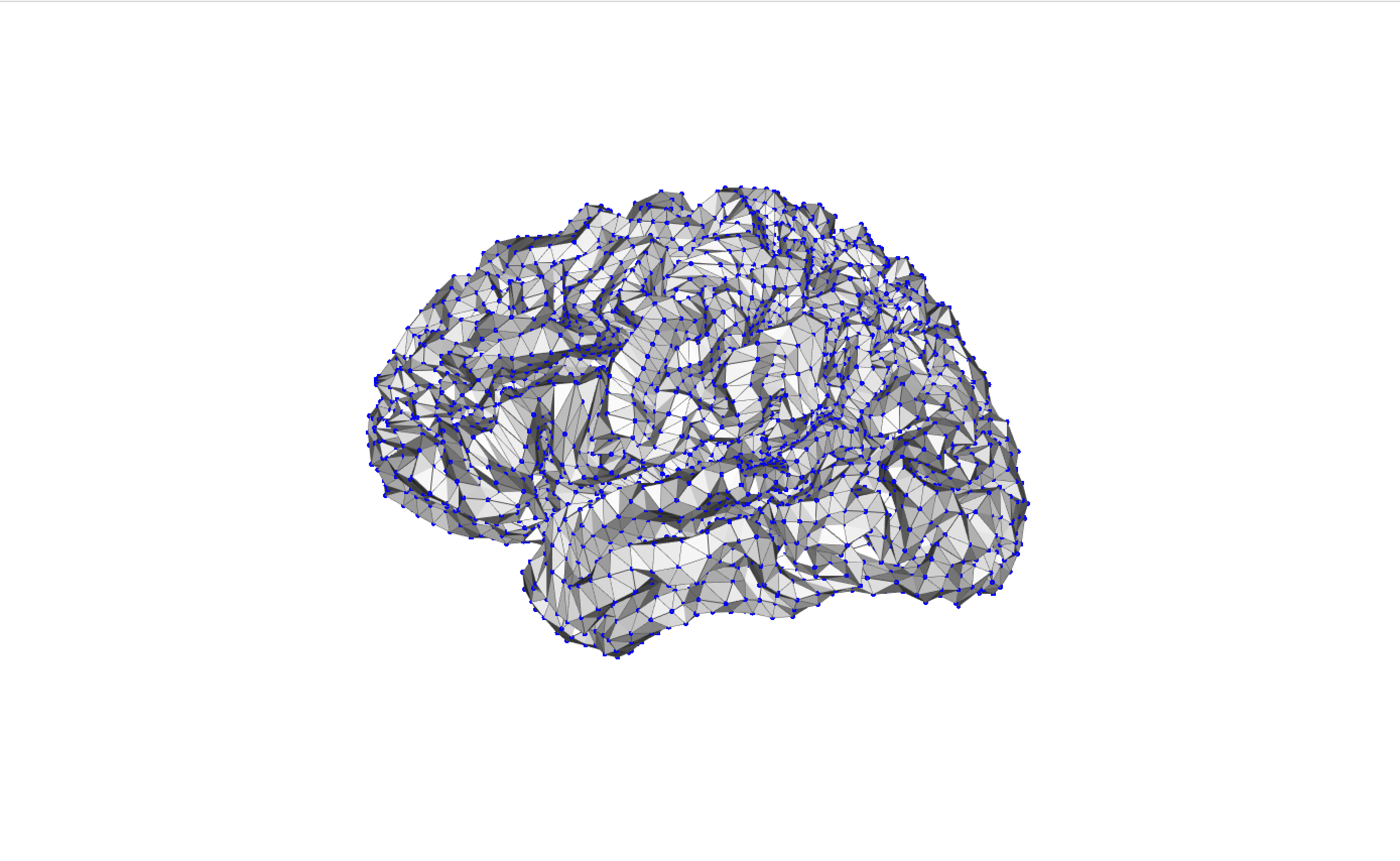} &
    \includegraphics[width=2.5in, trim=10cm 6cm 10cm 6cm, clip]{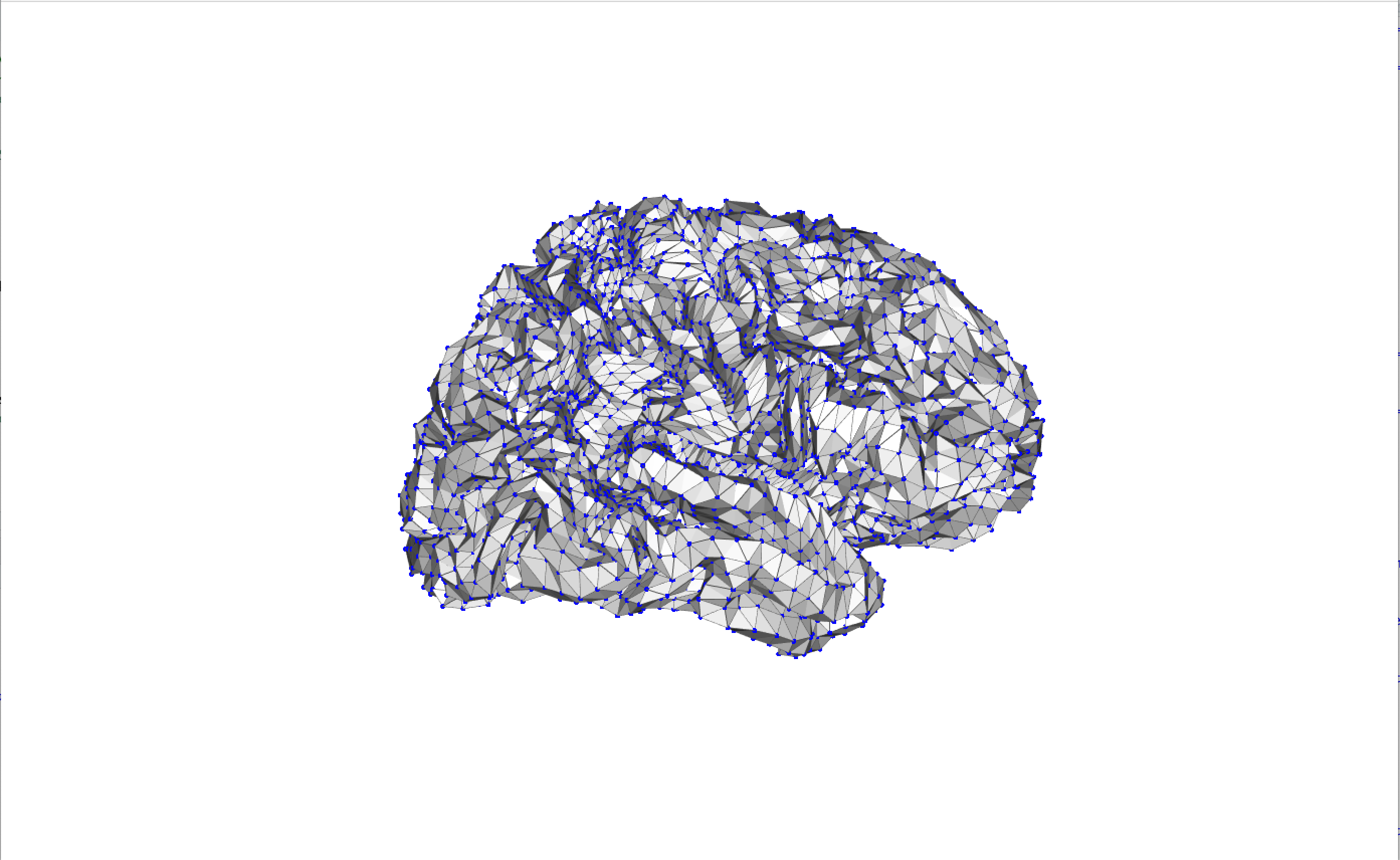} \\[4pt]
        \begin{picture}(0,120)\put(-5,60){\rotatebox[origin=c]{90}{Medial View}}\end{picture} & 
    \includegraphics[width=2.5in, trim=10cm 6cm 10cm 6cm, clip]{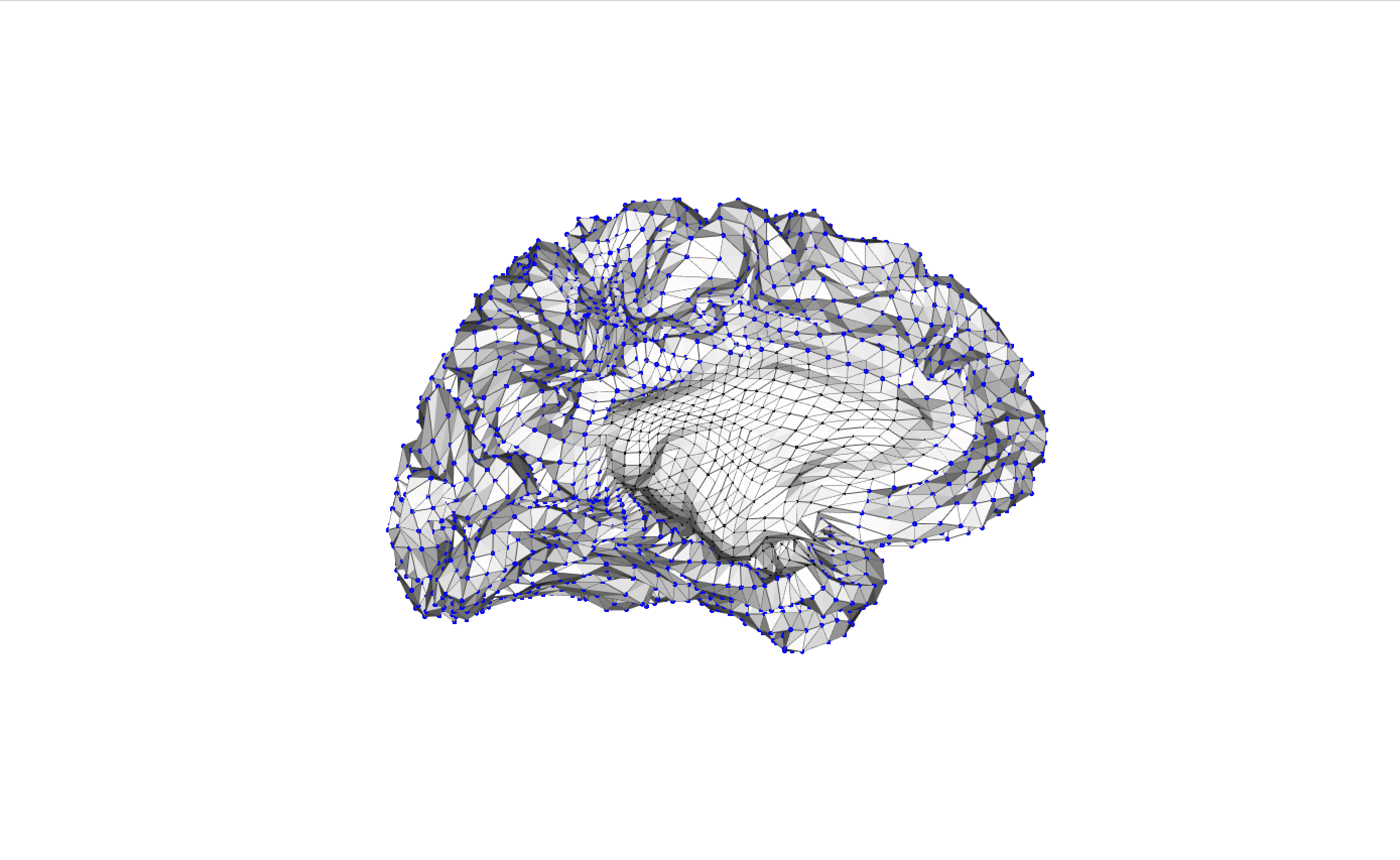} &
    \includegraphics[width=2.5in, trim=10cm 6cm 10cm 6cm, clip]{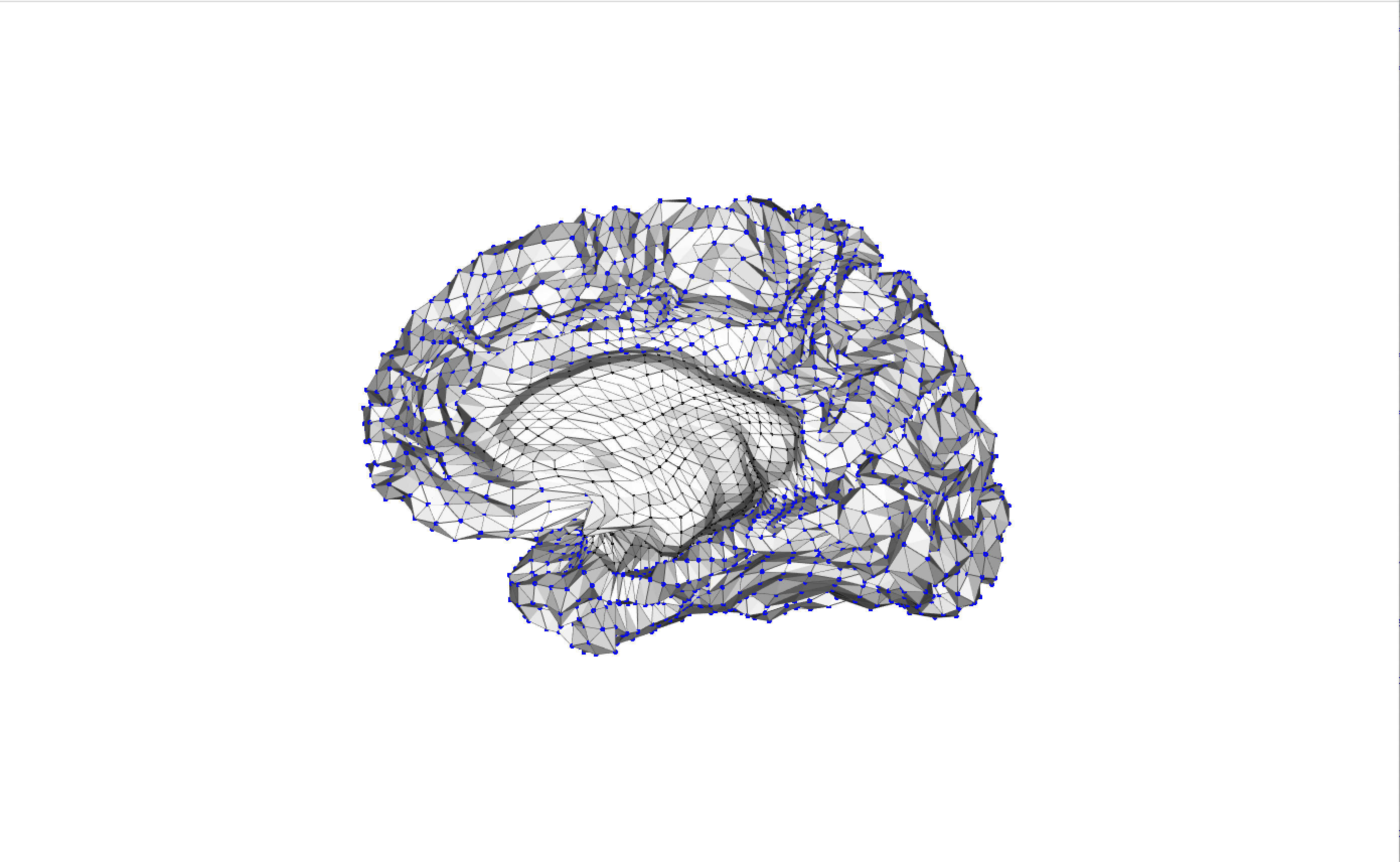} \\[4pt]
    \end{tabular}
    \caption{\small \textbf{Triangular mesh representing the midthickness surface of the selected subject.} The top row shows the lateral (exterior) view of each hemisphere; the bottom row shows the medial (interior) view.  Vertices corresponding to observed data locations are indicated with blue dots. Triangles connecting vertices form a neighborhood structure, which determines the spatial dependence structure in the SPDE priors assumed in the stICA model.  The medial wall or ``hole'' in the medial view of each hemisphere contains locations with missing data. These additional locations serve a similar role to the boundary layers of extra vertices added in the simulation study (see Figure \ref{fig:sim_mesh_comptime}a) by improving performance along the data boundary.}
    \label{fig:app_mesh}
\end{figure}

We apply the proposed and benchmark methods to resting-state fMRI data of one randomly selected subject (645450) from the Human Connectome Project (HCP) 900-subject release \citep{van2013wu}. The HCP is publicly available for download (\url{http://humanconnectome.org}) and includes fully processed fMRI data on the cortical surface (*.dtseries.nii) and corresponding surface model files (*.surf.gii). We use the midthickness surface model, which represents the midpoint in the cortical grey matter between the white matter (interior) surface and pial (exterior) surface. The triangular mesh representing this surface is shown in Figure \ref{fig:app_mesh}. 

The fMRI data was acquired using the left-to-right (LR) phase encoding, with a repetition time (TR) of 0.72 seconds between volumes and a total of $T = 1200$ volumes collected over approximately $15$ minutes \citep{van2013wu}. In addition to the minimal processing described in \cite{glasser2013minimal}, the fMRI data we analyze has been high-pass filtered and denoised using the ICA-FIX procedure \citep{salimi2014automatic, griffanti2014ica}. This version of the data is available as part of the HCP data release. Additionally, we use the Connectome Workbench \citep{marcus2011informatics} (available for download at \url{https://www.humanconnectome.org/software/connectome-workbench}) via the \texttt{ciftiTools} R package  \citep{ciftiTools} to resample the surface models and fMRI data from their original 32,000 vertex resolution to approximately 6,000 vertices per hemisphere. This greatly improves computational speed and feasibility with minimal loss of spatial resolution, as the surface area represented by each vertex remains much smaller than the spatial areas represented by different brain networks. Finally, we center the fMRI data across time and space, then scale the data by the square root of the average image variance.

\begin{figure}
    \centering
   \begin{tabular}{cc}
    Template Mean  & Template Variance \\[4pt]
    \fbox{\includegraphics[height=1.8in]{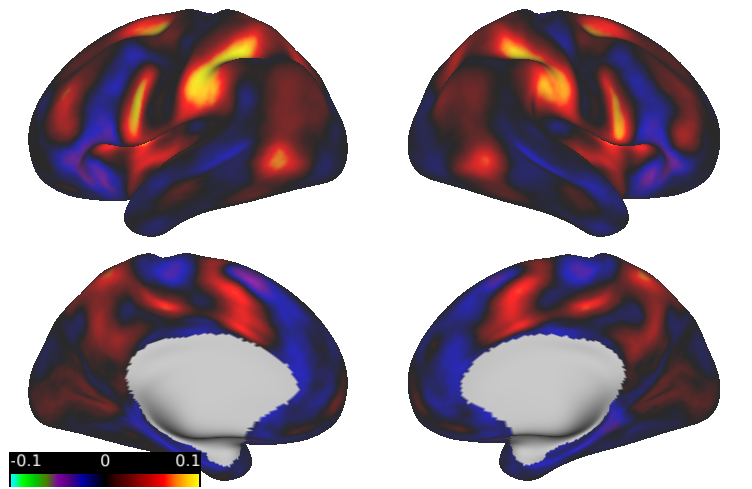}} &
    \fbox{\includegraphics[height=1.8in]{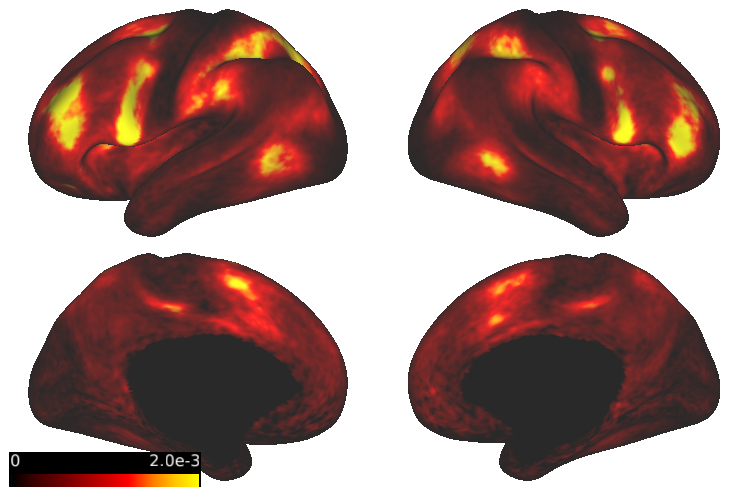}}  \\[4pt]
    \end{tabular}
    \caption{\small Template mean and variance for IC 10, an attention network. The template consists of a pair of such mean and variance maps for all 16 ICs.}
    \label{fig:app_template}
\end{figure}

For template estimation, we use resting-state fMRI data from 461 subjects in the HCP 500 subject release, following the procedure described in \citep{mejia2019template}. The data is processed in the same way as the focal dataset described above. More details of the dataset used for template estimation are as described in \cite{mejia2019template}.  Briefly, we select 16 ICs from the 25-component group ICA maps included in the HCP 500 subject release, which correspond to established brain networks. The exclusion of 9 group ICs also tests the ability of stICA and tICA to account for the presence of nuisance ICs, which may include these excluded group ICs along with subject-specific sources of neuronal and artifactual origin.  The estimated templates consist of a map of the mean and between-subject variance for each of the 16 ICs.  Figure \ref{fig:app_template} shows the mean and variance maps for one IC representing an attention network.

We apply the proposed EM algorithm to estimate the subject ICs and subject effects for each of the 16 template ICs. The model is fit within each hemisphere of the brain separately, since the triangular meshes representing each hemisphere do not intersect. We run the algorithm until convergence to a tolerance of 0.01. For comparison, we also obtain estimates using tICA (template ICA without spatial priors) and dual regression.  We then apply the excursions set approach described in Section \ref{sec:excursions} to identify areas of engagement in each IC, as well as areas of positive and negative deviation in each subject effect map.  These deviations represent differences between the individual subject's IC maps and the group average.  Positive deviations indicate that a particular area contributes more to a subject's brain network relative to the group, while negative deviations indicate that a particular area contributes less relative to the group.  For comparison, we also identify areas of engagement and deviation using tICA by performing a $t$-test at every vertex separately, then correcting for multiple comparisons using Bonferroni correction to control the family-wise error rate (FWER) at $\alpha=0.01$. This is analogous to achieving a joint posterior probability of $1-\alpha$ as in the stICA framework. We use an engagement threshold of $\gamma=0$ in each case. All computations were performed on a Linux computer with 64GB of RAM and 16 processors.  Computation time for stICA was approximately 6 hours 45 minutes per hemisphere. 
Computation time for identifying areas of engagement or deviation with stICA was approximately $45$ minutes per hemisphere for each IC or subject effect map.

To assess the ability of stICA and the benchmark methods to produce reliable results, we also perform the same analyses on a second session from the same subject, acquired and processed using the same techniques.  We quantify the reliability of subject-level features, namely the subject effect maps or deviations and the areas of positive and negative deviation.  We focus on these rather than on the ICs themselves, since reliability of IC maps could be spuriously inflated by over-shrinkage to the group mean. To quantify the reliability of the estimated subject effect maps, we compute the Pearson correlation between the maps based on each session of data.  To quantify the reliability of the areas of positive and negative deviation, we compute the overlap between the areas identified using each session of data. We consider two measures of overlap: the size (number of vertices) and the Dice coefficient, defined as the size of overlap divided by the average size of the two areas individually.  To determine the influence of the quantity of data available, we finally apply the same analyses to shorter scans of duration $T=400$ volumes (5 minutes) by truncating each session after the first $400$ volumes.  

\begin{figure}
    \centering
   \begin{tabular}{cccc}
    & stICA  & tICA  & Dual Regression \\[4pt]
    \begin{picture}(0,90)\put(-5,45){\rotatebox[origin=c]{90}{IC Estimates}}\end{picture} & 
    \fbox{\includegraphics[height=1.3in]{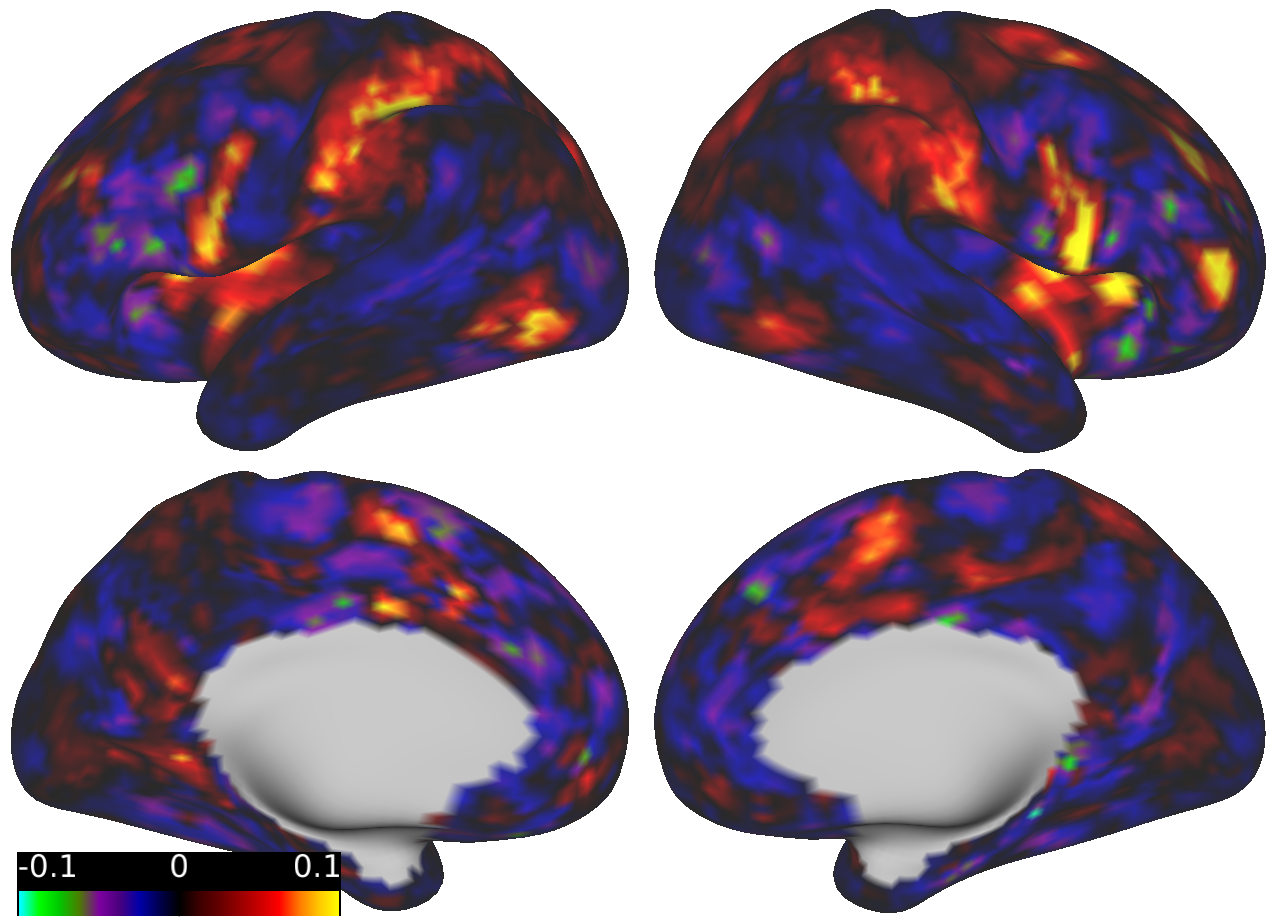}} &
    \fbox{\includegraphics[height=1.3in]{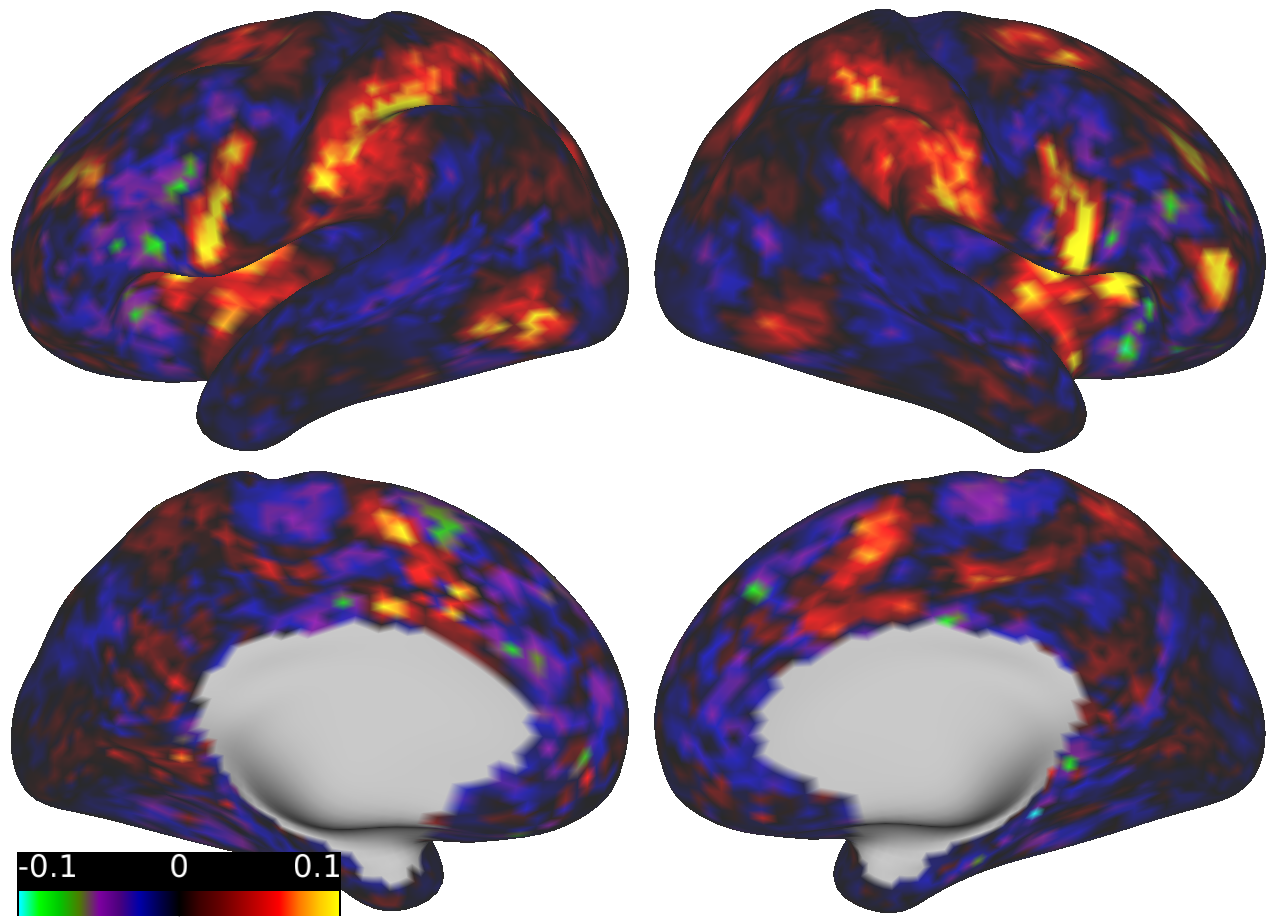}} &
    \fbox{\includegraphics[height=1.3in]{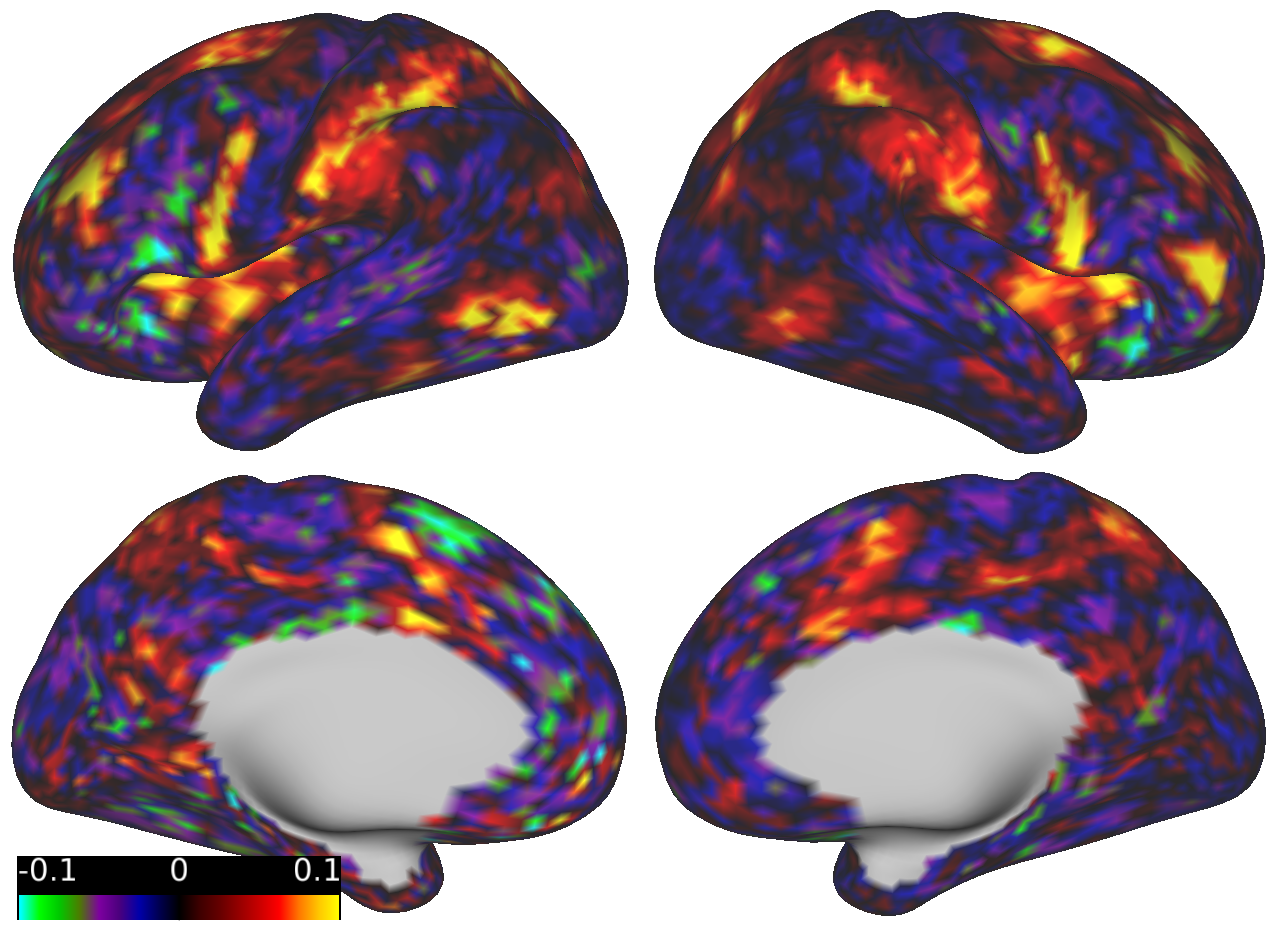}} \\[4pt]
    \begin{picture}(0,90)\put(-5,45){\rotatebox[origin=c]{90}{Marginal SD}}\end{picture} & 
    \fbox{\includegraphics[height=1.3in]{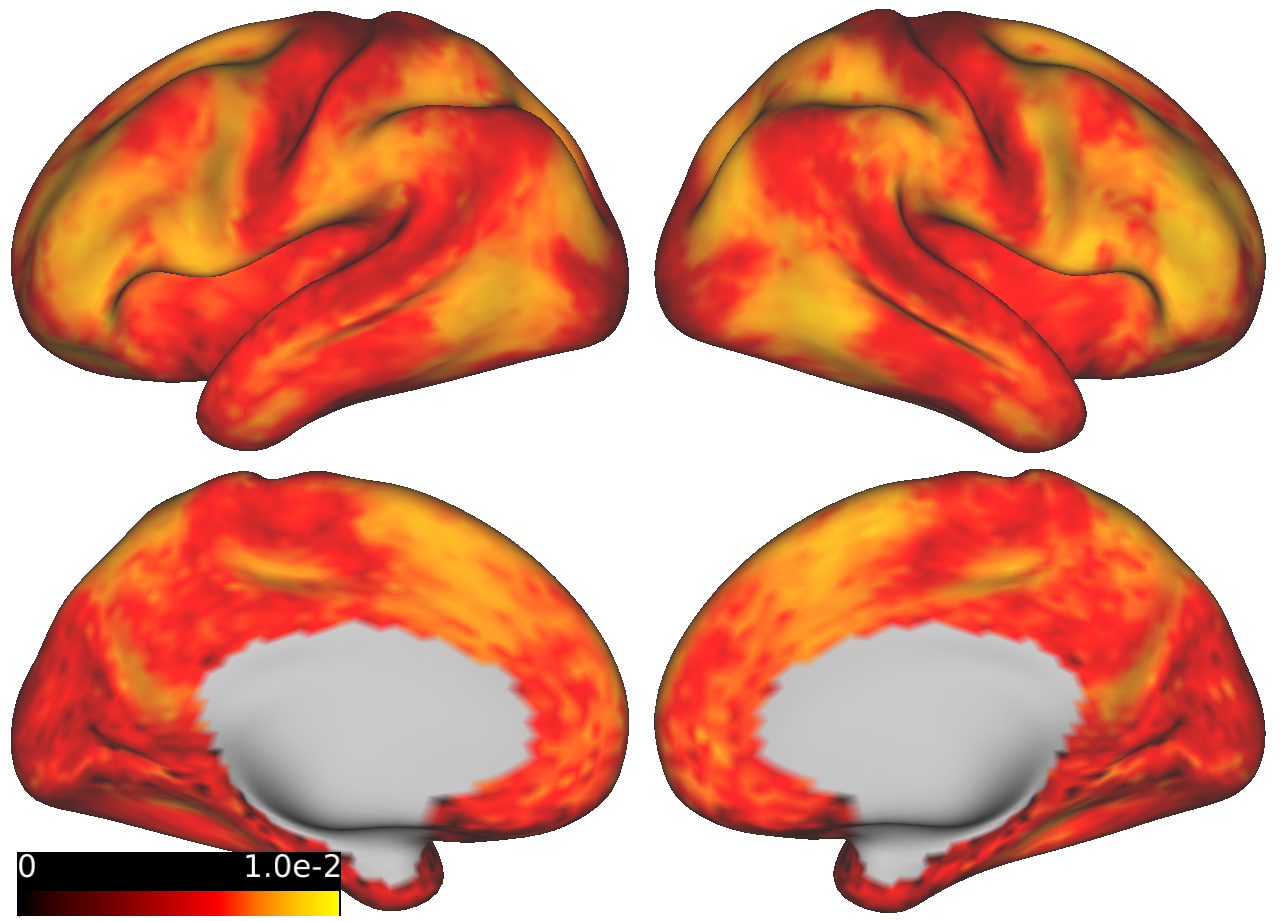}} &
    \fbox{\includegraphics[height=1.3in]{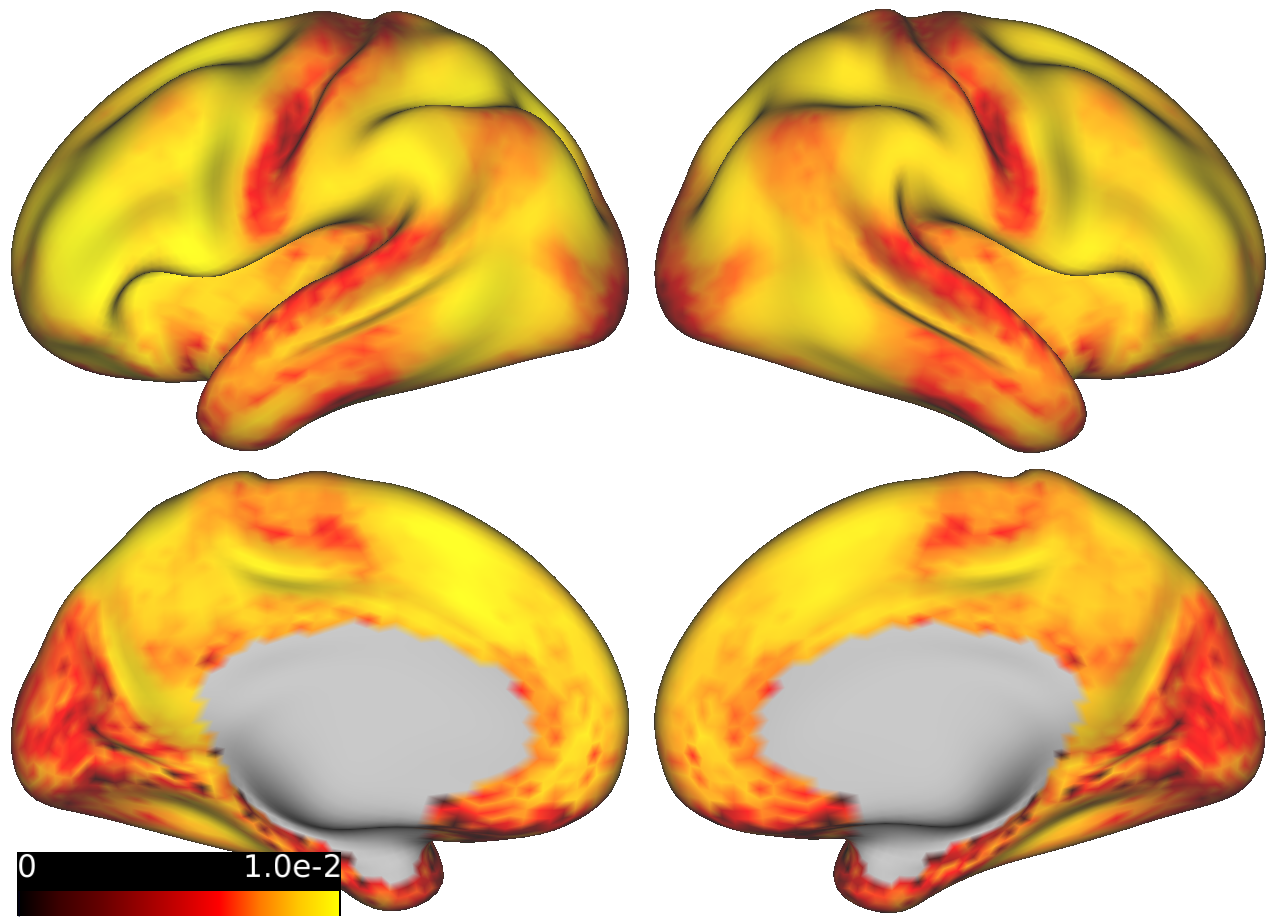}} &
    \fbox{\includegraphics[height=1.3in]{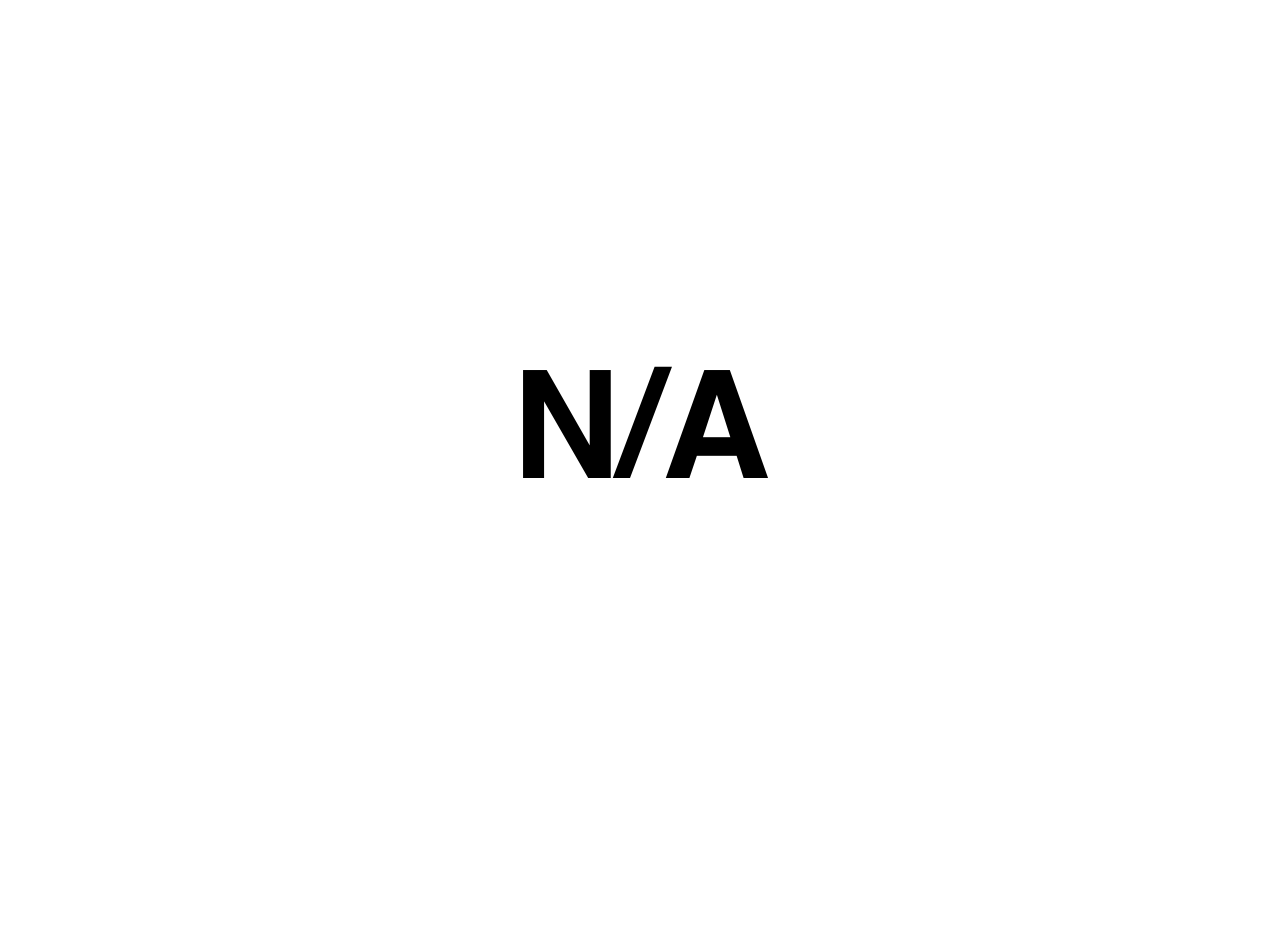}} \\[4pt]
    \begin{picture}(0,90)\put(-5,45){\rotatebox[origin=c]{90}{Areas of Engagement}}\end{picture} & 
    \fbox{\includegraphics[height=1.3in]{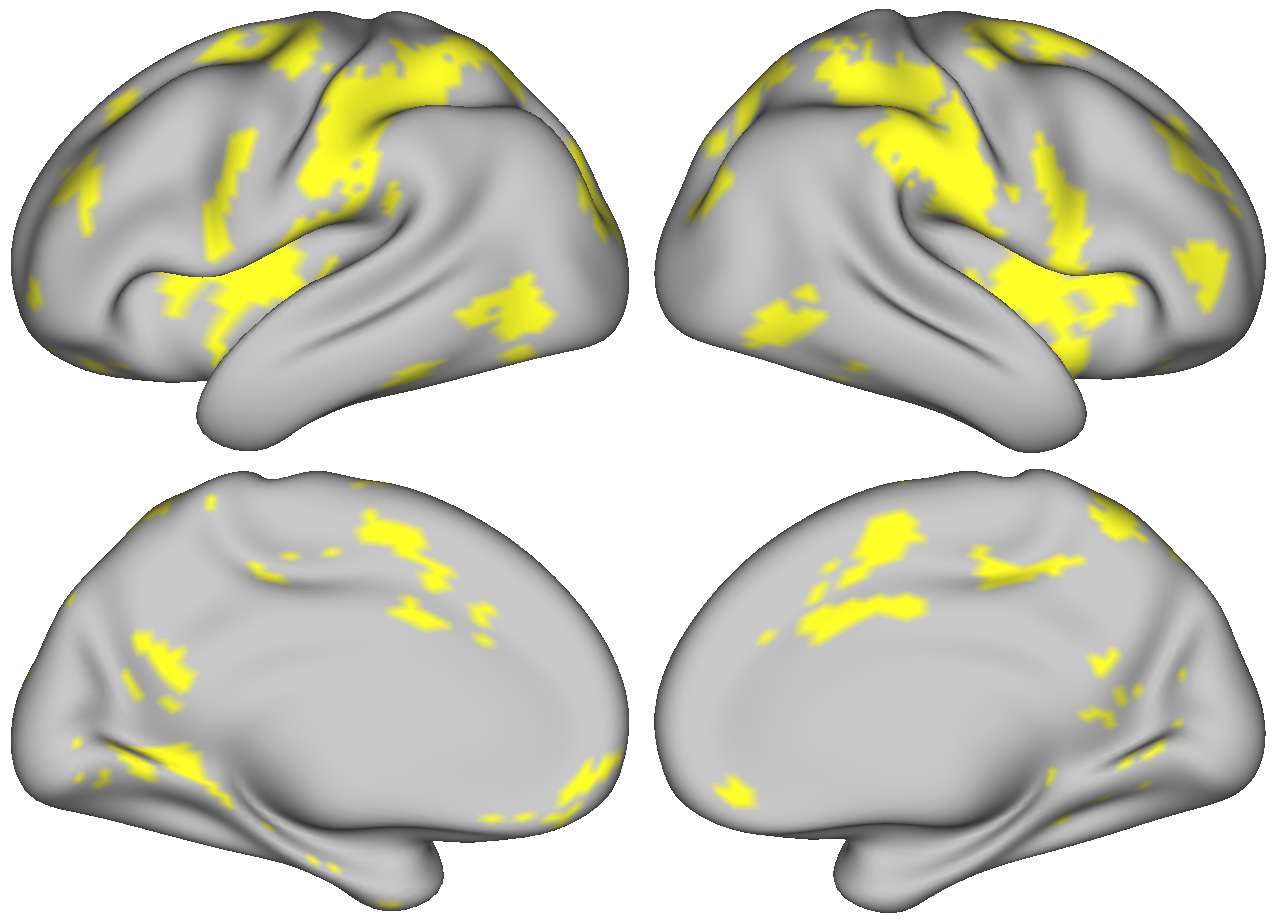}} &
    \fbox{\includegraphics[height=1.3in]{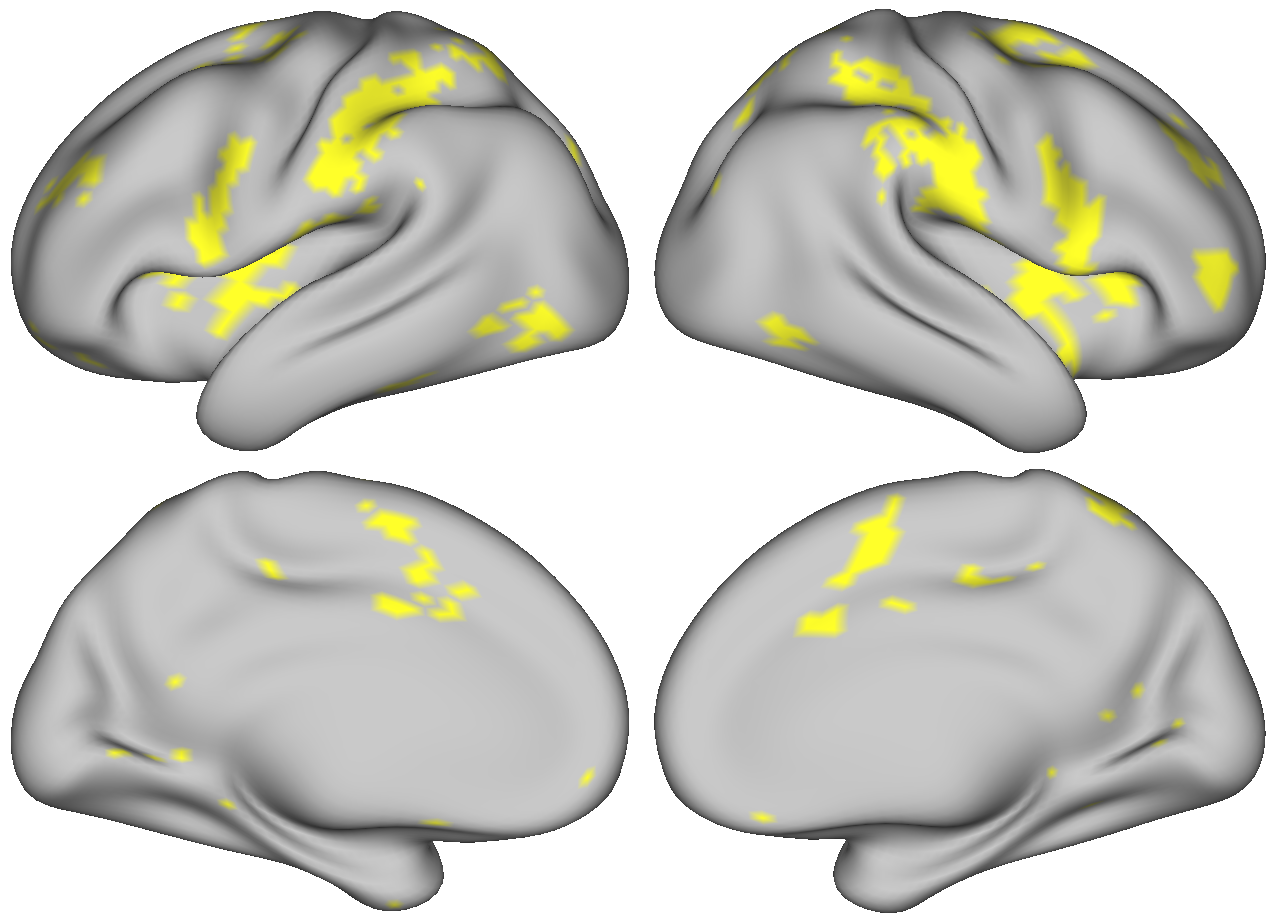}} &
    \fbox{\includegraphics[height=1.3in]{analysis/images/visit1_1200/DR_645450_IC10_SD.png}} \\[4pt]
    \end{tabular}
    \caption{\small Estimates, marginal standard deviations (SD) and areas of engagement for IC 10, an attention network. stICA areas of engagement are based on the excursions set approach and the joint posterior distribution across all vertex locations, as described in Section \ref{sec:excursions}.  tICA areas of engagement are based on performing a $t$-test at each location and correcting for multiple comparisons using Bonferroni correction to control the FWER.  An engagement threshold of $\gamma=0$ and a significance level of $\alpha=0.01$ was used for each procedure.  The areas of engagement based on stICA are larger than those based on tICA (both provide stringent false positive control), due to the smaller variances shown in the second row and the incorporation of spatial dependencies.}
    \label{fig:app_estimates}
\end{figure}

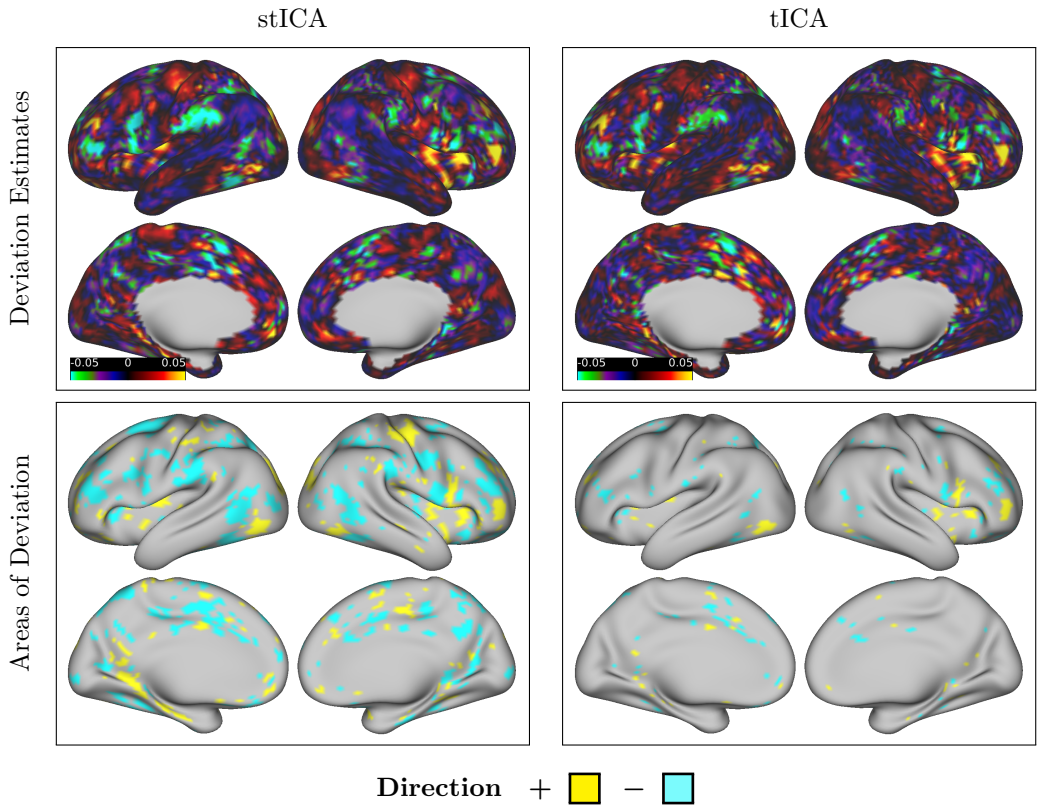
\begin{figure}
    \centering
   \begin{tabular}{ccc}
    & stICA  & tICA   \\[4pt]
    \begin{picture}(0,125)\put(-5,62){\rotatebox[origin=c]{90}{Deviation Estimates}}\end{picture} & 
    \fbox{\includegraphics[height=1.7in]{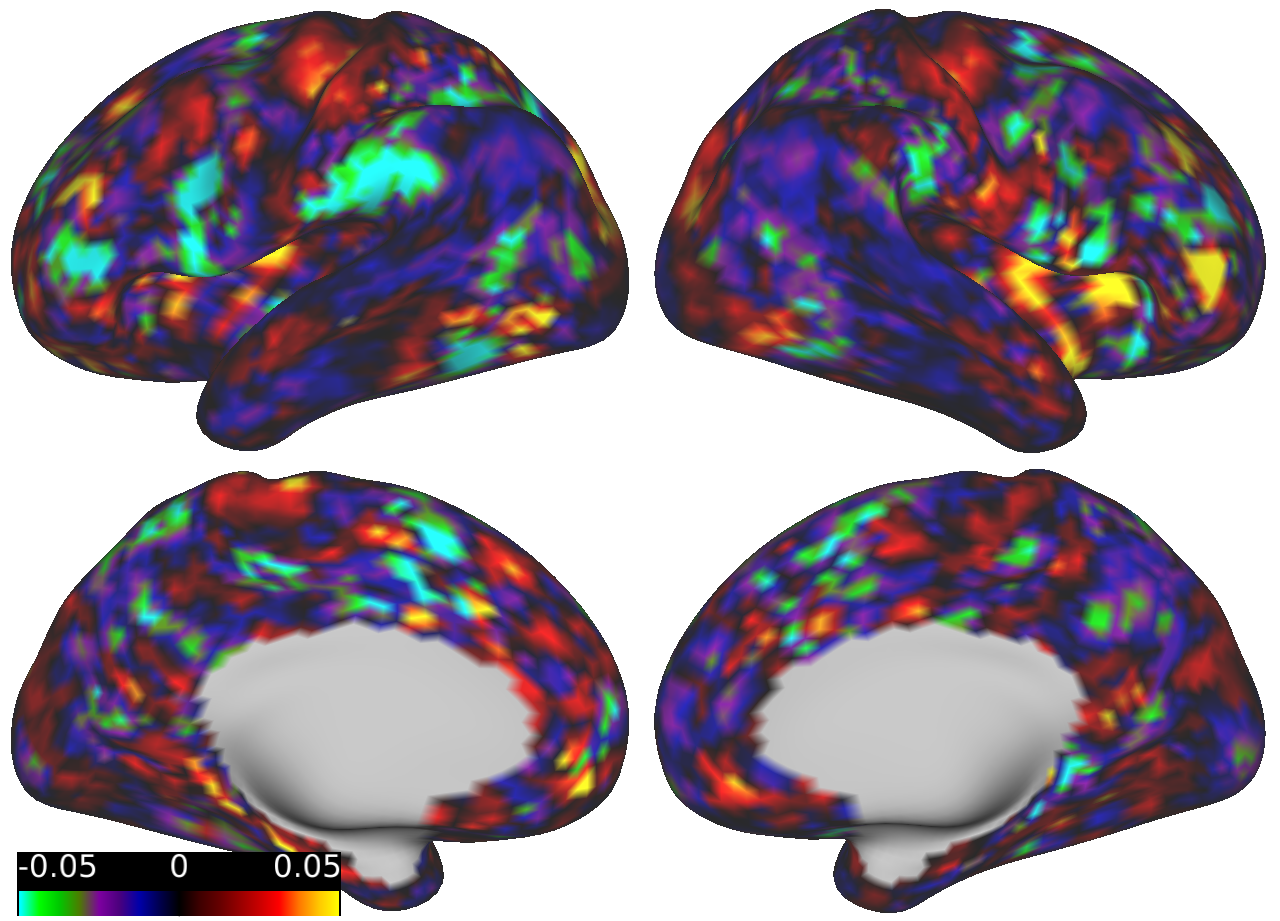}} &
    \fbox{\includegraphics[height=1.7in]{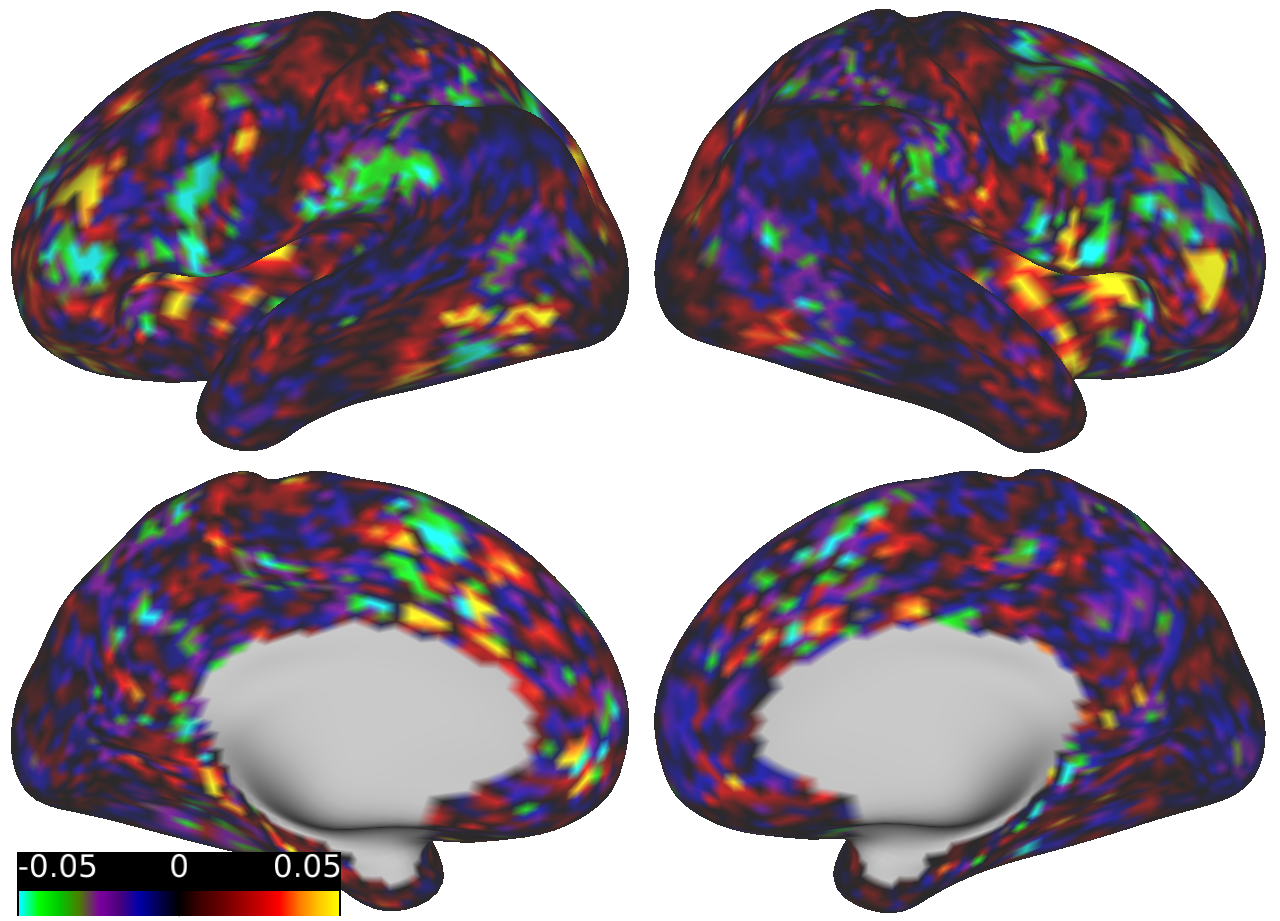}}  \\[4pt]
     \begin{picture}(0,125)\put(-5,62){\rotatebox[origin=c]{90}{Areas of Deviation}}\end{picture} & 
    \fbox{\includegraphics[height=1.7in]{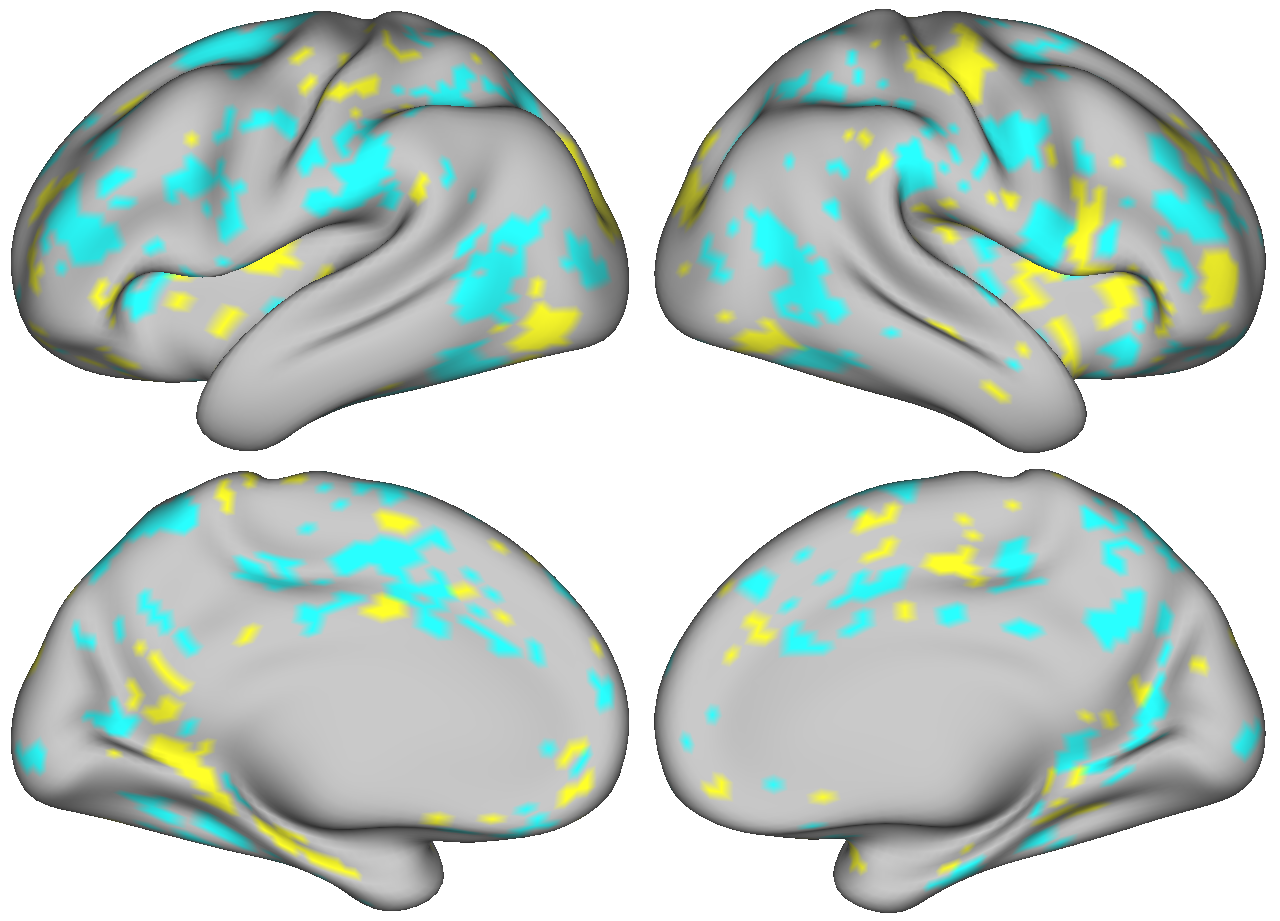}} &
    \fbox{\includegraphics[height=1.7in]{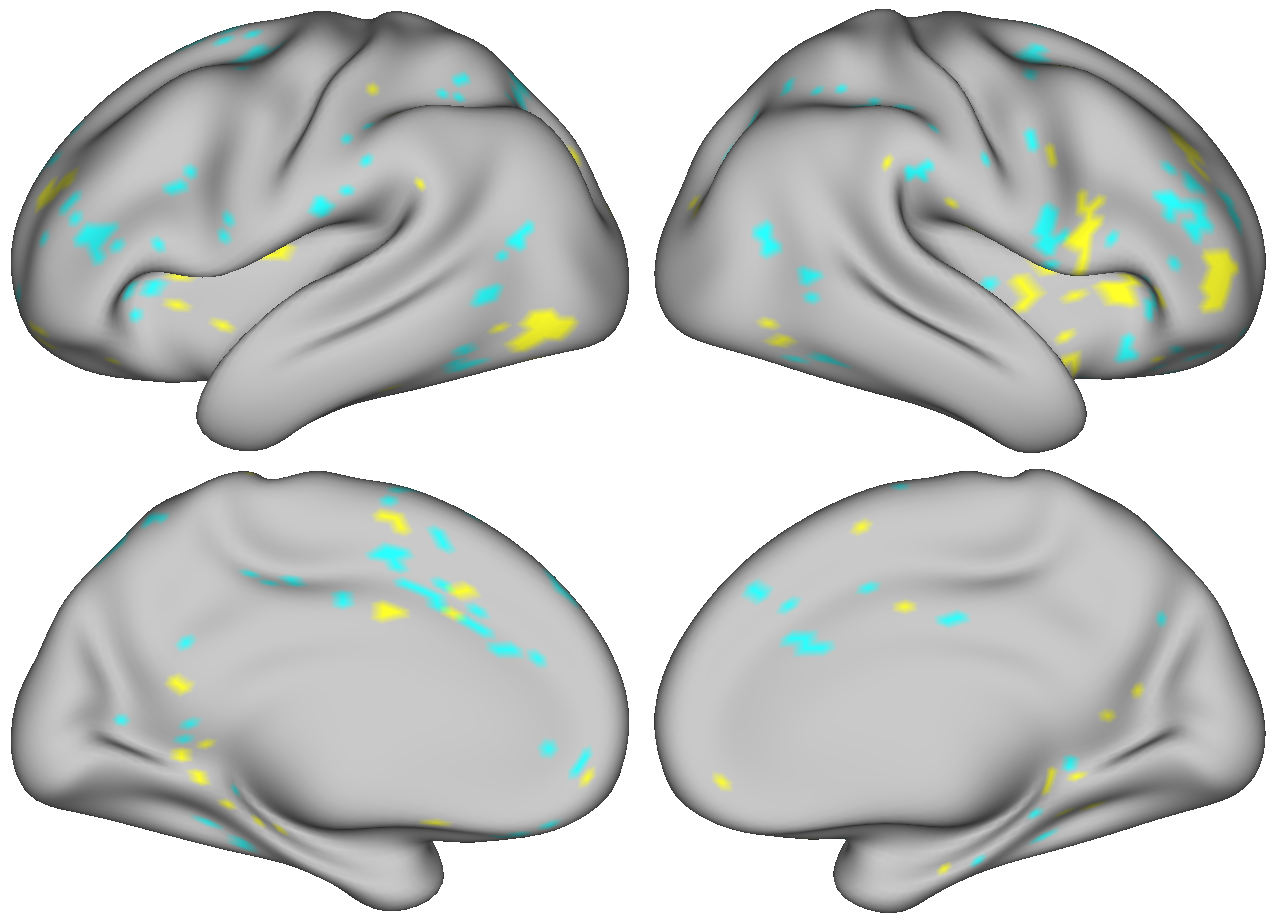}} \\[4pt]
    &  \multicolumn{2}{c}{ \textbf{Direction}\actlegendapplication }  \\[4pt]
    \end{tabular}    
    \caption{\textbf{Estimates and areas of deviation.} Deviations represent differences between the subject IC and the template mean.  Areas of deviation are determined in stICA through the joint posterior distribution of the deviations, based on the excursions set approach described in Section \ref{sec:excursions}.  In tICA, they are determined using the marginal posterior variance of the deviations at each location, followed by multiple comparisons correction using Bonferroni correction to control the FWER.  For each procedure, a threshold of $\gamma=0$ and a significance level of $\alpha=0.01$ was used.  The areas of deviation identified using stICA are larger and include more subtle deviations, whereas tICA is able to identify the most intense deviations seen in yellow (positive) and turquoise (negative).}
    \label{fig:app_deviations}
\end{figure}

Figures \ref{fig:app_estimates} and \ref{fig:app_deviations} shows results using stICA, tICA and dual regression for IC 10, an attention network.  Figure \ref{fig:app_estimates} displays estimates, marginal standard deviations (SD) and areas of engagement for the ICs.  Figure \ref{fig:app_deviations} displays estimates and areas of deviation for the subject effects.  In both figures, the estimates produced using stICA are noticeably smoother than those produced using tICA. Dual regression does not produce deviations or variance estimates, so the only mode of comparison is the IC estimates themselves, which appear noisier than those produced using either tICA or stICA.  Comparing the areas of engagement and deviation, those based on stICA appear larger and include more subtle deviations, whereas tICA is able to identify the most intense deviations seen in yellow (positive) and turquoise (negative).  As both stICA and tICA provide similar control over false positives, the difference in power is due to both the smaller marginal variance estimates, shown in the second row of Figure \ref{fig:app_estimates}, and accounting for spatial dependencies. 

\begin{figure}
    \centering
    \includegraphics[width=3in, trim = 2mm 5mm 0 10mm, clip]{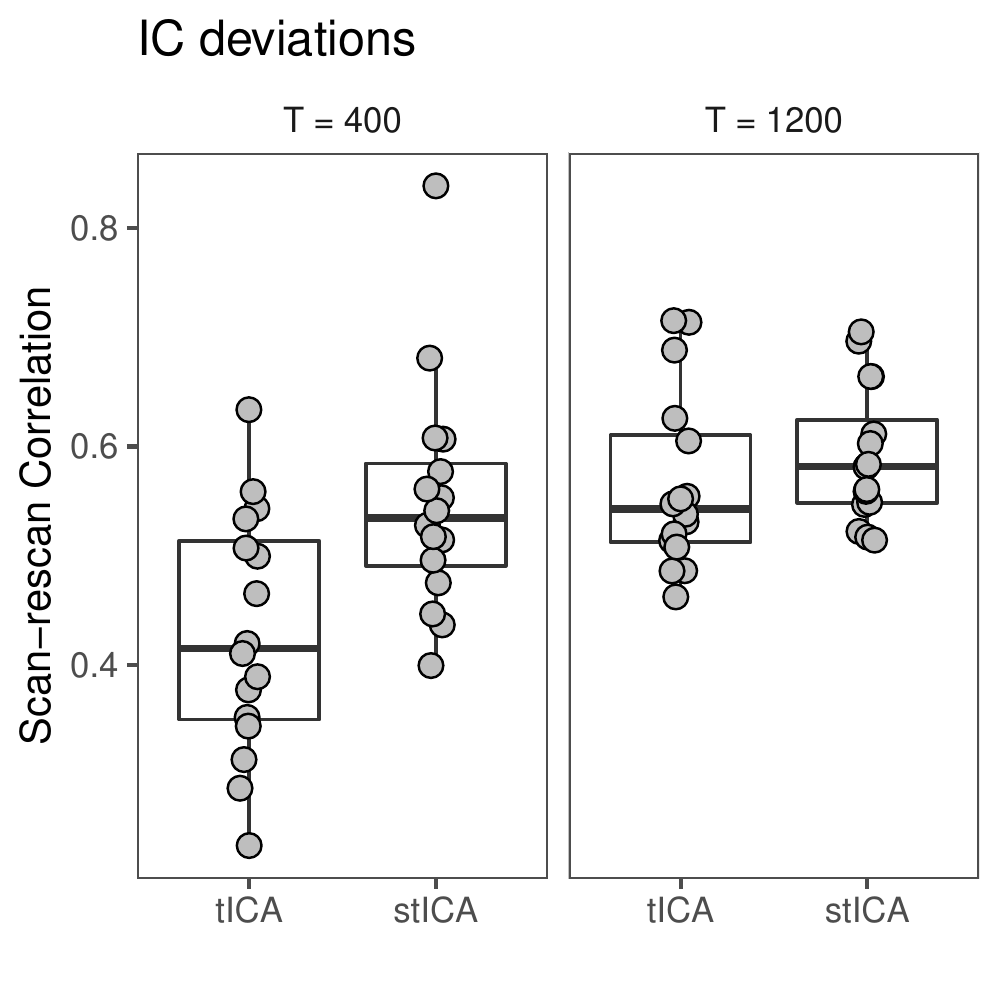}
    \caption{\textbf{Scan-rescan reliability of deviation estimates.} We assess the reliability of the estimates of deviations.  Deviations or subject effects represent differences between the subject-level ICs and the group average ICs.  We consider the effect of sample size by varying scan duration from $T=400$ volumes (5 minutes) to $T=1200$ volumes (15 minutes). Each point represents one of the 16 ICs, and boxplots represent the distribution over all ICs.  For shorter scans, stICA shows substantially more reliable deviation estimates than tICA. As the scan duration increases, the methods begin to converge, with stICA showing a slight improvement over tICA. This illustrates that the spatial priors improve accuracy of estimates the most when sample size is limited.  If more data is available, tICA may be sufficient to produce reliable estimates of subject effects.}
    \label{fig:app_reliability}
\end{figure}

\begin{figure}
    \centering
    \begin{subfigure}[b]{0.48\textwidth}
    \includegraphics[width=3in]{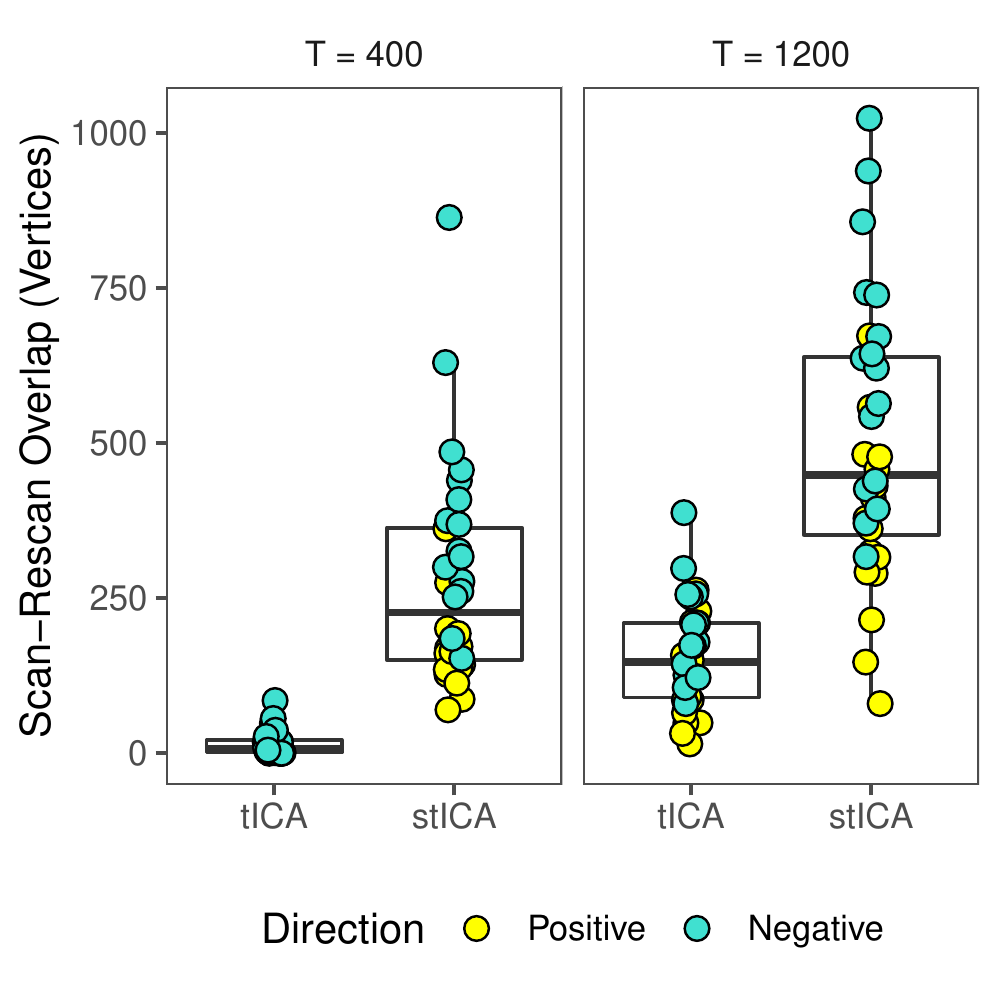}
    \caption{Size of overlap.}
    \end{subfigure}
    \begin{subfigure}[b]{0.48\textwidth}
    \includegraphics[width=3in]{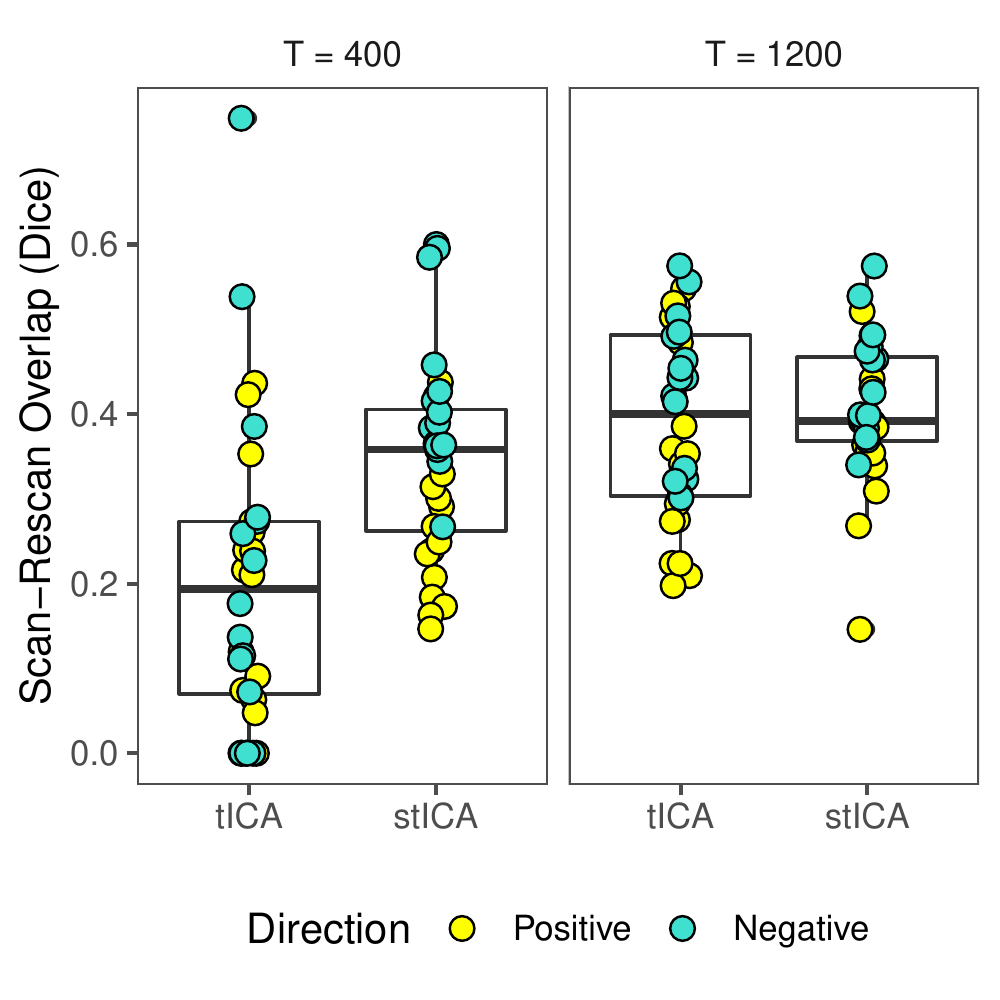}
    \caption{Dice coefficient of overlap.}
    \end{subfigure}
    \caption{\textbf{Reliability of areas of deviation.} We assess the scan-rescan reliability of the areas of positive and negative deviation identified with stICA and tICA. Deviations or subject effects represent differences between the subject-level ICs and the group average ICs.  Reliability of the areas of deviation across the two sessions is quantified in two ways: the number of overlapping vertices (a), and the Dice coefficient of the overlap, which ranges from 0 (no overlap) to 1 (perfect overlap) (b).  Each point represents the positive or negative deviations associated with one of the 16 ICs, and boxplots represent the distribution over all ICs in both directions. We consider the effect of sample size by varying scan duration from $T=400$ volumes (5 minutes) to $T=1200$ volumes (15 minutes).  As seen in panel (a), stICA identifies substantially larger overlapping areas of positive and negative deviation, regardless of scan duration. This reflects the fact that stICA has more power to detect true deviations than tICA.  Considering panel (b), first note that the denominator of the Dice coefficient is the average size of the two areas being considered; hence it becomes more difficult to achieve high Dice overlap as areas become larger to include more subtle deviations. Yet for short scan duration, stICA achieves greater Dice overlap, even while the areas of deviation are much larger than those identified with tICA.  For longer scan duration, the two methods appear to converge in terms of Dice overlap, although this is a greater achievement for stICA since the areas being identified are larger (a).}
    \label{fig:app_overlap}
\end{figure}

Figures \ref{fig:app_reliability} and \ref{fig:app_overlap} show the reliability of subject effects produced using stICA and tICA.  Figure \ref{fig:app_reliability} displays the correlation across sessions of the subject effect estimates produced by stICA and tICA, by scan duration.  Each point represents one of the 16 ICs, and boxplots display the distribution over all ICs.  For shorter scans, stICA shows substantially more reliable subject effect estimates than tICA. As the scan duration increases, the methods begin to converge, with stICA showing a slight improvement over tICA. This illustrates that the use of spatial priors improve accuracy of estimates the most when sample size is limited.  If more data is available, template ICA without spatial priors may be sufficient to produce reliable estimates of subject effects.  Figure \ref{fig:app_overlap} displays the overlap across sessions of the areas of positive and negative deviation produced by tICA and stICA, by scan duration.  Each point represents a single IC and direction of deviation (positive or negative), and boxplots represent the distribution over all ICs in both directions.  Figure \ref{fig:app_overlap}(a) illustrates that stICA identifies substantially larger reliable areas of positive and negative deviation, regardless of scan duration. This suggests that stICA has more power to detect true deviations than tICA, as observed in the simulation study.  Considering Figure \ref{fig:app_overlap}(b), first note that the denominator of the Dice coefficient is the average size of the two areas being considered; hence it becomes more difficult to achieve high Dice overlap as areas become larger to include more subtle deviations. Yet for short scan duration, stICA achieves greater Dice overlap, even while identifying much larger areas of deviation than tICA.  For longer scan duration, the two methods appear to converge in terms of Dice overlap, although this is a greater achievement for stICA since the areas of deviation are larger as seen in Figure 16(a).

\section{Discussion}
\label{sec:discussion}

In this paper, we have proposed a spatial template independent component analysis (stICA) model for subject-level brain network estimation with fMRI data.  Estimation can be performed through a computationally efficient expectation-maximization (EM) algorithm.  Joint posterior inference is possible using an efficient excursions set approach, which avoids the multiple comparisons problem. We have validated stICA through both simulation studies and a real fMRI data analysis. 

The proposed stICA approach has several benefits. First, accounting for spatial dependencies results in greater estimation efficiency and power to detect true effects, as demonstrated in the simulation studies and fMRI data analysis. Second, stICA can be used to directly investigate ``subject effects'', which represent differences between a subject brain network and the population average.  These subject effects may be used to investigate any number of phenomena, including effects of disease progression and treatment, typical and atypical development, aging and cognitive decline, and changes in brain state.  Such investigation of subject effects is not possible with commonly used ad-hoc approaches like dual regression.  Third, stICA can produce thresholded maps in a principled manner, which is often of interest in practice. Finally, the proposed EM algorithm is quite computationally feasible.

In our simulation studies and real data analysis, we have compared the performance of stICA with two existing approaches: template ICA (tICA) without spatial priors and dual regression. Both stICA and tICA clearly outperform dual regression, but the difference in performance between tICA and stICA is more nuanced.  In the simulation studies and fMRI analysis with shorter scan duration (approximately 5 minutes), stICA outperforms tICA in all measures.  But in the fMRI analysis with longer scan duration (approximately 15 minutes), stICA and tICA show somewhat similar performance in terms of producing reliable subject effects and areas of deviation, though stICA still exhibits greater power to detect effects.  This suggests that the use of spatial priors may be most helpful when data quantity is limited, while tICA may be sufficient if more data is available.  This is convenient, since the computational demands of stICA grow with scan duration and may become very burdensome for very long scans, while tICA is quite fast even for large datasets.  If, however, one wants to maximize power to detect true effects, stICA should always be used, since by accounting for spatial dependencies it has greater power to detect engagements and deviations, regardless of scan duration. 

Several limitations should also be noted. First, the proposed EM algorithm is an empirical Bayes approach and does not account for uncertainty in the unknown model parameters. This could result in underestimation of the posterior variances and lead to inflated false positive rates when identifying areas of engagement or deviation. While we do not observe any evidence of this in our simulation studies, it is possible that it may be a more salient issue in challenging settings where parameter estimates may be less accurate.  A possible future topic of research is to develop a fully Bayesian approach with priors on the model parameters to better account for uncertainty. A second limitation of the stICA model is that it is fit within each hemisphere separately.  A unified model for both hemispheres, while more computationally demanding, would improve the ability to estimate shared model parameters.  Finally, here we apply the stICA model to data on the cortical surface; we do not consider subcortical and cerebellar volumetric gray matter, which may also be of interest in some applications.  While it is possible in theory to construct a mesh for each subcortical region and apply the stICA model, the resulting number of vertices may be large in some regions. Therefore, model estimation may be computationally challenging unless the model is stratified across regions.  A whole-brain spatial modeling approach incorporating both cortical surfaces and all subcortical grey matter regions is an important topic for future research.


\bibliography{main}
\bibliographystyle{apalike}

\appendix

\section{Computation of $\bfR_\ell^{-1}$}
\label{app:inverse}

Assume without loss of generality that the $N$ mesh locations are ordered so that the $V$ original data locations appear first, followed by the additional locations added to form the mesh.  Then (suppressing the $\ell$ notation for simplicity), the SPDE covariance $\bfQ^{-1}$ can be written as a block matrix 
$$
\bfQ^{-1} = \begin{bmatrix} (\bfQ^{-1})_{11}  & (\bfQ^{-1})_{12} \\
(\bfQ^{-1})_{12}' & (\bfQ^{-1})_{22} 
\end{bmatrix},
$$
where $(\bfQ^{-1})_{11} = \bfA\bfQ^{-1}\bfA'$.  Therefore, the inverse spatial correlation of the data locations is $\bfR_\ell^{-1} = [(\bfQ^{-1})_{11}]^{-1}$.  Note that $\bfQ$ is a known and sparse matrix, and adopt the same block matrix notation for $\bfQ$.  Note that the $(\bfQ^{-1})_{11}$ can be written as $(\bfQ_{11}-\bfQ_{12}\bfQ_{22}^{-1}\bfQ_{12}')^{-1}$.  Then, its inverse is simply $\bfR_\ell^{-1} = [(\bfQ^{-1})_{11}]^{-1}=\bfQ_{11}-\bfQ_{12}\bfQ_{22}^{-1}\bfQ_{12}'$. $\bfQ_{22}^{-1}$ can be computed through Cholesky factorization of $\bfQ_{22}$, a sparse matrix, which will be efficient as long as $N-V$, the number of additional mesh locations, is reasonably small.

\section{Computation of $\hat\bfM$}
\label{app:Mhat}

Recall that the MLE of the mixing matrix $\bfM$ is given as in equation (\ref{eqn:Mhat}) by 
$$
{\hat\bfM} = \Big(\sum_{v=1}^V\bfy(v)\bft(v)'\Big)\Big(\sum_{v=1}^V \bfT(v,v) \Big)^{-1},
$$ 
where $\bft=\bfP E[\bfs|\bfy,\Thetahatk]$, $\bfT=\bfP E[\bfs\bfs'|\bfy,\Thetahatk]\bfP'$, $\bft(v)$ is the $v$th block of size $L$ of $\bft$, and $\bfT(v,v)$ is the $v$th diagonal $L\times L$ block of $\bfT$. The first term can be computed based on the value of $E[\bfs|\bfy,\Thetahatk]=\bfmu_{s|y}$, computed as described in Section \ref{sec:E_step}.  For computation of the second term involving $E[\bfs\bfs'|\bfy,\Thetahatk]$, we can rewrite 
$$
\bfT=\bfP\bfSigma_{s|y}\bfP' + \bfP E\left[\bfs|\bfy,\Thetahatk\right]E\left[\bfs|\bfy,\Thetahatk\right]'\bfP'=\bfP\bfSigma_{s|y}\bfP'+\bft\bft'=:\bfT_1+\bfT_2
$$  
and
$$
\bfT_1 = \bfP\bfD\bfOmega^{-1}\bfD\bfP'=\bfP\bfD\bfP'\bfP\bfOmega^{-1}\bfP'\bfP\bfD\bfP'=:\bfD_{PP}(\bfOmega_{PP})^{-1}\bfD_{PP},
$$
where $\bfD_{PP}$ is a diagonal matrix and $\bfOmega_{PP}=\bfP\bfOmega\bfP'$. Therefore, $\bfT(v,v)$ can be computed as 
$$
\bfT(v,v) = \bfT_1(v,v)+\bfT_2(v,v) =\bfD_{PP}(v,v)\bfOmega_{PP}^{-1}(v,v)\bfD_{PP}(v,v) + \bft(v)\bft(v)'.
$$ 

Therefore, only a small number of the elements of $\bfOmega_{PP}^{-1}$ are needed, namely the $V$ diagonal blocks of size $L\times L$.  We therefore compute only the elements of $\bfOmega_{PP}^{-1}$ corresponding to the non-sparse elements of $\bfOmega_{PP}$ and the additional required elements.  This can be done efficiently using the Takahashi equations \citep{takahashi1973formation, erisman1975computing, rue2007approximate} implemented in the \texttt{inla.qinv} function in \texttt{R-INLA} \citep{lindgren2015bayesian} with the \texttt{pardiso} parallel computation library \citep{schenk2004solving}. 

\section{Computation of $\hat\kappa_\ell$}
\label{app:kappa-hat}

In the M-step of the EM algorithm, to obtain the MLE of $\kappa_\ell$ we need to maximize $f_\ell(\kappa_\ell|\Thetahatk)$, defined in equation (\ref{eqn:opt_kappa_l}).  This involves computing $\bfmu_{s_\ell|y}$ and two trace terms, specifically $\text{Tr}(\bfR_\ell^{-1}(\hat\bfOmega^{-1})_{\ell,\ell})$ and $\text{Tr}(\bfR_\ell^{-1}\hat\bfW_{\ell,\ell})$, where $\bfR_\ell$ is sparse, and its sparsity pattern does not depend on $\kappa_\ell$.  Recall that $\hat\bfW=\hat\bfOmega^{-1}\hat\bfm\hat\bfm'\hat\bfOmega^{-1}$, and $\hat\bfOmega^{-1}\hat\bfm$ was computed in the E-step.  Therefore, we can easily compute $\hat\bfW_{\ell,\ell}=[\hat\bfOmega^{-1}\hat\bfm]_\ell [\hat\bfOmega^{-1}\hat\bfm]_\ell'$. 

Recall that $\bfOmega=\bfR^{-1}+\bfD\bfP'\bfM_{\otimes}'(\nu_0^2\bfC_{\otimes})^{-1}\bfM_{\otimes}\bfP\bfD$, $\hat\bfOmega$ is its value given the parameter estimates $\Thetahatk$, and $(\hat\bfOmega^{-1})_{\ell,\ell}$ is the $\ell$th diagonal block of size $V\times V$ of $\hat\bfOmega^{-1}$.  Following the same strategy outlined in Appendix \ref{app:Mhat}, we can avoid computing the large number of off-block diagonal elements of $\hat\bfOmega^{-1}$.  Namely, we use the \texttt{inla.qinv} function with the \texttt{pardiso} parallel computation library to compute only the elements of $\hat\bfOmega^{-1}$ corresponding to the non-sparse locations in $\hat\bfOmega$, plus the $V\times V$ block diagonal elements.  

We can simplify computation further, because only the diagonal elements of $\bfR_\ell^{-1}(\hat\bfOmega^{-1}){\ell,\ell}$ and $\bfR_\ell^{-1}\hat\bfW_{\ell,\ell}$ are needed for the traces. As $\bfR_\ell^{-1}$ is sparse and symmetric, the diagonal elements are equivalent to the column sums of the element-wise product of $\bfR_\ell^{-1}$ with $(\hat\bfOmega^{-1})_{\ell,\ell}$ and $\hat\bfW_{\ell,\ell}$, respectively.  Therefore, the only elements of $(\hat\bfOmega^{-1})_{\ell,\ell}$ and $\hat\bfW_{\ell,\ell}$ that we need to estimate are those corresponding to the non-zero entries of $\bfR_\ell^{-1}$, which is identical for all $\ell=1,\dots,L$.

\section{Choosing initial values of $\kappa$}
\label{app:initial}

Given an initial estimate of an IC, e.g. from dual regression or standard template ICA, we can compute the MLE of $\kappa_\ell$ to use as an initial value for $\kappa_\ell$ in the EM algorithm.  Given an estimate $\hat\bfdelta_\ell$, we consider the model 
$$
\hat\bfdelta_\ell = \bfdelta_\ell + \bfe_\ell = \bfD_\ell\bfA\bfx_\ell + \bfe_\ell,\quad \bfe_\ell\sim N(\bfzero, \sigma^2\bfI_V)
$$
$$
\bfx_\ell\sim N(\bfzero, \bfQ_\ell^{-1}),
$$
where $\bfA$ ($V\times N$) is the mesh projection matrix, $\bfD_\ell$ is known and $\bfQ_\ell=c_1(\kappa_\ell^2\bfF + 2\bfG + \kappa_\ell^{-2}\bfG\bfF^{-1}\bfG)$ (see Section 1.1).  Then, we can assume a prior directly on $\bfdelta_\ell$: 
$$
\bfdelta_\ell \sim N(\bfzero, \bfD_\ell\bfA\bfQ_\ell^{-1}\bfA'\bfD_\ell = \bfD_\ell\bfR_\ell\bfD_\ell),
$$
where $\bfR_\ell$ can be computed as described in Appendix \ref{app:inverse}. The posterior of $\bfdelta_\ell$ is given by 
\begin{align*}
\bfdelta_\ell|\hat\bfdelta_\ell &\sim N(\bfmu_\ell,\bfOmega_\ell^{-1}), \\ 
\bfmu_\ell&=\sigma^{-2}\bfOmega_\ell^{-1}\hat\bfdelta_\ell \\ 
\bfOmega_\ell&=(\bfD_\ell\bfR_\ell\bfD_\ell)^{-1}+\sigma^{-2} \\ 
&= \bfD_\ell^{-1}\left[\bfR_\ell^{-1}+ \sigma^{-2}\bfD_\ell^2\right]\bfD_\ell^{-1} \\
&=: \bfD_\ell^{-1}\bfK_\ell\bfD_\ell^{-1},
\end{align*}
where $\bfK_\ell=\bfR_\ell^{-1}+ \sigma^{-2}\bfD_\ell^2$. Hence, $\bfmu_\ell=\sigma^{-2}\bfD_\ell\bfK_\ell^{-1}\bfD_\ell\hat\bfdelta_\ell$. Note that $\bfK_\ell=\bfR_\ell^{-1}+ \sigma^{-2}\bfD_\ell^2$ is a sparse matrix and does not involve $\bfD_\ell^{-1}$, which may be computationally unstable due to template variance values close to or equalling zero.

The marginal data log-likelihood is proportional to
\begin{align*}
-&\log|\bfOmega_\ell|-\log|\bfD_\ell\bfR_\ell\bfD_\ell|-V\log(\sigma^2)-
\left\{\bfmu_\ell'(\bfD_\ell\bfR_\ell\bfD_\ell)^{-1}\bfmu_\ell + \sigma^{-2}\|\hat\bfdelta_\ell-\bfmu_\ell\|^2\right\} \\
\propto &-\log|\bfK_\ell|-\log|\bfR_\ell|-V\log(\sigma^2)-
\left\{\bfmu_\ell'(\bfD_\ell\bfR_\ell\bfD_\ell)^{-1}\bfmu_\ell + \sigma^{-2}\|\hat\bfdelta_\ell-\bfmu_\ell\|^2\right\} \\
= &-\log|\bfK_\ell|+\log|\bfR_\ell^{-1}|-V\log(\sigma^2)-
\left\{\sigma^{-4}\hat\bfdelta_\ell'\bfD_\ell\bfK_\ell^{-1}\bfR_\ell^{-1}\bfK_\ell^{-1}\bfD_\ell\hat\bfdelta_\ell+ \sigma^{-2}\|\hat\bfdelta_\ell-\sigma^{-2}\bfD_\ell\bfK_\ell^{-1}\bfD_\ell\hat\bfdelta_\ell\|^2\right\} \\
= &-\log|\bfK_\ell|+\log|\bfR_\ell^{-1}|-V\log(\sigma^2)-
\left\{\sigma^{-4}\hat\bfdelta_\ell'\bfD_\ell\bfK_\ell^{-1}\bfR_\ell^{-1}\bfK_\ell^{-1}\bfD_\ell\hat\bfdelta_\ell+ \sigma^{-2}\hat\bfdelta_\ell'(\bfI-\sigma^{-2}\bfD_\ell\bfK_\ell^{-1}\bfD_\ell)^2\hat\bfdelta_\ell\right\} \\
= &-\log|\bfK_\ell|+\log|\bfR_\ell^{-1}|-V\log(\sigma^2)-
\left\{\sigma^{-2}\hat\bfdelta_\ell'\left[\sigma^{-2}\bfD_\ell\bfK_\ell^{-1}\bfR_\ell^{-1}\bfK_\ell^{-1}\bfD_\ell+ (\bfI-\sigma^{-2}\bfD_\ell\bfK_\ell^{-1}\bfD_\ell)^2\right]\hat\bfdelta_\ell\right\} \\
= &-\log|\bfK_\ell|+\log|\bfR_\ell^{-1}|-V\log(\sigma^2)-
\left\{\sigma^{-2}\hat\bfdelta_\ell'\bfE_\ell\hat\bfdelta_\ell\right\},
\end{align*}
where we can simplify
\begin{align*}
\bfE_\ell &= \sigma^{-2}\bfD_\ell\bfK_\ell^{-1}\bfR_\ell^{-1}\bfK_\ell^{-1}\bfD_\ell+ (\bfI-\sigma^{-2}\bfD_\ell\bfK_\ell^{-1}\bfD_\ell)^2 \\
&= \sigma^{-2}\bfD_\ell\bfK_\ell^{-1}\bfR_\ell^{-1}\bfK_\ell^{-1}\bfD_\ell+ \bfI-2\sigma^{-2}\bfD_\ell\bfK_\ell^{-1}\bfD_\ell + \sigma^{-4}\bfD_\ell\bfK_\ell^{-1}\bfD_\ell^2\bfK_\ell^{-1}\bfD_\ell \\
&= \bfI + \sigma^{-2}\bfD_\ell\bfK_\ell^{-1}\left[\bfR_\ell^{-1}-2\bfK_\ell+\sigma^{-2}\bfD_\ell^2\right] \bfK_\ell^{-1}\bfD_\ell\\ 
&= \bfI - \sigma^{-2}\bfD_\ell\bfK_\ell^{-1}\bfD_\ell
\end{align*}

The marginal log-likelihood is therefore given by
$$
-\log|\bfK_\ell|+\log|\bfR_\ell^{-1}|-V\log(\sigma^2)
-\sigma^{-2}\hat\bfdelta_\ell'\hat\bfdelta_\ell 
+ \sigma^{-4}\hat\bfdelta_\ell'\bfD_\ell\bfK_\ell^{-1}\bfD_\ell\hat\bfdelta_\ell,
$$

where ${\bfK_\ell}^{-1}\bfD_\ell\hat\bfdelta_\ell$ can be computed without explicit inversion of $\bfK_\ell$, a $V\times V$ matrix, by solving a system of linear equations.

It is trivial to generalize to the common smoothness case, where $\kappa_\ell=\kappa$: we simply replace $\bfD_\ell$ and $\bfR_\ell$ with $\bfD=\text{diag}\{\bfD_\ell\}$ and $\bfR=\text{diag}\{\bfR_\ell\}$, respectively.  The only other change to the marginal data log-likelihood is that $-V\log(\sigma^2)$ is replaced by $-VL\log(\sigma^2)$. 

\end{document}

%% file: Macros.tex
\def\bfA{\mathbf A}
\def\bfB{\mathbf B}
\def\bfC{\mathbf C}
\def\bfD{\mathbf D}
\def\bfE{\mathbf E}
\def\bfF{\mathbf F}
\def\bfG{\mathbf G}
\def\bfH{\mathbf H}
\def\bfI{\mathbf I}

\def\bfK{\mathbf K}

\def\bfM{\mathbf M}

\def\bfP{\mathbf P}
\def\bfQ{\mathbf Q}
\def\bfR{\mathbf R}

\def\bfT{\mathbf T}
\def\bfU{\mathbf U}

\def\bfW{\mathbf W}

\def\bfY{\mathbf Y}

\def\bfb{\mathbf b}

\def\bfe{\mathbf e}

\def\bfm{\mathbf m}

\def\bfs{\mathbf s}
\def\bft{\mathbf t}
\def\bfu{\mathbf u}
\def\bfv{\mathbf v}

\def\bfx{\mathbf x}
\def\bfy{\mathbf y}

\def\bfdelta  {\bm \delta}

\def\bfmu     {\bm \mu}

\def\bfDelta  {\mathbf \Delta}

\def\bfSigma  {\mathbf \Sigma}

\def\bfOmega  {\mathbf \Omega}

\newcommand{\bfzero}{{\mathbf 0}}

\def\boldfacefake#1{\kern-4pt
    \hbox{ \mathsurround=0pt
    \hbox to 0.4pt{$#1$\hss}\hbox to 0.4pt{$#1$\hss}\hbox {$#1$}}}

\newcommand{\be}{\begin{eqnarray}}
\newcommand{\ee}{\end{eqnarray}}
\newcommand{\ba}{\begin{eqnarray*}}
\newcommand{\ea}{\end{eqnarray*}}
\newcommand{\bc}{\begin{center}}
\newcommand{\ec}{\end{center}}
\newcommand{\btab}[1]{\begin{tabular}{#1}}
\newcommand{\etab}{\end{tabular}}

\newcommand{\reals}{\mbox{\rm I\kern-.20em R}}
\newcommand{\sreals}{\mbox{\small \rm I\kern-.20em R}}